\documentclass[preprint2]{aastex}

\def\kms{km\,s$^{-1}$}

\def\msun{M$_{\odot}$}

\def\Rsun{R$_{\odot}$}

\def\Ks{$K\mathrm{s}$}

\def\Lp{$L'$}

\def\gtrsim{\mathrel{\hbox{\rlap{\hbox{\lower3pt\hbox{$\sim$}}}\hbox{\raise2pt\hbox{$>$}}}}}
\def\lesssim{\mathrel{\hbox{\rlap{\hbox{\lower3pt\hbox{$\sim$}}}\hbox{\raise2pt\hbox{$<$}}}}}
\def\smash{{\sc sm}a{\sc sh}$+$}

\newcommand{\mas}{\ensuremath{\mathrm{\, mas}}}

\newcommand{\IWA}{\ensuremath{\mathrm{IWA}}}
\newcommand{\OWA}{\ensuremath{\mathrm{OWA}}}
\newcommand{\VS}{\ensuremath{V^2}}
\newcommand{\closure}{\ensuremath{\phi}}

\shorttitle{SMASH+: Observational Campaign and Companion Detection}
\shortauthors{Sana et al.}

\begin{document}


\title{Southern Massive Stars at High Angular Resolution:\\
    Observational Campaign and Companion Detection}


\author{H. Sana\altaffilmark{1,2}, 
J.-B. Le Bouquin\altaffilmark{3,4},
S. Lacour\altaffilmark{5},
J.-P. Berger\altaffilmark{6},
G. Duvert\altaffilmark{3,4}, 
L. Gauchet\altaffilmark{5},\\
B. Norris\altaffilmark{7}, 
J. Olofsson\altaffilmark{8}, 
D. Pickel\altaffilmark{5}, 
G. Zins\altaffilmark{3,4},
O. Absil\altaffilmark{9,16}, 
A. de Koter\altaffilmark{10,11},\\ 
K. Kratter\altaffilmark{12,13}, 
O. Schnurr\altaffilmark{14}, 
H. Zinnecker\altaffilmark{15}}

\altaffiltext{1}{European Space Agency / Space Telescope Science Institute, 3700 San Martin Drive, Baltimore, MD 21218, United States}
\altaffiltext{2}{E-mail: hsana@stsci.edu}
\altaffiltext{3}{Univ. Grenoble Alpes, IPAG, F-38000 Grenoble, France}
\altaffiltext{4}{CNRS, IPAG, F-38000 Grenoble, France}
\altaffiltext{5}{LESIA, Observatoire de Paris, CNRS, UPMC, Universit\'e Paris-Diderot, Paris Sciences et Lettres, 5 Place Jules Janssen, 92195 Meudon, France}
\altaffiltext{6}{European Southern Observatory, Schwarzschild-Str. 2, 85748 Garching bei M\"unchen, Germany} 
\altaffiltext{7}{Sydney Institute for Astronomy, Institute for Photonics and Optical Science, School of Physics, University of Sydney, NSW 2006, Australia}
\altaffiltext{8}{Max-Planck-Institut f\"ur Astronomie, Koenigstuhl 17, 69117 Heidelberg, Germany}
\altaffiltext{9}{D\'epartement d'Astrophysique, G\'eophysique et Oc\'eanographie, Universit\'e de Li\`ege, 17 All\'ee du Six Ao\^ut, 4000 Li\`ege, Belgium}
\altaffiltext{10}{Astrophysical Institute Anton Pannekoek, Universiteit van Amsterdam, Science Park 904, 1098XH Amsterdam, The Netherlands}
\altaffiltext{11}{Instituut voor Sterrenkunde, Universiteit Leuven, Celestijnenlaan 200 D, 3001, Leuven, Belgium}
\altaffiltext{12}{Hubble Fellow, JILA, 440 UCB, University of Colorado, Boulder, CO 80309-0440, United States}
\altaffiltext{13}{Steward Observatory / Dept. of Astronomy, University of Arizona, 933 N Cherry Ave, Tucson, AZ, 85721}
\altaffiltext{14}{Leibniz-Institut f\"ur Astrophysik Potsdam, An der Sternwarte 16, 14482 Potsdam, Germany}
\altaffiltext{15}{Deutsches SOFIA Institute, SOFIA Science Center, NASA Ames Research Center, M.S, N232-12, Moffett Field CA 94035, United States}
\altaffiltext{16}{F.R.S.-FNRS Research Associate}

\begin{abstract}
Multiplicity is one of the most fundamental observable  properties of massive O-type stars and offers a promising way to discriminate between massive star formation theories.
Nevertheless, companions at separations between
1 and 100 milli-arcsec (mas) remain mostly unknown due to intrinsic observational limitations. At a typical distance of 2~kpc, this corresponds to projected physical separations of 2-200~AU.
The Southern MAssive Stars at High angular resolution survey (\smash) was designed to fill this gap by providing the first systematic interferometric survey of Galactic massive stars. We observed 117 O-type stars with VLTI/PIONIER and 162 O-type stars with NACO/SAM, respectively probing the separation ranges 1-45 and 30-250~mas and brightness contrasts of $\Delta H < 4$  and $\Delta H < 5$. Taking advantage of NACO's field-of-view, we further uniformly searched for visual companions in an 8\arcsec-radius  down to $\Delta H = 8$. 
This paper describes the observations and  data analysis, reports the discovery of almost 200 new companions in the separation range from 1~mas to 8\arcsec\ and presents the catalog of detections, including the first resolved measurements  of over a dozen known long-period spectroscopic binaries.

Excluding known runaway stars for which no companions are detected, 96 objects in our main sample ($\delta < 0$\degr; $H<7.5$) were observed both with  PIONIER and NACO/SAM. The fraction of these stars with at least one resolved companion within 200~mas is 0.53. Accounting for known but unresolved spectroscopic or eclipsing companions, the multiplicity fraction at separation $\rho < 8$\arcsec\ increases to $f_\mathrm{m}=0.91\pm0.03$.  The fraction of luminosity class V stars that have a bound companion reaches 100\%\ at 30~mas while their average number of physically connected companions within 8\arcsec\ is $f_\mathrm{c}=2.2\pm0.3$. 
This  demonstrates that massive stars form nearly exclusively in multiple systems. The nine non-thermal radio emitters observed by \smash\ are all resolved, including the newly discovered pairs HD\,168112 and CPD$-$47\degr2963.  This lends strong support to the universality of the wind-wind collision scenario to explain the non-thermal emission from O-type stars. 
\end{abstract}


\section{Introduction} \label{intro}

One of the most striking properties of massive stars is their high degree of multiplicity. In clusters and associations, 75\%\ of the O-type objects have at least one companion detected either through spectroscopy or through imaging techniques \citep{MHG09}. This rate has not been corrected for observational biases so that the true multiplicity fraction might very well come close to 100\%. Typically, the detected companions have a mass 1 to 5 times smaller than the primary mass and are mostly O and B stars. We can thus postulate that the typical end product of massive star formation is not a single star but a multiple system, with at least one and possibly several massive companions (e.g. \citealt{KM06,Kru12}).  The properties of the binary population, for example the period and mass ratio distributions, can then serve as a useful diagnostics to discriminate between different formation models.  Different massive star formation theories do indeed have different expectations for multiplicity properties \citep[for recent reviews, see][]{ZiY07, TBC14}. Unfortunately, observations have so far failed to provide a comprehensive view of the O star multiplicity  over the full separation range relevant for massive star formation and evolution, leaving us with a strongly biased view towards tight (physical separation $d < 1$~AU) and wide ($d > 10^3$~AU) companions.

Binary detection through spectroscopy is typically limited to systems with  mass ratios $M_1/M_2$ up to  5 to 15 (for double- and single-lined binaries respectively) and to separations up to a few AU (corresponding to periods of about 1~yr). Most imaging techniques suffer from a brightness contrast vs.\ separation bias \citep{TBR08, SaE11} which limits the detection of moderate brightness companions to separations larger than several 0.1\arcsec\ at best. Separations below 0.1\arcsec\ have most successfully been probed through various flavors of interferometry, such as speckle, aperture masking and long baseline interferometry, although very few observations have been able to probe the regime of highest contrasts ($\Delta \mathrm{mag} > 2$) and closest angular separations \citep[$\rho < 75$~mas; for a review, see][]{SaE11}.

This paper introduces the Southern MAssive Stars at High angular resolution survey (\smash), an interferometric survey of over 100 Galactic O-type stars designed to systematically explore the separation range between 1 and 200~mas. The {\it Sparse Aperture Masking} mode \citep[SAM,][]{LTI11} of NACO at the {\it Very Large Telescope} (VLT) has allowed us to resolve massive binaries with separations in the range of 30-250~mas \citep[e.g.][]{SLLB12}. Angular separations smaller than 30~mas require the use of long baseline interferometry. Until now, it has been impossible to observe a sufficiently large sample because of the low  magnitude limit,  restricting the  number of observable objects, and because of the typically large execution time needed to achieve a reasonable detection rate, i.e.\ to sufficiently cover the $uv$ plane \citep{SaLB10}. The advent of  the four beam combiner PIONIER \citep{Le-Bouquin:2011} at the {\it VLT Interferometer} \citep[VLTI,][]{HAA08, HAB10}, that combines the light of four telescopes, critically changed the situation, by opening the 1-45 mas angular resolution window to a survey approach. 
 
In this paper, we report on the first observational results of our survey. Bias correction and detailed theoretical implications will be addressed in subsequent papers in this series. This paper is organized as follows. Section~\ref{sect: obs} describes the sample selection,  observational campaign and instrumental setups. Section~\ref{sect: data_analysis}  presents the data analysis and binary detection algorithms. The \smash\ constraints on the multiplicity properties of our sample stars are presented in Sect.~\ref{sect: mult}. Section~\ref{sect: discuss} discusses our results and Sect.~\ref{sect: ccl} summarizes our main findings. Finally, Appendix~\ref{sect: notes} and \ref{sect: notes_supp} compile notes on individual objects and provide finding charts for  systems with more than three companions detected in the NACO field-of-view.

\section{Observations}\label{sect: obs}

\subsection{Observational sample} \label{sect: sample}

The sample selection has been driven by the need to observe a sufficiently large number of O stars to derive meaningful statistical constraints and by the observational constraints imposed by PIONIER. The size of the sample defines the precision at which one will constrain the multiplicity rate. The statistical uncertainty ($\sigma_{f_m}$) on the measured multiplicity fraction ($f_\mathrm{m}$) in the considered range depends both on $f_\mathrm{m}$ and on the sample size $N$ \citep{SGE09}. It is given by 
\begin{equation}
\sigma_{f_m}(f_\mathrm{m},N) = \sqrt{f_\mathrm{m} (1-f_\mathrm{m})/N}. \label{eq: bin}
\end{equation}
For a given sample size, $\sigma_{f_m}$ peaks at $f_\mathrm{m} = 0.5$, so that $\sigma_{f_m}(f_\mathrm{m},N) \leq \sigma_{f_m}(0.5,N)$. Observing a sample of $N=100$ is thus required to obtain a precision of $\sigma_{f_m} < 0.05$ for any $f_\mathrm{m}$. 

Following PIONIER observational constraints, the \smash\ survey has been designed as a magnitude- and declination-limited survey. The practical limiting magnitude of PIONIER in its small spectral dispersion mode is $H = 7.5$. The limiting magnitude from the fast guiding systems of the auxiliary telescopes (STRAP) allowing for a proper injection of the beams into the instruments fibers is $V = 11$. The practical range of accessible declinations ($\delta < 0$\degr) is limited by observability constraints of the auxiliary telescopes in the intermediate and large configurations. The Galactic O Star Catalog \citep[GOSC-v2,][]{SMAW08} lists 147 O-type stars fulfilling these criteria. Rejecting the Orion stars that have already been observed by the VLTI \citep{GPR13}, we are left with 138 possible targets. Twelve of these are flagged as runaway stars in the GOSC and are handled separately from the main target list. 

Tables~\ref{tab: targets} and \ref{tab: rw} list the properties of stars in our main list of targets  and in the runaway list.  Columns 1 and 2 indicate whether the object has been observed with PIONIER and NACO. Columns 3 and 4 provide the main identifier used in our survey (HD number if available, BD/CPD identifiers otherwise) and alternative names commonly used in the literature. Columns 5 to 12 indicate the spectral classification, coordinates (J2000.0), $H$-, \Ks- and $V$-band magnitudes. 

The bulk of the \smash\ observations has been obtained in the course of a {\it European Southern Observatory} (ESO) large program (189.C-0644) which was granted 20 VLTI nights over the period April 2012 -- March 2013 and three NACO/SAM nights in June 2013. The NACO observations are  complemented by a 2011 pilot program and additional programs in 2012 and 2013 for a total of 13 VLT/UT4 nights (see Table~\ref{tab: campaign} for an overview). Thirteen stars have further been observed as backup targets of various PIONIER runs from Dec 2013 to Aug 2014. All in all, 102 stars (81\%) from our main target list  have been observed with PIONIER and 120 (95\%) with NACO/SAM. 96 stars (76\%) have both PIONIER and NACO/SAM observations and only the runaway star HD\,157857 has not been observed. Figures~\ref{fig: spt} and \ref{fig: spt_histo2D} provide an overview of the distributions of spectral sub-types, luminosity classes and magnitudes for the stars in our main sample. 

We also observed 37 O stars outside our main list of targets. Table~\ref{tab: sup} summarizes the main properties of the supplementary targets in a format identical to that of Table~\ref{tab: targets}.  These supplementary targets are either northern stars, stars just above our magnitude cutoff, or stars taken from the GOSC-v2 supplements.  Among these additional stars, BN Gem is a known runaway that has $H$ and $V$ magnitudes within our magnitude limits but is a Northern star. We list it along with the other runaway objects in  Table~\ref{tab: rw}.   In total, we observed 174 different stars. 162 have NACO observations and 117 have PIONIER ones. 105 stars have both types of observations. 

\subsection{Observational biases}

As a consequence of our magnitude-limited approach, our sample contains several built-in biases. While it is not our intent to perform detailed bias corrections in this initial paper, we describe here several aspects that need to be kept in mind while directly interpreting the  observational  results of the \smash\ survey.

As for all magnitude-limited surveys, the brightness selection criterion favors nearby stars as well as intrinsically brighter objects. We used the absolute $H$-band magnitude of O stars listed in \citet{MaP06} to estimate the maximum distance at which an isolated O star can be located for its apparent magnitude to be brighter than our cut-off of $H=7.5$.  Neglecting the effect of extinction, Fig.~\ref{fig: dist} shows the obtained maximum distances as a function of spectral sub-type for the various luminosity classes considered in \citet{MaP06}. Early-type O dwarfs can be located up to 3.3~kpc away, while late-type O dwarfs need to be closer than 1.5~kpc to belong to our sample. Early- and late-type giants need to be at a distance of 3.8 and 2.5~kpc at the most while supergiants may reside up to 4.2~kpc away. Extinction will probably not affect these distance estimates by much more than a few 100~pc given that its effect in the $H$-band is rather limited. 

The GOSC catalog is complete down to $B=8$, roughly corresponding to $V=8.3$ and $H=9.0$ in the absence of reddening. Our initial target list is thus dominated by our magnitude cut-off at $H=7.5$, but for stars that have a $B$-band extinction larger than 1.5~mag. GOSC further does not contain many stars more distant than the Carina nebula, i.e.\ more than $\approx 3-3.5$~kpc away. In that sense the sample of supergiants and, to some extent, the sample of giant stars, are more reminiscent of a volume limited sample. 

Close to the magnitude cut-off, our approach also favors multiple objects which receive an apparent brightness boost through their unresolved companions, while similar isolated objects may have been left out of the sample, falling short of the magnitude cut-off \citep[for further discussion of the effects of magnitude cutoff on the measured binary fraction, see][]{SdKdM13}. Equal brightness binaries can  be observed up to a distance larger by 600~pc compared to distances shown in Fig.~\ref{fig: dist}.

Because of the effects described above, our sample contains a larger fraction of supergiants, a larger fraction of hot stars and a larger fraction of multiple systems than a distance limited sample would. The first two effects can be mitigated by discussing our observational results as a function of spectral type and luminosity class. Given proper bias corrections, the latter aspect may be viewed as advantageous as it implies that telescope time is spent on objects that we have more chance to resolve as multiple.

\subsection{Long baseline interferometry}

\subsubsection{Observational setup and calibration}

All long baseline interferometric data were obtained with the PIONIER combiner \citep[][]{Le-Bouquin:2011,Le-Bouquin:2012a} and the four auxiliary telescopes of the VLTI. We used the widest configurations offered by the auxiliary telescopes: A0-K0-GI-I1 in period P89 (Apr - Sep 2012) and A0-K0-G1-I3 in period P90 (Oct 2012 - Mar 2013). Data were dispersed over three spectral channels across the $H$ band (1.50 - 1.80\,$\mu$m), providing a spectral resolving power of $R \approx 15$. As discussed in Sect.~\ref{sec:iowa}, this is the best compromise between sensitivity and the size of the interferometric field-of-view (FOV).

Data were reduced and calibrated with the \texttt{pndrs} package described in \citet[][]{Le-Bouquin:2011}. Each observation block (OB) provides five consecutive files within a few minutes. Each file contains six squared visibilities \VS{} and four phase closures \closure{} dispersed over the three spectral channels. Whenever possible, the five files were averaged together to reduce the final amount of data to be analyzed and to increase the signal-to-noise ratio. The statistical uncertainties typically range from 0.5 to 10 degrees for the phase closures and $2.5$ to $20$\%\ for the squared visibilities, depending on target brightness and atmospheric conditions.

Each observation sequence of one of our \smash\ targets was immediately followed by the observation of a calibration star in order to master the instrumental and atmospheric response. \citet{Le-Bouquin:2012a} have shown that this calibration star should be chosen close to the science object both in terms of position (within a few degrees) and magnitude (within $\pm 1.5$\,mag). We were unable to use the pre-computed JMMC Stellar Diameters Catalog (JSDC\footnote{\anchor{http://www.jmmc.fr/catalogue\_jsdc}{http://www.jmmc.fr/catalogue\_jsdc}}) to look for calibration stars as this catalog only contains a suitable calibrator density down to a magnitude $H\approx6$. Instead we used the tool \texttt{SearchCal}\footnote{\anchor{http://www.jmmc.fr/searchcal}{http://www.jmmc.fr/searchcal}} in its FAINT mode \citep{Bonneau:2011a} to identify at least one suitable calibration star within a radius of $3$\degr\ of each object observed within our sample. 

Most of our objects are grouped into clusters in the sky. Consequently the instrumental response could be cross-checked between various calibration stars. This allowed us to unveil a few previously unknown binaries among the calibration stars. These have been reported to the bad calibrator list\footnote{\anchor{http://apps.jmmc.fr/badcal}{http://apps.jmmc.fr/badcal}} maintained by the IAU and the Jean Marie Mariotti Center\footnote{\anchor{http://www.jmmc.fr}{http://www.jmmc.fr}}. We estimated the typical calibration accuracy to be 1.5\degr\ for the phase closures and $5$\%\ for the squared visibilities. 

A critical point for the final accuracy on the binary separation is the calibration of the effective wavelengths. In PIONIER this calibration is performed routinely in the course of the observation using the optical path modulation as a Fourier transform spectrometer of the internal source. The typical accuracy is $2\,\%$ \citep{Le-Bouquin:2011}. Finally, the on-the-sky orientation of PIONIER has been checked several times and is consistent with the definition of \citet{Pauls:2005fk}.

\subsubsection{PIONIER field-of-view and dynamics}
\label{sec:iowa}

Long baseline interferometric observations are only sensitive to binaries in a specific range of separations, that can be approximately defined by an inner and an outer working angle. The inner working angle (IWA), i.e.\ the maximum angular resolution, is defined by the typical length $B$ of the interferometric baselines and the wavelength $\lambda$ of the observations:
\begin{equation}
  \IWA=\frac{\lambda}{2B} \approx1.5\mas. \label{eq: iwa}
\end{equation}
The spatial frequency smearing across one spectral channel induced by the low spectral resolving power $R \approx 15$ of the  PIONIER observations is the main limiting factor for the  outer working angle (OWA):
\begin{equation}
  \OWA=R\,\frac{\lambda}{B} \approx 45\mas. \label{eq: owa_pio}
\end{equation}

We checked that neither the temporal averaging over several minutes nor the spatial frequency smearing over the telescope pupil  impact the expected \OWA{}. Contrary to the \IWA{}, the \OWA{} is not a hard limit. Pairs with wider separations still leave a strong signature in the interferometric observables. However, estimating properly their separation and flux ratio becomes challenging. These pairs are better studied with complementary techniques, such as speckle interferometry, aperture masking or adaptive optics.

In addition, the single-mode optical fibers of PIONIER theoretically restrict the FOV to the Airy disk of the individual apertures. This corresponds to 180~\mas\ when using the auxiliary telescopes. However, this limit is much less clear when considering the effect of the atmospheric turbulence. For sure, our PIONIER survey is blind to binaries with separation larger than 500~mas.

Even within the range $1.5-45$~\mas, the depth to which a companion can be detected depends on the relative orientation of the companion and the interferometric baselines. This is due to the sparse structure of the point spread function associated with the diluted aperture of an interferometer. Consequently, the sensitivity limit should be defined for a given completeness level. Considering an accuracy of 1.5\degr\ on the phase closures and three OBs per target, we found that our survey should provide a 90\%\ coverage of the separation regime between $1.5$ and $45$~\mas\ for a flux ratio dynamics of $1:20$, equivalent to a magnitude difference of $\Delta H=3.25$ \citep[see the middle panel of figure~2 in][]{Le-Bouquin:2012}.

\subsubsection{PIONIER observations}

The bulk of the observations were obtained during 20 nights of visitor-mode spread over ESO periods 89 and 90 (Table~\ref{tab: campaign}). Thirteen stars were further observed as backup targets from Dec 2013 to Apr 2014. Raw and reduced data in OIFITS format are stored in the PIONIER archive and are available upon request. The time lost to  weather amounted to approximately $20$\%, and is largely due to wind speeds larger than $10$~m~s$^{-1}$. The amount of technical losses was approximately $10$\%, dominated by issues on the auxiliary telescopes and the delay lines. Two nights in August 2012 have been used to unveil and characterize the polarization behavior of the VLTI optical train.

As described earlier, VLTI observations were obtained for 117 objects from the initial target selection and for six supplementary targets. Of the observed sample, 73\%\ of the objects have magnitude $H>6.0$ (Fig.~\ref{fig: spt}), which is the limiting magnitude of the VLTI/AMBER instrument in service mode. Observing such a large number of faint objects was only made possible thanks to the  sensitivity and efficiency offered by the PIONIER instrument.

\subsection{Aperture masking and AO observations}

\subsubsection{Observational setup and calibration}

All aperture masking data have been obtained with the NACO instrument on
the VLT/UT4 telescope.  In most cases, four to eight targets were grouped by magnitude and angular proximity in the sky in a single observing sequence. 
Targets in a given group were observed sequentially using the {\it star hopping}  mode \citep{LTI11}. In this approach,
we froze the adaptive optics (AO) configuration on the first target and  fast switched
between targets using telescope offsets, without neither AO re-acquisition nor
optimization on the subsequent targets in the series. For long science sequences, no calibrators were observed.
Instead, we used the scientific objects that turned out to be point sources
as calibrators. The advantage of the star hopping mode lies in its high efficiency. It approximately doubles the observing time spent on scientific targets compared to the classical approach of using science-calibrator sequences of observations. Whenever stars could not be grouped together,  a  K~III stellar calibrator, with similar magnitude and a nearby position on the sky, was observed immediately before or after the scientific object.

The NACO/SAM observations made use of the 7-hole mask \citep{TLA10} and, for the vast majority of our targets, were repeated using at least two different broadband filters. Most of our targets were observed with  the $H$ and \Ks\ filter and the visible wave-front sensor. Depending on the weather conditions and instrumental/operational constraints, some targets were observed with the \Lp\ broadband filter and/or the AO correction made use of the near-infrared (NIR) wave-front sensor. 

We used either the S27 camera and a 512$\times$512 pixel windowing or the S13 camera in full frame mode. These choices result in an effective FOV of 13\arcsec\ $\times$ 13\arcsec. For a given object, a typical observation  consists generally of a set of eight data cubes of 100 frames with individual integration times ranging from 100 to 250~ms, depending on the stellar brightness. Each data cube was taken with the object at different positions on the detector (dithering). The standard reduction comprises flat fielding, bad pixel correction
and background subtraction. The background was estimated using the
median value of the eight data cubes. An example of reduced and stacked NACO images is shown in Fig.~\ref{fig: naco_img} and further images are provided in the appendix.

\subsubsection{NACO/SAM field-of-view and dynamics} \label{sect: naco_fov}
As seen in Fig.~\ref{fig: naco_img}, the point spread function (PSF) of a star appears as a complex fringe pattern that results from Fizeau
interference between the holes of the aperture mask. The size of the PSF is given by the Airy
disk of a single hole ($\approx 400$~mas in the \Ks-band). 

The NACO/SAM data result from the combination of aperture masking and adaptive optics techniques. They allow us to investigate two complementary separation regimes. At small working angles, the analysis of the Fizeau interference pattern produced by the masked aperture enables us to search for companions within each object's PSF.

The IWA of this technique is obtained from Eq.~(\ref{eq: iwa}) with $B$ taken to be the maximum separation between holes, i.e.\ $\approx 7$~m. This yields about 30~mas. 

The OWA is limited by the size of the PSF as we fit the data with sines and cosines weighted by the
Airy pattern. The weighting is the culprit, however necessary to have a
good fit of the data and avoid being hampered by detector noise
outside the diffraction pattern. The OWA is therefore $1.22 \lambda / d_\mathrm{hole} \approx 300$~mas, in the $H$-band, where $d_\mathrm{hole}=1.2$~m is the diameter of a hole in our adopted aperture mask. 

As for PIONIER observations the NACO/SAM OWA is not clear cut, but a progressive decrease in
sensitivity down to zero outside the first Airy lobe. The maximum brightness contrast that can be achieved depends on the signal-to-noise of our observations, and typically reaches 5~mag.

\subsubsection{NACO FOV and dynamics} \label{sect: naco_ao}
At larger working angles, i.e.\ outside the PSF of the individual objects ($\rho > 300$~mas), the 13\arcsec\ $\times$ 13\arcsec\ FOV of our NACO observations provides us with an AO-corrected image of the surrounding field (Fig.~\ref{fig: naco_img}).  The IWA is limited by the blurring of the extended aperture masking PSF, hence to $1.22 \lambda / d_\mathrm{hole} \approx 0.3$\arcsec. We further limited our search to a field of 8\arcsec\ around the target.

The PSF of each companion in the NACO FOV also results from the Fizeau interference pattern produced by the masked aperture, so that our AO images are not as deep as one could expect from similar exposure images obtained on an 8m-class telescope. Yet they are often the deepest AO-corrected images ever obtained around our stars and allow us to search for companions with separations between 0.3\arcsec\ and 8\arcsec\ and with a brightness contrast of up to 8~mag.

\subsubsection{NACO/SAM observations}

We observed a total of 162 targets during five observing runs spread from March 2011 to July 2013 (Table~\ref{tab: campaign}), for a total of almost 12 nights. One third of the time was lost due to bad weather. Detector issues during our Feb 2012 run restricted the effective FOV to a 6.5\arcsec\ $\times$ 13\arcsec\ area but had no further impact on the aperture masking observations. Most of the run was anyway lost to poor weather, with only six objects observed in the course of three nights.


\section{Companion detection} \label{sect: data_analysis}

In this section, we describe the algorithms adopted to search for companions.
Owing to the different nature of data collected by the \smash\ survey, different approaches were used for the PIONIER, the NACO/SAM and the NACO FOV data.

For the long baseline interferometric data obtained by PIONIER (Sect.~\ref{sect: pio_analysis}), we fit both a single star and a binary model to the squared visibility and phase closures and compare the obtained $\chi^2$ to decide which model fits best.
The analysis of the NACO data is split in two parts, according to the separation regime considered. At small working angles ($\rho \lesssim 250$~mas, NACO/SAM), i.e.\ within the diffraction pattern of the NACO PSF, we perform an interferometric analysis of the Fizeau interference pattern produced by the aperture mask to search for companions in Fourier space (Sect.~\ref{sect: SAM-SWA}). At larger working angles ($\rho \gtrsim 250$~mas, NACO FOV), i.e.\ outside the object PSF, we use a cross-correlation technique to search for (mostly faint) companions in a 8\arcsec-radius from the central object (Sect.~\ref{sect: SAM-LWA}).

\subsection{PIONIER data analysis} \label{sect: pio_analysis}
The calibrated interferometric data were analyzed following the approach detailed in section~3.2 of \citet{Absil:2011}. The underlying idea is to test whether an observation is compatible with that of a single star model. The main differences with \citeauthor{Absil:2011}  are:
\begin{itemize}
\item[-] We do not re-normalize the $\chi^2$ with the best-fit binary model. This is because of the limited size of the data set obtained for each individual object (typically 2 OBs).
\item[-] The analysis is performed using the phase closures and the squared visibilities jointly.
\item[-] The stellar surfaces are considered to be unresolved, which is a realistic assumption for our  early-type objects observed with 100m baselines.
\end{itemize}
Consequently in our analysis, the probability $P_1$ for the data to be compatible with the single-star model is:
\begin{equation} \label{eq:proba_single}
  P_1 = 1 - \mathrm{CDF}_\nu(\chi^2)
\end{equation}
with
\begin{equation} \label{eq:chi2_single}
 \chi^2 = \sum \frac{(\VS - 1)^2}{\sigma_{\VS}^2}\;+\; \sum \frac{\closure^2}{\sigma_\closure^2}.
\end{equation}
$\mathrm{CDF}_\nu$ is the $\chi^2$ cumulative probability distribution function with $\nu{}$ degrees of freedom ($\nu$ being the total number of \VS{} and \closure{}  minus the number of parameters in the model). The distribution of the computed $\chi^2$ values is shown in Fig.~\ref{fig:histo_chi2}.

If the probability $P_1$ in Eq.~(\ref{eq:proba_single}) is higher than an adopted threshold, the dataset is considered to be compatible with the single-star model. For these objects, we derived a two-dimensional map of sensitivity limits as detailed in section 3.3 of \citet{Absil:2011}. We then computed an annular sensitivity limit for a completeness of 90\%. That is, for each radius, we identified the dynamic for which a companion would have been detected over 90\%\ of the annular region. 

If the probability $P_1$ in Eq.~(\ref{eq:proba_single}) is below the adopted threshold,  the detection of spatial complexity in the object is considered significant and we  reject the single-star model. In this case we perform a least-square fit of the data with a binary model. We incorporate in the model a first-order correction to account for the bandwidth smearing. The complex visibility $V$, hence the squared visibilities and phase closures, of our binary model is defined as :
\begin{equation}
  V = \frac{1\;+\;f\,\exp(-2i \pi x)\;\mathrm{sinc}(\pi x R)}{1\;+\;f},
\end{equation}
with
\begin{equation}
  x = \rho\;(u \sin\theta + v \cos\theta),
\end{equation}
where $f$, $\rho{}$ and $\theta{}$ are the flux ratio, the angular separation and the position angle of the binary. The latter is defined as the orientation of the secondary measured from North to East. The vector $(u,v)$ is the spatial frequency of the observation \citep{Pauls:2005fk} and $R$ is the spectral resolving power. For the few objects that were observed several times, we performed the fit with the binary model independently for each epoch.

In some cases, the best-fit binary model still does not provide a satisfactory reduced $\chi^2$. This indicates that the object shows some additional spatial complexity that is not  properly reproduced by  the binary model. In particular, this situation occurs for the seven detected pairs whose separations are larger than the PIONIER \OWA{} (see Fig.~\ref{fig:chi2_sep}). For these objects, our model does not hold anymore because of the limited validity of the bandwidth smearing correction. Fortunately, most of these objects were observed with NACO/SAM, allowing us to confirm the tentative PIONIER detection in each case.

For the PIONIER companion detection, we adopted a $P_1$ threshold of 0.9973 (corresponding to $3\sigma$ for a Gaussian distribution). The probability of false detection is thus lower than $0.27\%$ (Eq.~\ref{eq:proba_single}), hence  less than one object given our sample size. 
A total of 42 objects were flagged with positive detection and separations within the PIONIER OWA, i.e.\ 45~mas. 
We visually inspected all data sets (detections and non-detections). One object with positive detection was removed ($\mu$~Nor) because it shows an inconsistent signal between epochs as well as a poor fit with a binary model. 

The properties of the resolved systems are summarized in Table~\ref{tab: pio_detect}. Column~1 indicates the target name. Columns~2 and 3 identify the pair and the instrument setup. Column 4 gives the epoch of observations in Besselian years (b.y.). Columns~5 to 7 provide the position angle, projected separation and $H$-band magnitude difference between the two companions. Columns 8 ($\Delta K\mathrm{s}$) and 9 ($\Delta L'$) are not used for the PIONIER detections.

\subsection{SAM interferometric data analysis} \label{sect: SAM-SWA}

As mentioned in Sect.~\ref{sect: naco_fov}, the interferometric analysis of the NACO/SAM data corresponds to a search for a
stellar companion within the diffraction pattern of the PSF.
We used the SAMP pipeline presented in \citet{LTA11}. In short, the
PSF is modeled as a sum of spatial frequencies, modulated by the Airy
pattern caused by the diffraction of a single hole. To each pair of
holes corresponds a baseline vector and a spatial frequency. The
individual frames are projected onto that set of spatial frequencies.
The bispectrum is obtained by multiplication of the complex values
extracted from a triangle of holes, hence three spatially closing
frequencies. The phase closures are then extracted from the argument of
the bispectrum. The final calibration is made by subtracting the
average value of all point-like stars observed within the same OB.

Detection is then obtained as in \citet{LTA11}, similarly to what
we have done for the PIONIER data: the phase closures are adjusted by models for either an unresolved object or  
a resolved binary system.  All SAM data sets of a given target are fitted 
simultaneously. Combining data obtained with the different filters allows us 
to lift the degeneracy on the position of the global optimum in the $\chi^2$ map that 
results from the periodic sampling of the $uv$-plane.
In a few cases, data were obtained at different epochs. These were still combined together given that we do not expect significant changes in the position of the companions over the 2.5~yr maximum baseline of our observations. The one exception to this rule is HD\,93129\,AaAb, for which our three observational epochs are handled separately.

We distinguish  three outcomes of the fitting procedure: (i)
non-detection: the phase closures are compatible with zero within the
uncertainties; (ii) clear detection: the phase closures are
compatible with a binary model; (iii) tentative detection: the phase closures are not compatible with a point source, but the binary model does
not fit well either. 
 In the following we only report the clear detections (case ii).

The properties of the resolved systems are summarized in Table~\ref{tab: pio_detect} to allow for a direct comparison with PIONIER measurements. Columns~8 and 9 indicates \Ks\ and \Lp\ magnitude differences between the central object and the detected companion(s).

\subsection{NACO field-of-view analysis } \label{sect: SAM-LWA}

The second analysis of the NACO data aims to search for stellar companions
outside the diffraction pattern of the SAM PSF.
After correction of the detector defects,  each frame and each data cube
is shifted to center the target on a reference point. Each cube
is then collapsed, and the central PSF is extracted for reference.
This PSF is then cross-correlated over the entire detector. Last, all the
cross-correlated images -- one for each data cube -- are derotated according to the parallactic angle
(SAM observations are done in pupil tracking) and averaged. For each
image we looked for companions by searching for local maxima in the cross-correlation function,
independently of filters or epochs.

Properties of the detected companions are listed in  Table~\ref{tab: naco_detect}, in a layout similar to that of Table~\ref{tab: pio_detect}. Column 8 gives the probability of spurious detection $P_\mathrm{spur}$ obtained in Sect.~\ref{sect: spurious}.
 Whenever several companions are detected, their properties are listed in Cols.~2-8 on subsequent lines.
 Each companion in Table~\ref{tab: naco_detect}  either corresponds to a clear detection, or to a detection confirmed at several epochs and/or in several filters.

\subsection{Detected companions and internal consistency}

PIONIER resolved 53 companions in the sample of 117 objects (42 have $\rho < 45$~mas and 11 have larger separations). 48 of these companions are resolved for the first time. Their separations range from $\approx$ 1~mas to $>100$~mas. 22 of the pairs fall in the sensitivity regime of NACO/SAM. In practice, all the companions detected by PIONIER with $24< \rho < 120$~mas are also detected by NACO/SAM. The PIONIER accuracy remains higher than that of  SAM up to its OWA, i.e.\ about 45~mas. Interestingly, all tentative PIONIER detections outside its OWA are confirmed by NACO/SAM up to the mentioned separation of 120~mas. PIONIER is hardly sensitive to any binaries with $\rho > 150$~mas. These properties line up very well with the expected sensitivity range discussed in Sect.~\ref{sect: obs}, illustrating the excellent internal consistency of the \smash\ detections at small angular resolutions.

The positions of the detected companions in the separation vs.\ brightness-contrast plane are displayed in Fig.~\ref{fig: rho_f}, together with the median sensitivity limit of all our observations. The latter show that we have an excellent coverage of the parameter space except in the 200-500~mas range, corresponding to the transition between NACO/SAM and NACO-FOV detections. These results will be discussed more extensively in Sect.~\ref{sect: discuss}, but two interesting comments can already be made:
(i) the density of similar brightness pairs ($\Delta H < 1$) drops significantly at separations larger than 50~mas;
(ii) there seems to be a lack of fainter companions ($\Delta H > 3$) in the range 10-30~mas and 50-150~mas even though our first estimate of the detection limit extends down to $\Delta H = 4$ and 5~mag, respectively. Further investigations on the accuracy of our detection limit estimates will allow to verify this result.

\section{Constraints on the multiplicity properties} \label{sect: mult} 

In this section, we present the  statistical constraints  on the multiplicity properties of massive stars. Section~\ref{sect: spurious} investigates spurious associations. Section~\ref{sect: earlier} compares our new detections with previous knowledge in the regime of separations investigated by the \smash\ survey. The observed multiplicity fraction and average number of companions per star are described in Sect.~\ref{sect: fractions}. Finally, Sect.~\ref{sect: lc} investigates how the multiplicity properties change with the luminosity class.

\subsection{Spurious associations} \label{sect: spurious} 

Given the detection limits adopted in the previous section, all the companions that we report are, to a very large degree of confidence, real objects. The components  of some of the detected pairs may however not have any physical relation with one another.
 In this section, we estimate the probability $P_\mathrm{spur}$ of spurious association that would result from background or foreground objects or from line-of-sight alignment in a cluster environment. For each central object, we queried the 2MASS catalog to look for the number $N_\mathrm{obj}$ of stars brighter than $\Delta H<5$~mag within a radius of $r=120$\arcsec. This contrast threshold is representative of our SAM observations and a conservative value for PIONIER since our faintest PIONIER detection is at $\Delta H = 3.2$~mag. The local density was then converted into a probability of spurious detection due to chance alignment by conservatively assuming that all our  PIONIER and SAM observations are sensitive to separations up to $\rho=0.2$\arcsec:
\begin{equation}
  P_\mathrm{spur} = N_\mathrm{obj} \times \left(\frac{\rho}{r}\right)^2. \label{eq: Pspur}
\end{equation}
We found that $P_\mathrm{spur}$ is always smaller than 0.001\%. Consequently our interferometric survey is virtually free from spurious detections. All the companions detected by PIONIER and SAM are very likely to be physically related to their central object.

We performed a similar test for the companions detected in the NACO FOV, but using the actual magnitude difference to retrieve $N_\mathrm{obj}$ from the 2MASS catalog as well as the measured projected separation $\rho$ in Eq.~(\ref{eq: Pspur}). Results are presented in  Fig.~\ref{fig: Pspur} and listed in Table~\ref{tab: naco_detect}.
 
All companions with $\rho < 1$\arcsec, 2\arcsec\ and 5\arcsec\ have  a probability $1-P_\mathrm{spur}$ of physical connection better than 99, 90 and 50\%\ respectively. At large separations, only the brightest companions are likely physically connected to their central object while at closest separations ($\rho < 1$\arcsec), all companions are likely bound. This is in line with the conclusions reached by \citet{MAp10} for AstraLux observations.

\subsection{The \smash\ observational window} \label{sect: earlier}

Most of the objects in our sample have been previously observed by various high-resolution imaging campaigns. We compiled the astrometric results of \citeauthor{MGH98}\ (\citeyear{MGH98, MHG09}), \citet{NWW04}, \citet{TBR08}, \citet{TMH10} and \citet{MAp10} together with the spectroscopic status from the GOSC-v3 \citep{MASM13} in a single database in order to obtain the most complete view of the multiplicity properties of our sample stars. The astrometric data and naming conventions were cross checked against those of the Washington Double Star catalog \citep[WDS][]{MWH01}. The SB status from the GOSC-v3 was further complemented by  results of various published spectroscopic surveys \citep{SGN08, SGE09, SJG11, SdMdK12,CHN12}, by early results from the spectroscopic survey of Galactic O and WN stars \citep[OWN, ][]{BGA10} described in \citet{SMAM14} and by individual papers on various objects (see individual notes in Appendix A). Regarding the results of \citet{CHN12}, we only accepted SB status for V-III class stars. Radial velocity measurements of II-I stars may indeed be affected by atmospheric variability and, unless an orbital period was available for these objects, we conservatively ignore a potential spectroscopic companion.

 In  Fig.~\ref{fig: sep_cdf}, we compare the cumulative number distribution of companion separations before and after \smash. As expected, our survey is the first to resolve a significant number of systems with separations smaller than 50~mas (only two companions were known out of 52 detected now). Moreover \smash\  contributes to the companionship census at larger separations.  In total, our survey has increased the number of resolved companions within 100 mas roughly by a factor of 17 (from 4 to 66) and within 8\arcsec\ roughly by a factor of 4 (from 64 to 260).

Fig.~\ref{fig: sep_cdf} also shows two clear trends, although the physical interpretation remains unclear.  First, there is an apparent concentration of companions at separations of 30-50~mas, which corresponds to the transition between the PIONIER and the SAM samples. The larger sample and the higher sensitivity of the SAM observations seems to only account for about half the  increase in the cumulative number density, while the other half seems to be genuine (Fig.~\ref{fig: sep_cdf_main}); however, an appropriate correction for  observational biases is needed for confirmation. Second, the companion distribution function increases linearly with the logarithm of the separation above 50-60 mas, but this increase is almost entirely due  to relatively faint companions ($\Delta H > 5$). At closer separations, these faint companions are below the detection threshold of all the previous surveys, including \smash. It is thus not possible to provide observational constraints as to whether such faint companions exist at smaller separations.

\subsection{Multiplicity fraction and number of companions} \label{sect: fractions}
Here we derive the observed (uncorrected) multiplicity and companion fractions (Sects.~\ref{sect: mf} and \ref{sect: nf}) obtained in our survey. We start by defining these quantities as some confusion has arisen in the literature on the use of these terms.

\subsubsection{Definitions}
The {\it number of  multiple systems} $N_\mathrm{m}$ is the number of observed central objects with at least one companion.

The {\it fraction of multiple systems}, or {\it multiplicity fraction} $f_\mathrm{m}$, is the ratio of the number of multiple systems  $N_\mathrm{m}$ to the sample size $N$. 

The {\it number of  companions} $N_\mathrm{c}$ is the total number of companions  observed around a given sample of central objects.

The {\it fraction of companions}  $f_\mathrm{c}$ is the average number of companions per central objects, i.e.\ the ratio of the total number of companions $N_\mathrm{c}$ to the sample size $N$.  

These quantities will occasionally be restricted to sub-categories, such as resolved (R) or unresolved eclipsing or spectroscopic (E/SB) systems, or to specific separation ranges. 

The uncertainties on the multiplicity fractions follow binomial statistics as described in Sect.~\ref{sect: sample}. The uncertainties $\sigma_{f_c}$ on the fraction of companions $f_\mathrm{c}$ follow Poisson statistics and can be estimated as
\begin{equation}
\sigma_{f_c} = \sqrt{N_\mathrm{c}}/N. \label{eq: pois}
\end{equation}

Note that, Eq.\ \ref{eq: bin} (resp.\ Eq.\ \ref{eq: pois}) only provides accurate confidence intervals for $f_\mathrm{m}$ significantly different than 0 or 1 (resp.\ for $f_\mathrm{c}$ significantly different than 0). Because the values of  $f_\mathrm{m}$  and $f_\mathrm{c}$ that we derive do not always meet these criteria, we  estimated the 68\%-confidence intervals using Monte Carlo simulations that take into account the realization probability and sample size of each (sub)sample and allow for asymmetric boundaries. The values of $f_\mathrm{m}$, $f_\mathrm{c}$ and their uncertainties are provided in Table~\ref{tab: fraction}.

\subsubsection{Fraction of multiple systems}\label{sect: mf}

PIONIER resolved 42 stars with at least one companion closer than 45~mas, hence 36\%\ of the 117 stars observed with PIONIER. SAM detects 23 companions in the range 45-250~mas, hence 14\%\ of the 162-star SAM sample and 8 companions (5\%\ of the SAM sample) with separations in the range 30-45~mas, but too faint to be detected by PIONIER. In total 40\%\ of the total number of stars with either PIONIER or NACO observations have at least one resolved companion within 250~mas.  The uncertainty on the parent multiplicity fraction is 4\%. There is a remarkable uniformity in the fraction of multiple resolved systems at early- and mid-O spectral sub-types as well as a function of their NIR brightness (Fig.~\ref{fig: spt_kmag}).

Restricting ourselves to the 96  stars of our main sample that have been observed both by PIONIER and SAM, the fraction of multiple systems resolved  by PIONIER (at $\rho < 45$~mas) and by NACO/SAM (either at $\rho > 45$~mas or too faint to be detected by PIONIER) rises to $f_\mathrm{m}^\mathrm{pio}=0.39$ and $f_\mathrm{m}^\mathrm{sam}=0.17$, respectively. In total, $f_\mathrm{m}^\mathrm{1-200mas}=0.53$ of our main sample has at least one detected companion in the 1-200~mas range. The uncertainty on the observable parent multiplicity fraction is 0.05.

Accounting for the resolved systems\footnote{In the following, we only considered resolved companion in the separation range 1-8000~mas, thus $f_\mathrm{m}^\mathrm{R} = f_\mathrm{m}^\mathrm{1-8000mas}$.} (R; 51 systems) and for the known eclipsing (E) or spectroscopic (S) binaries (47 systems), we now obtain a total of 87 systems with at least one companion within 8\arcsec. The fraction of multiple systems is thus $f_\mathrm{m}^\mathrm{RES}=91\pm3$\%. Fig.~\ref{fig: triple} shows the cumulative fraction of multiples $f_\mathrm{m}$ as a function of the angular separation $\rho$. The $f_\mathrm{m}(\rho)$ curve is plotted for different minimum multiplicity degrees, from at least one companion (double systems) to at least four companions (quintuple systems). Including the spectroscopic companions, about one quarter of our sample contains three or more stars within 250~mas and are hierarchical triple (or higher multiplicity) systems. 

The total fraction of multiple systems $f_\mathrm{m}^\mathrm{RES}$ that we compute is dominated by close companions (either unresolved E/SB or resolved with separations $\rho < 250$~mas). Hence they are unaffected by spurious detections due to chance alignment.

\subsubsection{Fraction of companions} \label{sect: nf}

The number of resolved companions per central object varies from 0 to 6 (Fig.~\ref{fig: Ncomp}). However, most of the  systems with more than one resolved companion have their additional companion(s) found outside a 250~mas radius. This may reflect a limitation of our snapshot approach as the sparse $uv$ coverage and the modeling approach described in Sect.~\ref{sect: data_analysis} may not easily allow detection of more than one companion (although see the case of HD\,93160). Alternatively, it may reflect a stability criterion for hierarchical systems.  Dynamical stability of a triple system indeed requires that the inner binary and outer companion have semi-major axes that are different by a factor of at least 3 to 5 depending on mass ratio and eccentricity  \citep[e.g.][]{Tok04,VaK06}. This possibly restricts the range of systems hosting more than one companion in the 1-250~mas range.

Limiting ourselves to the main sample and to companions with $\Delta H<5$ and excluding (resp.\ including) the spectroscopic  or eclipsing companions, the total number of resolved companions is 84 (resp.\ 134), yielding an average fraction of companions $f_\mathrm{c}^\mathrm{R}(\Delta H<5)$ of 0.9 (resp.\ $f_\mathrm{c}^\mathrm{RES}(\Delta H<5)=1.4$) within an 8\arcsec\ radius (Fig.~\ref{fig: sep_cdf2}). Lifting the $\Delta H$ criterion, the  fraction of resolved (resp. resolved and E/SB) companions rises to $f_\mathrm{c}^\mathrm{R}=1.7$ (resp.\ $f_\mathrm{c}^\mathrm{RES}=2.3$).

After statistical correction for spurious detections due to chance alignment, the averaged fraction of resolved companions is $f_\mathrm{c}^\mathrm{R}=1.5$. Including the unresolved E/SB companions, the fraction becomes $f_\mathrm{c}^\mathrm{RES}=2.1$ (Table~\ref{tab: fraction}).  This value is larger than the value of 1.5 obtained by \citet{PBH99} for a sample of 14 OB stars in the Orion Nebula cluster. Both values however agree within errors when restricting \citeauthor{PBH99} results to the only four O-type objects in their sample. Furthermore, our fraction of companion is larger than the bias-corrected value of 1.35 obtained for B-type stars in the Sco-Cen OB association \citep{RIR13}, suggesting again that the fraction of companion increases with  spectral type, hence with stellar mass.

\subsection{Luminosity classes} \label{sect: lc}

Fig.~\ref{fig: spt_kmag} and Table~\ref{tab: fraction} present the fraction of resolved systems for the different luminosity classes (LCs). As for the overall sample, the overall multiplicity fractions $f_\mathrm{m}^\mathrm{RES}$ of the individual luminosity classes reach their maximum value before 200~mas (Fig.~\ref{fig: sep_cdf_lc}). These multiplicity fractions are thus dominated by close companions  and are unaffected by spurious detections.

\begin{figure}
\includegraphics[width=8cm]{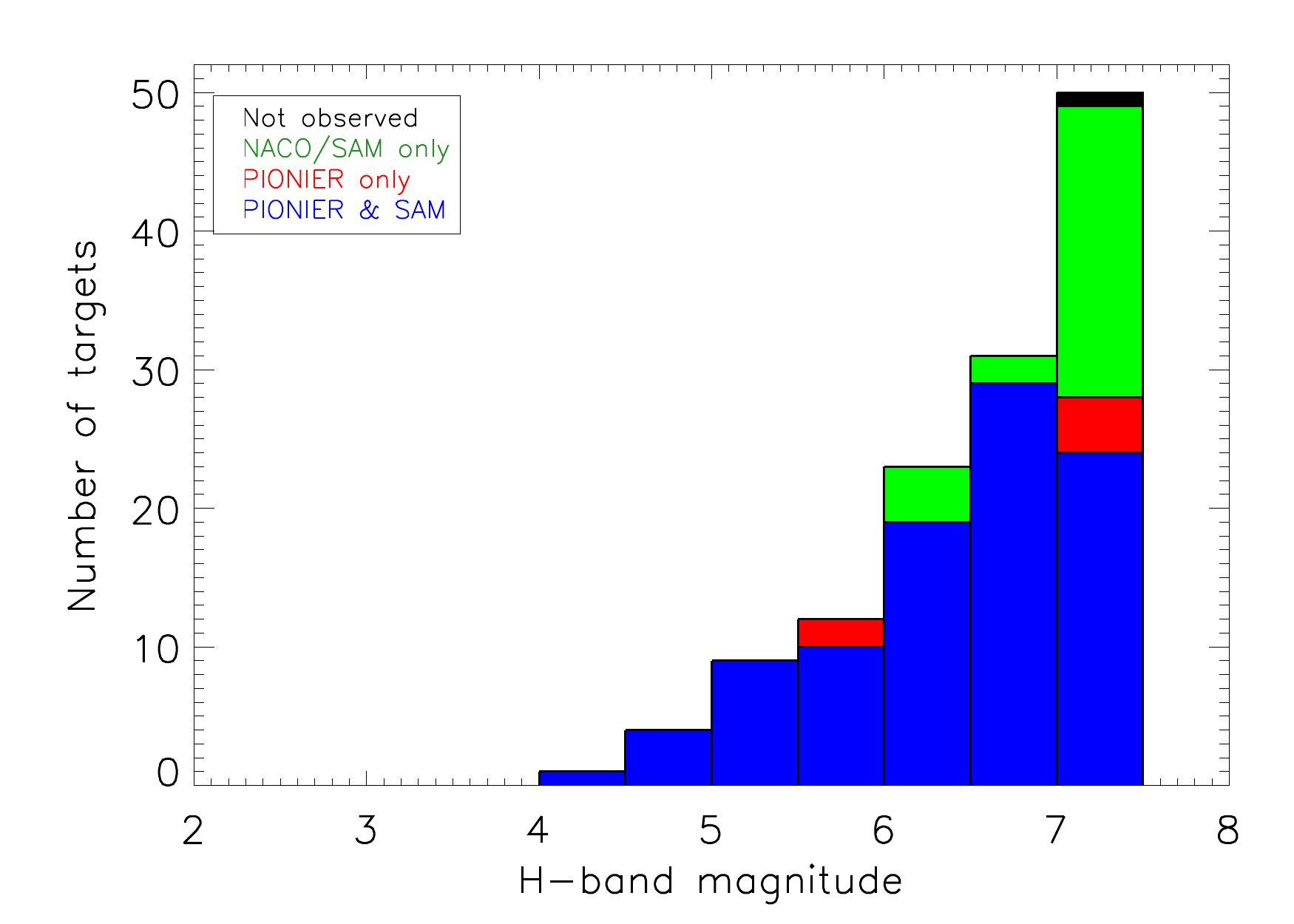}
\includegraphics[width=8cm]{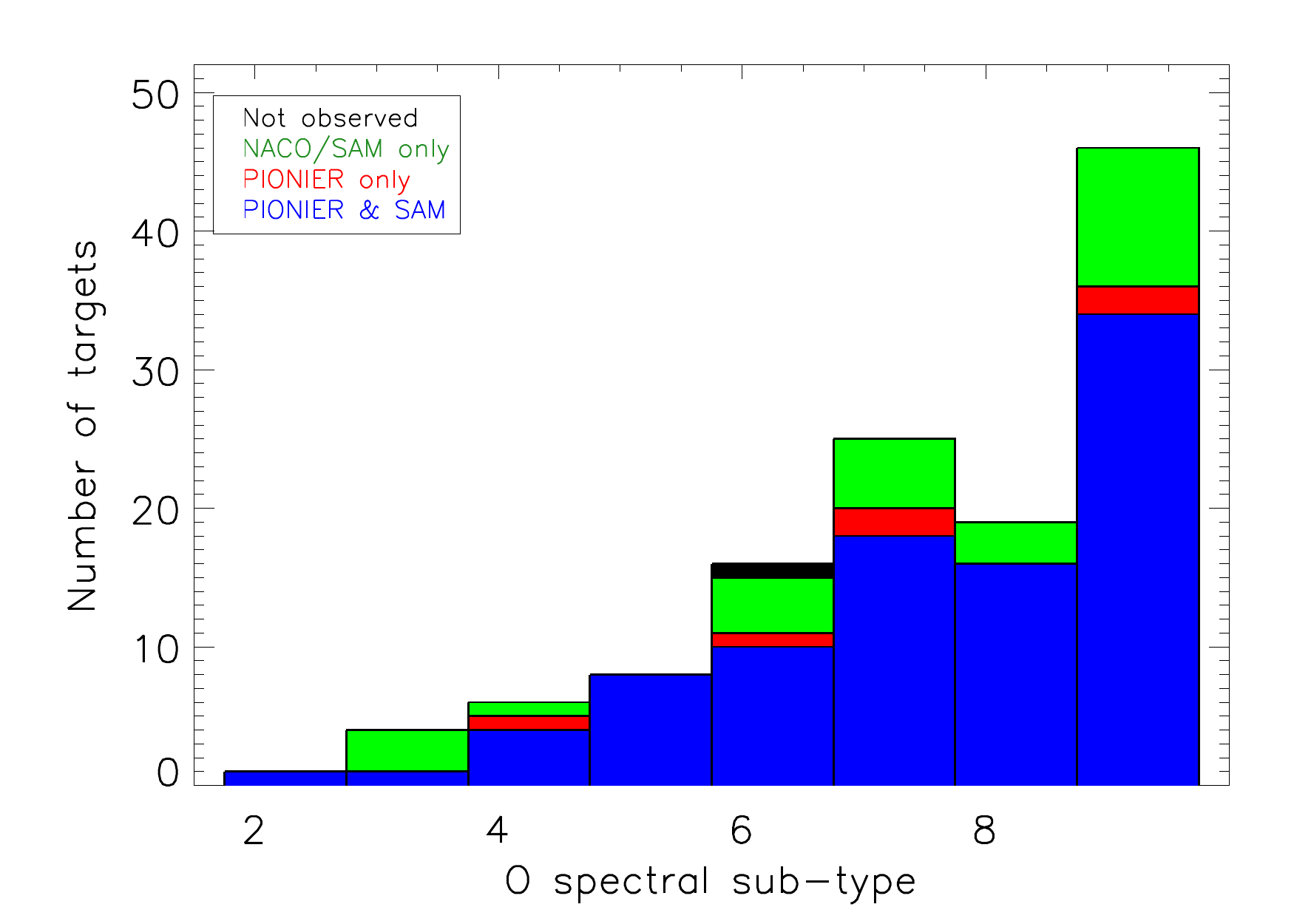}
\includegraphics[width=8cm]{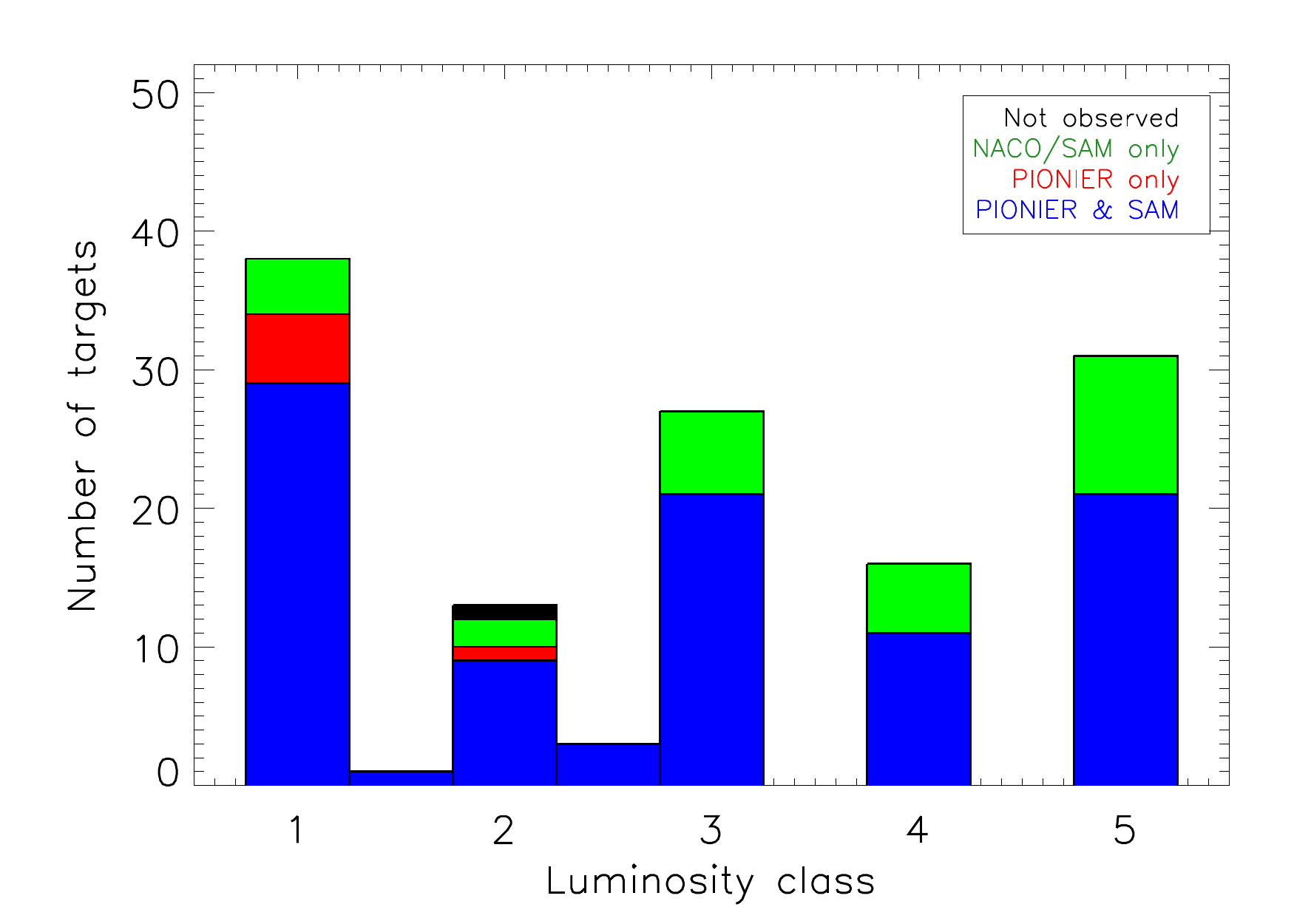}
\caption{ Distributions of $H$-band magnitudes ({\it upper panel)}, spectral sub-types {\it (middle panel)} and luminosity classes  {\it (lower panel)} of \smash\ targets in our main sample. Fractional luminosity classes indicate uncertain classification between the two neighboring classes.}
\label{fig: spt}
\end{figure}

\begin{figure}
\includegraphics[width=8cm]{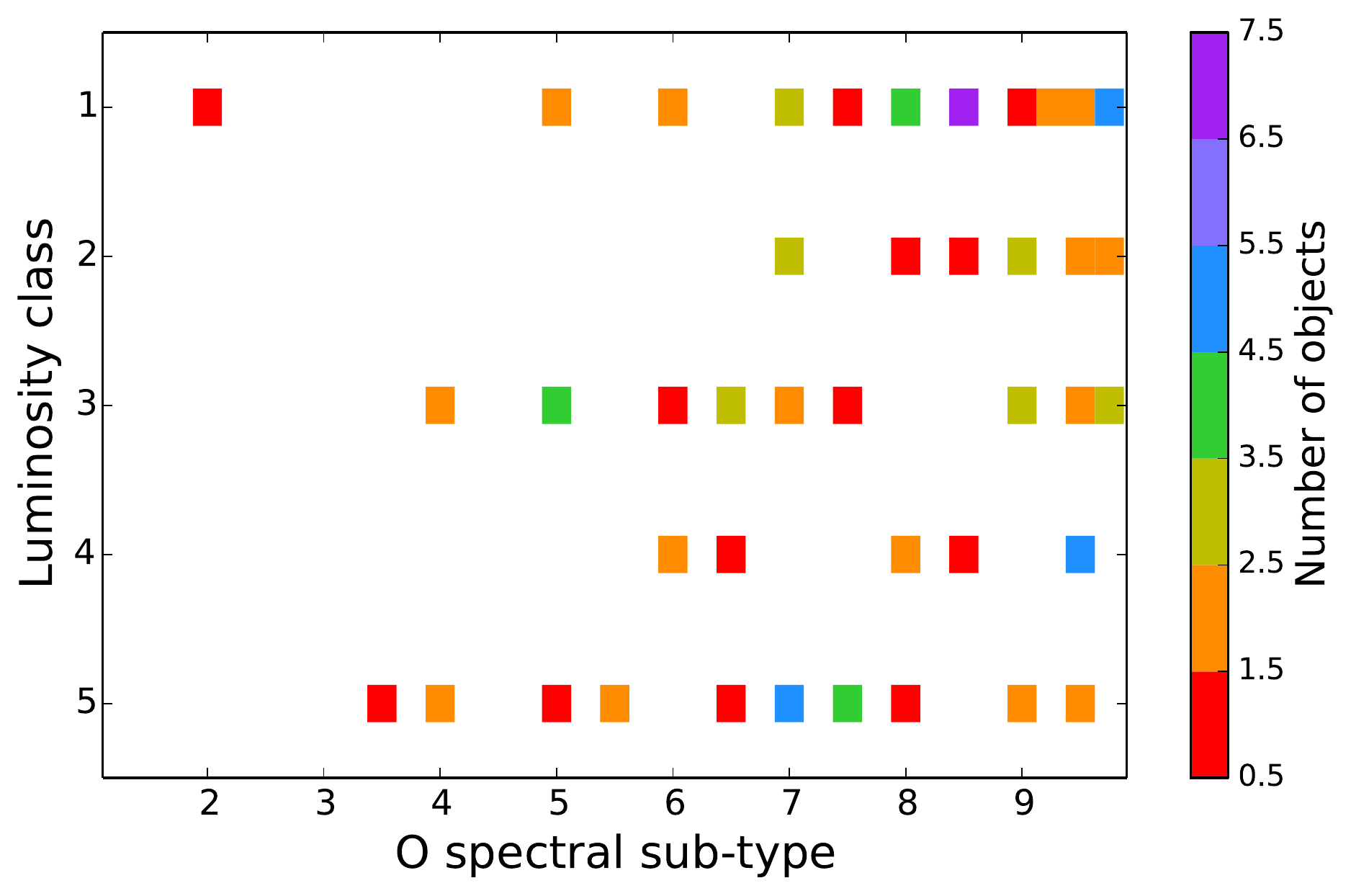}
\caption{ Distributions of luminosity classes  vs.\ spectral sub-types  of \smash\ targets in our main sample. }
\label{fig: spt_histo2D}
\end{figure}

\begin{figure}
\centering
\includegraphics[width=8cm]{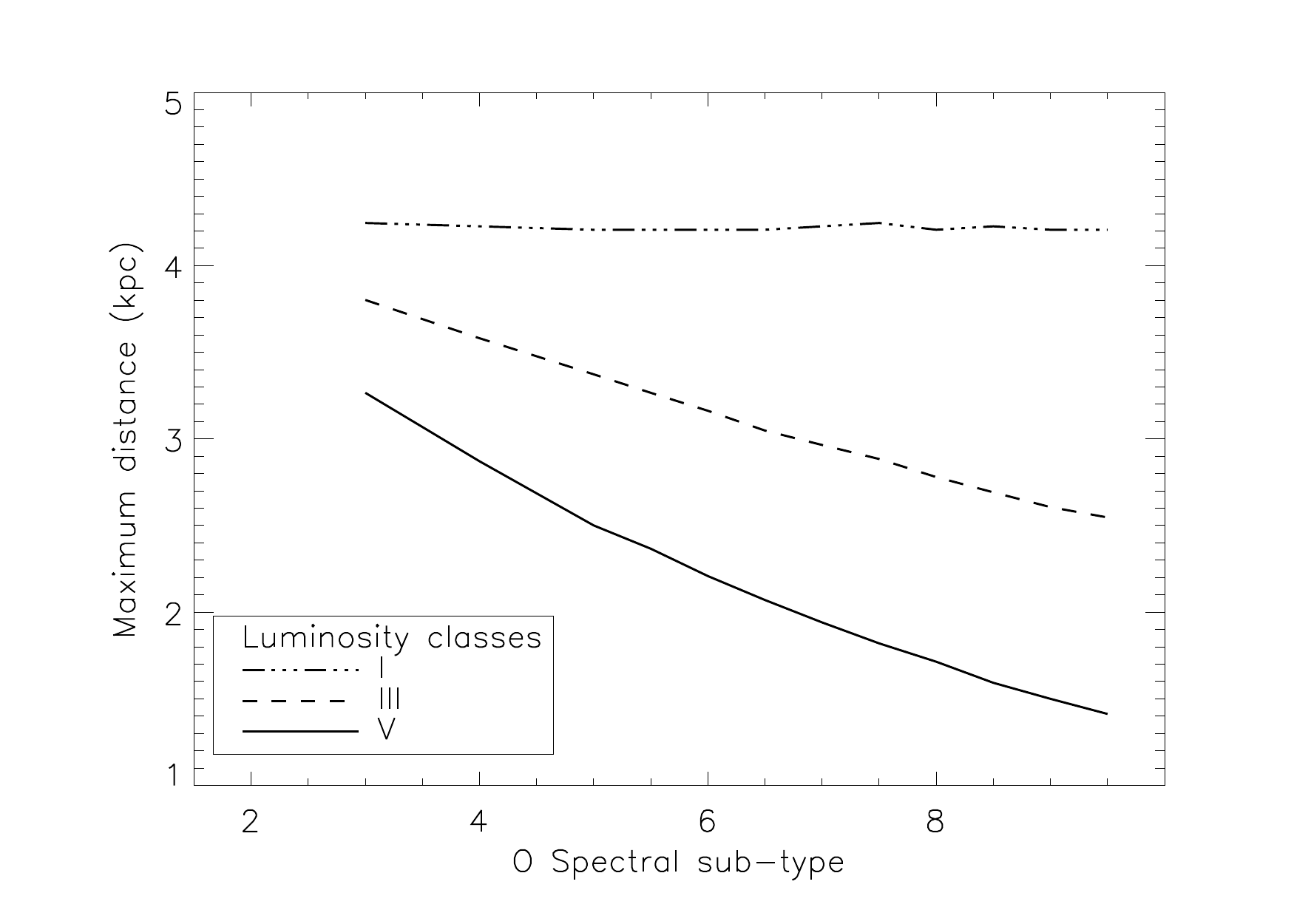}
\caption{ Maximum distance of a single star for its $H$-band apparent magnitude to be brighter than the \smash\ cutoff magnitude ($H=7.5$)  as a function of spectral sub-type and luminosity class. The figure ignores the effects of extinction and multiplicity, which act in opposite directions.}
\label{fig: dist}
\end{figure}

\begin{figure}
  \centering
  \includegraphics[width=0.47\textwidth]{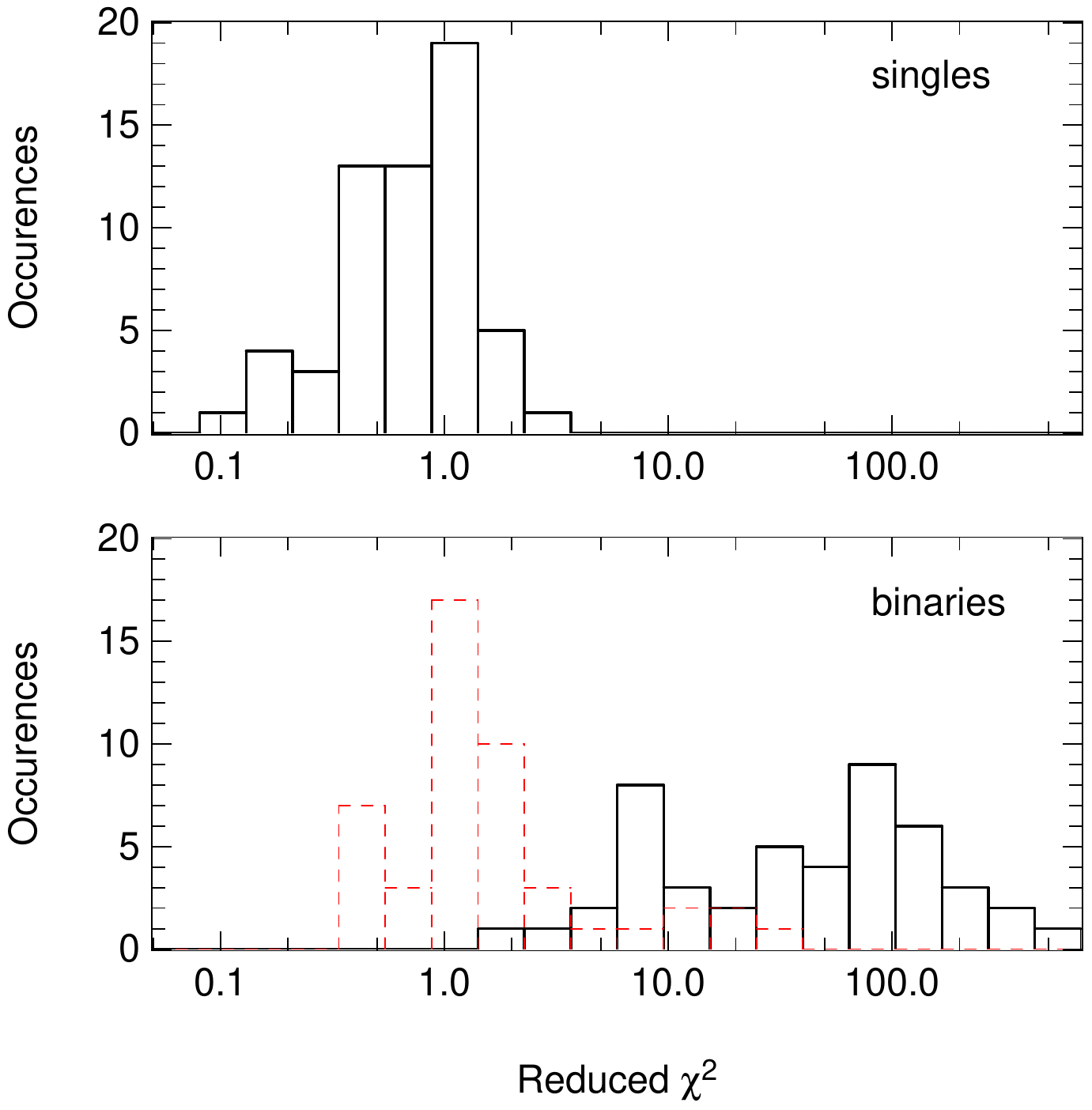}
  \caption{Distribution of the PIONIER reduced $\chi^2$ obtained with the single-star model (Eq.~\ref{eq:chi2_single}; solid line) for the unresolved targets ({\it upper panel}) and the resolved pairs ({\it lower panel}). The dashed line gives the distribution of the reduced $\chi^2$ obtained with the best-fit binary model.}
  \label{fig:histo_chi2}
\end{figure}

\begin{figure}
  \centering
  \includegraphics[width=0.47\textwidth]{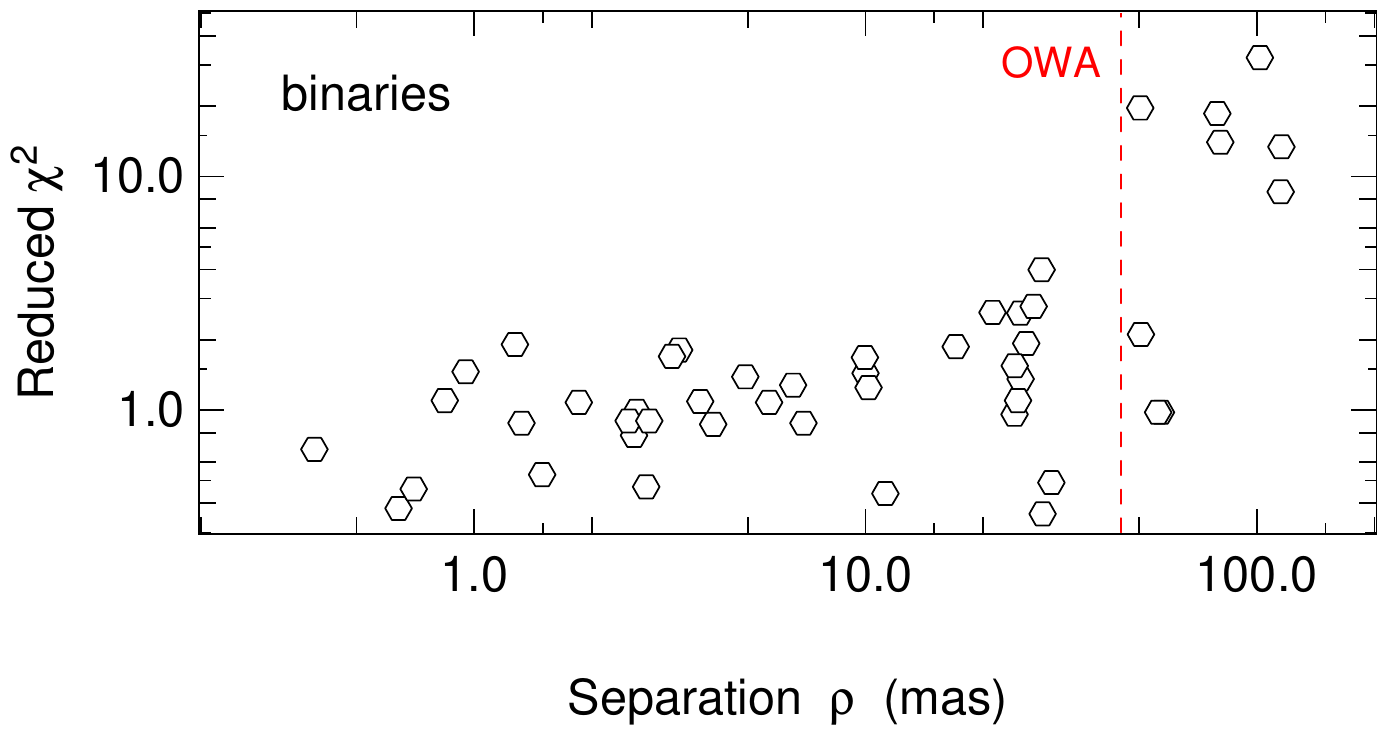}
  \caption{Reduced $\chi^2$ plotted against the separation for the best-fit binary models. Poor fits are only observed outside the outer working angle (OWA).}
  \label{fig:chi2_sep}
\end{figure}

\begin{figure*}
  \centering
  \includegraphics[width=0.48\textwidth]{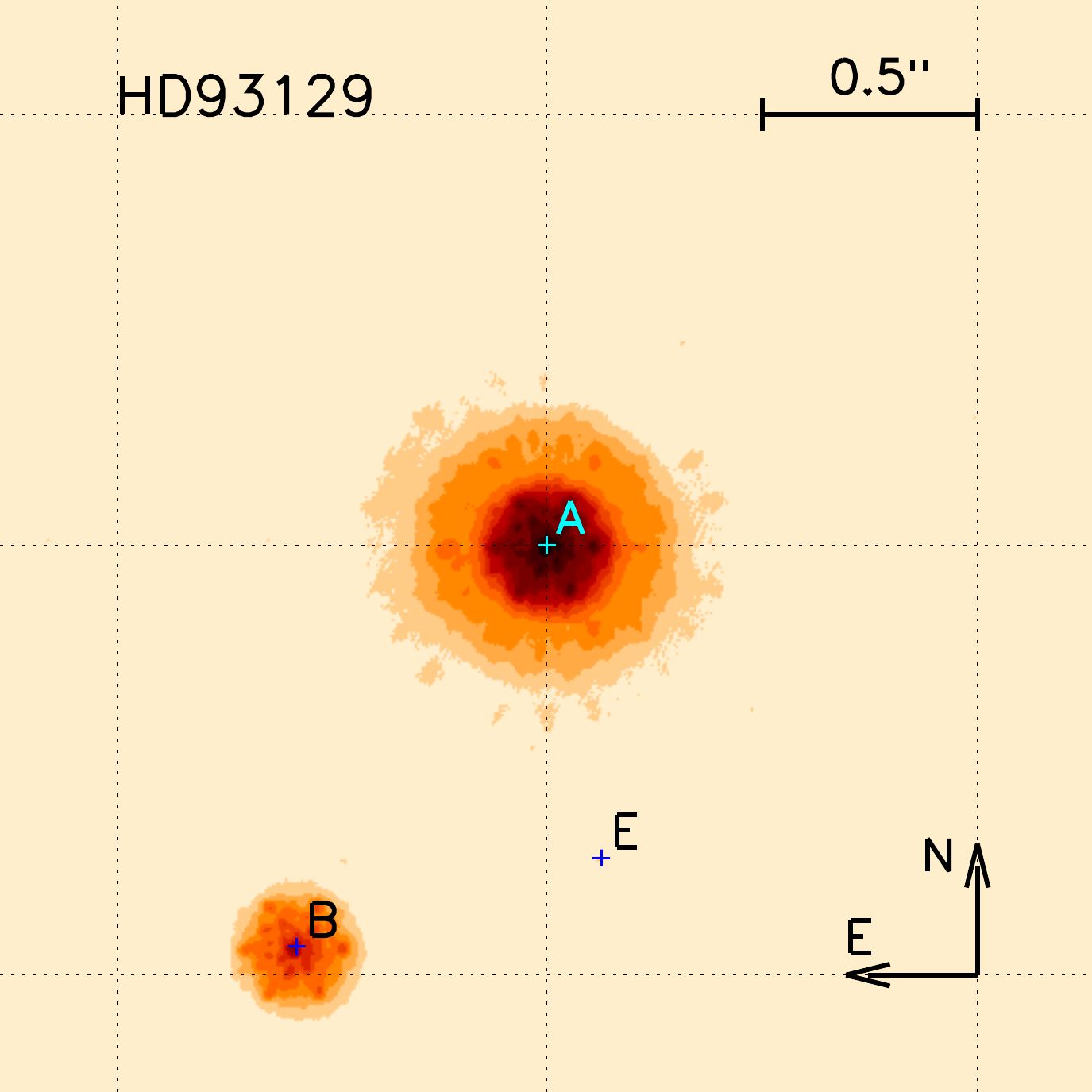}
  \includegraphics[width=0.48\textwidth]{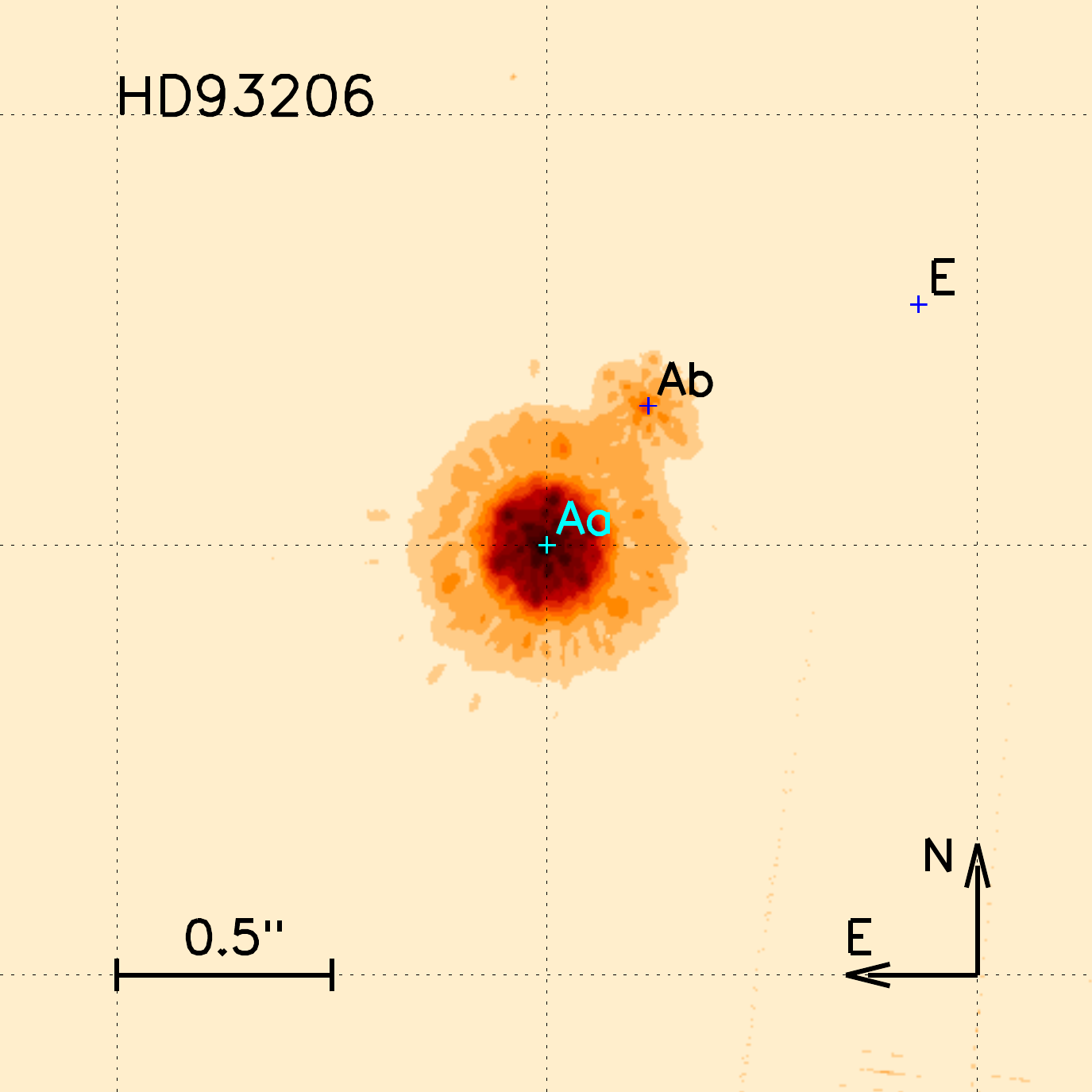}
  \includegraphics[width=0.48\textwidth]{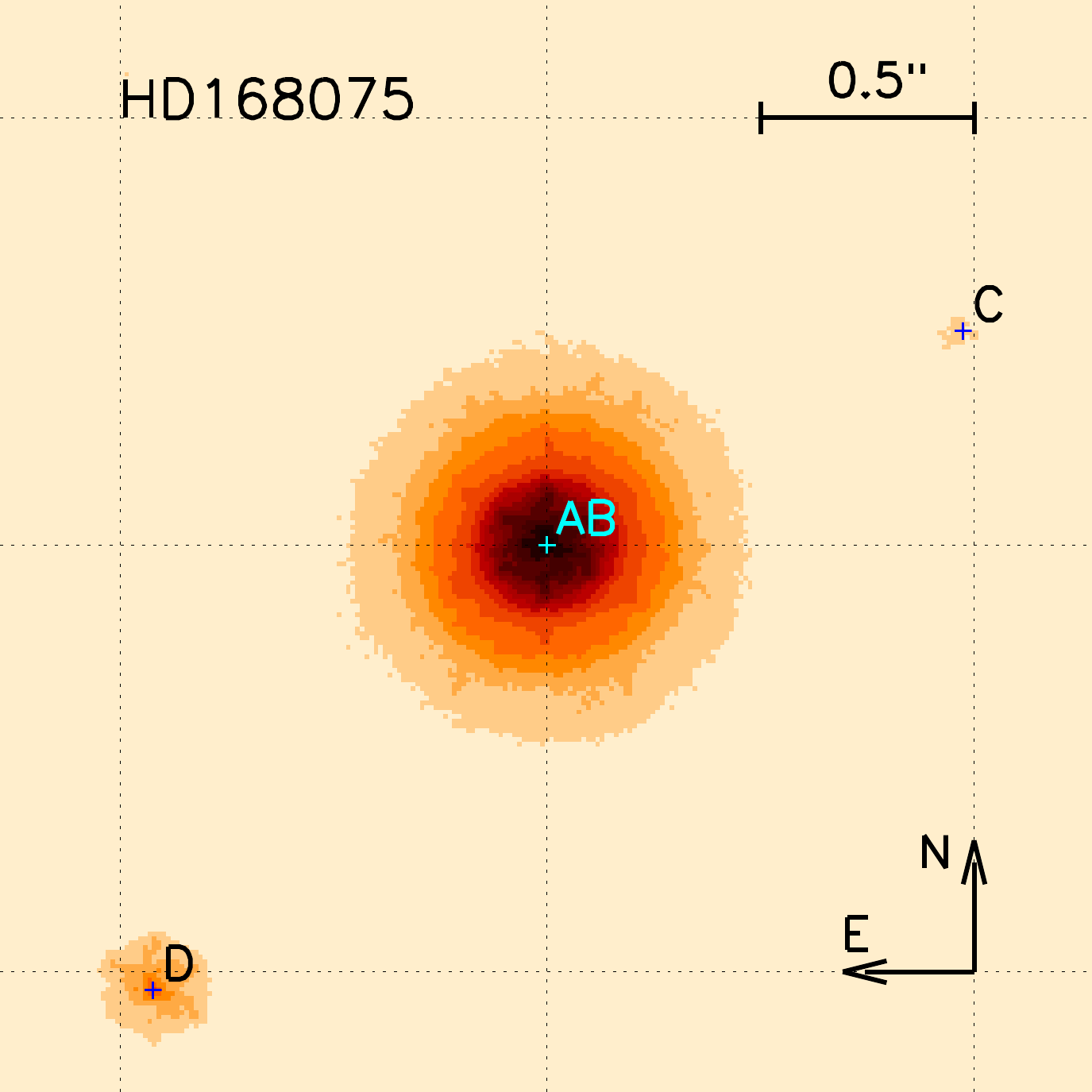}
  \includegraphics[width=0.48\textwidth]{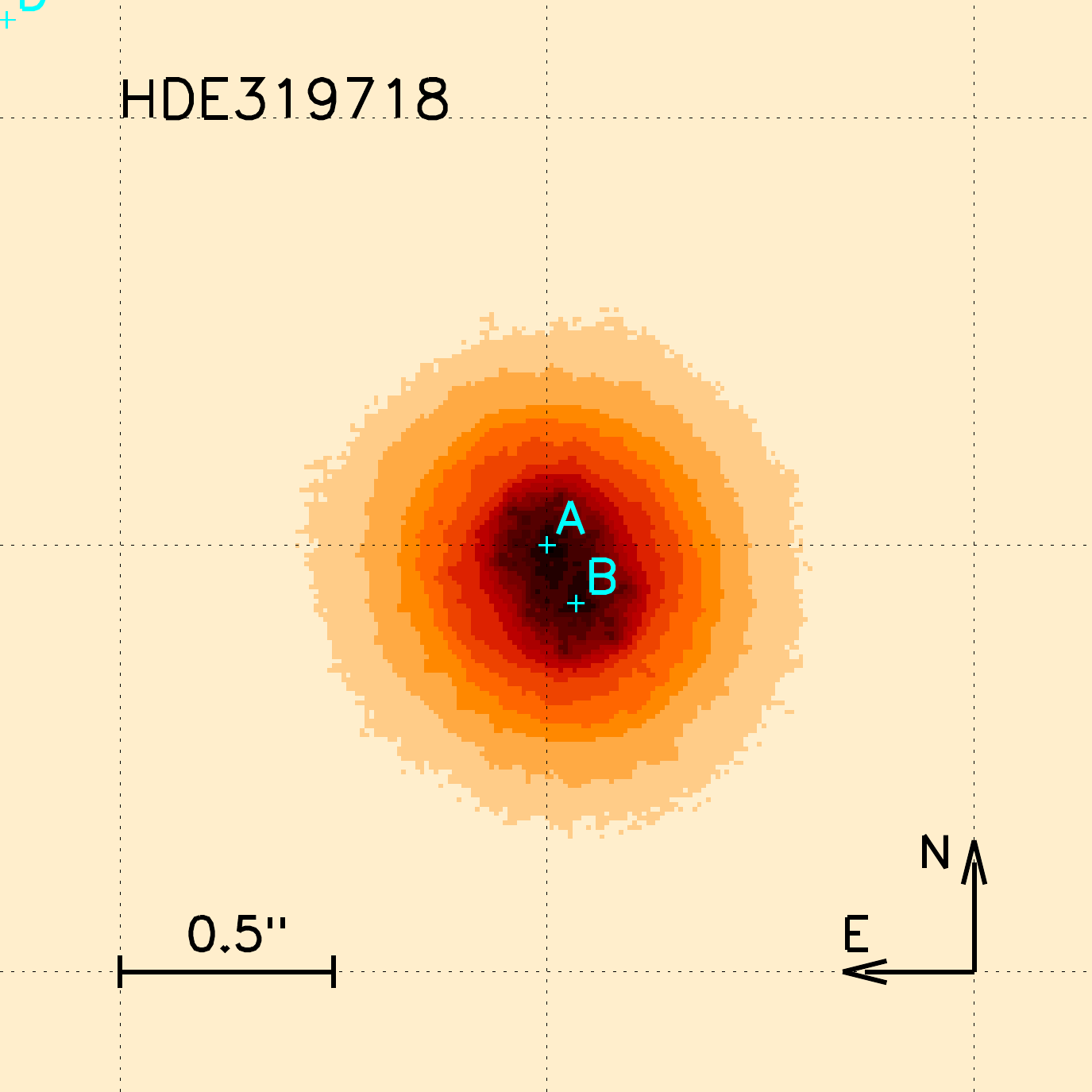}
  \caption{Examples of NACO data sets featuring the multiple systems HD\,93129, HD\,93206, HD\,168075 and HD\,319718. Only the central 2\arcsec of the NACO FOV are shown. The faint E components of HD\,93129 and HD\,93206 are not visible with the adopted cut but their positions are marked. }
  \label{fig: naco_img}
\end{figure*}

While the statistical accuracy is more limited due to the smaller sample sizes (Fig.~\ref{fig: spt}),  the fraction of resolved systems with companions within 200~mas seems smaller among supergiants than among dwarfs. We hardly identify any trends with spectral type. Inspection  of the cumulative distribution of the angular separations for different LCs (Fig.~\ref{fig: sep_cdf_lc}) confirms the larger fraction and the smaller separations of the companions observed for dwarfs: half  of our dwarf sample has a resolved companion within 20~mas and 76\%\ within 100~\mas. Equivalent fractions for giants and supergiants are about 33\%\ and 17\%\ and about 43\%\ and 41\%, respectively, suggesting a smooth transition from LCs V to I.  This conclusion is left unaffected by the inclusion of the spectroscopic companions. A similar trend is observed in the averaged fraction of companions which decreases from $f_\mathrm{c}^\mathrm{RES}=2.3 \pm 0.3$ to $1.9 \pm 0.3$ for LCs V to I. 

The decreasing multiplicity and companion fractions from LCs V to I may indicate that companions are lost over time, either as a result of disrupting dynamical interactions or because of binary evolution (coalescence). Alternatively, it may reflect an observational bias. Giants and supergiants are intrinsically brighter than dwarfs. This results in an increased contrast between the central star and its companion(s), so that the fainter ones may end up beyond our current contrast limits. 

\begin{figure*}
  \centering
  \includegraphics[width=\textwidth]{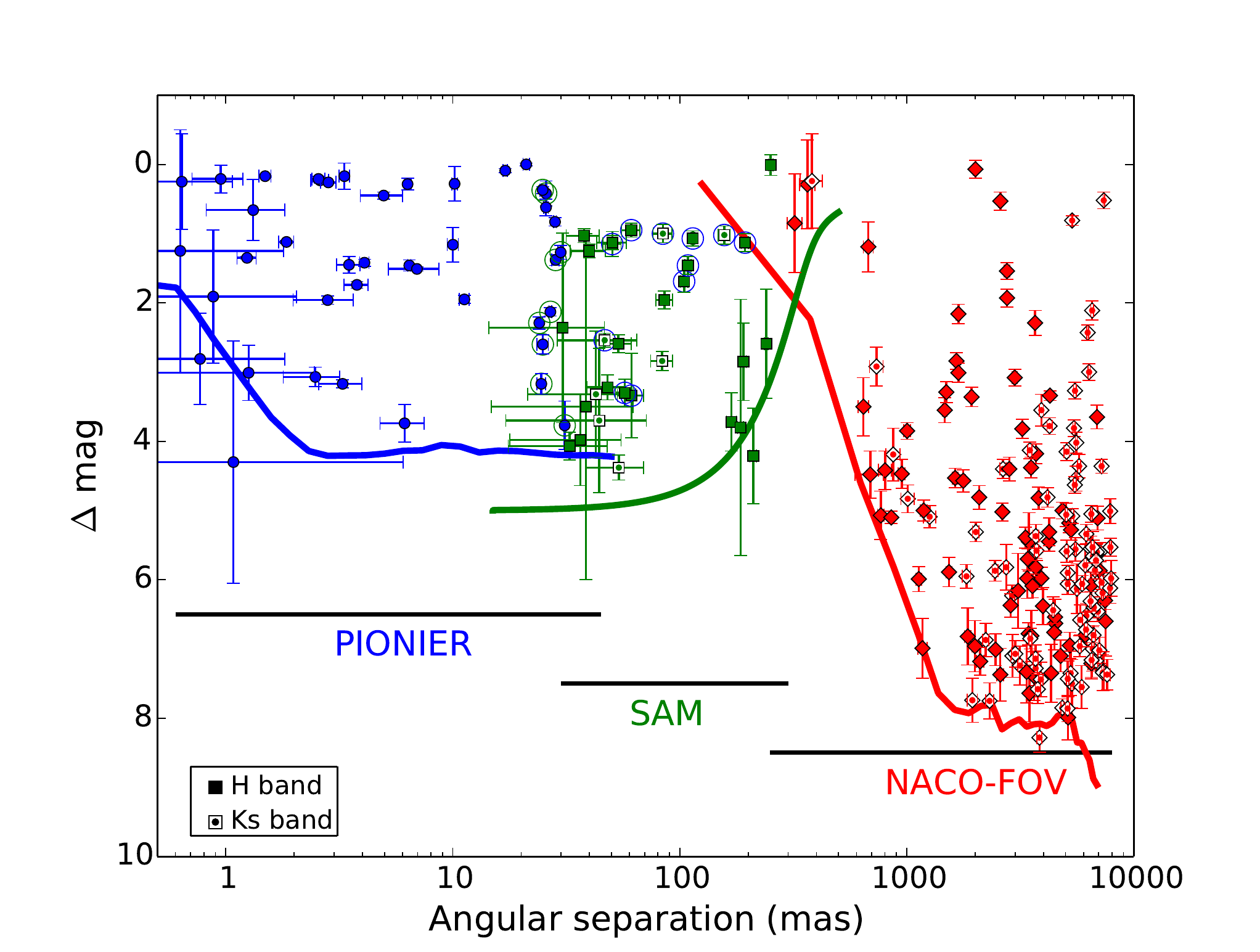}
  \caption{Plot of  the magnitude difference ($\Delta$ mag) vs.\ angular separations ($\rho$) for the detected pairs. Only one detection per object has been considered, and the $H$-band has been preferred whenever available. The solid lines indicate the median $H$-band sensitivity of our survey across the different separation ranges. The $K\mathrm{s}$ sensitivity curves are similar. Different colors indicate observations with different  instrumental configurations (PIONIER: blue, NACO/SAM: green, NACO FOV: red) while different symbols indicate different observational bands  ($H$: filled, $Ks$: open).  Large circles indicate objects detected by both SAM and PIONIER.}

  \label{fig: rho_f}
\end{figure*}

\begin{figure}
  \centering
  \includegraphics[width=0.49\textwidth]{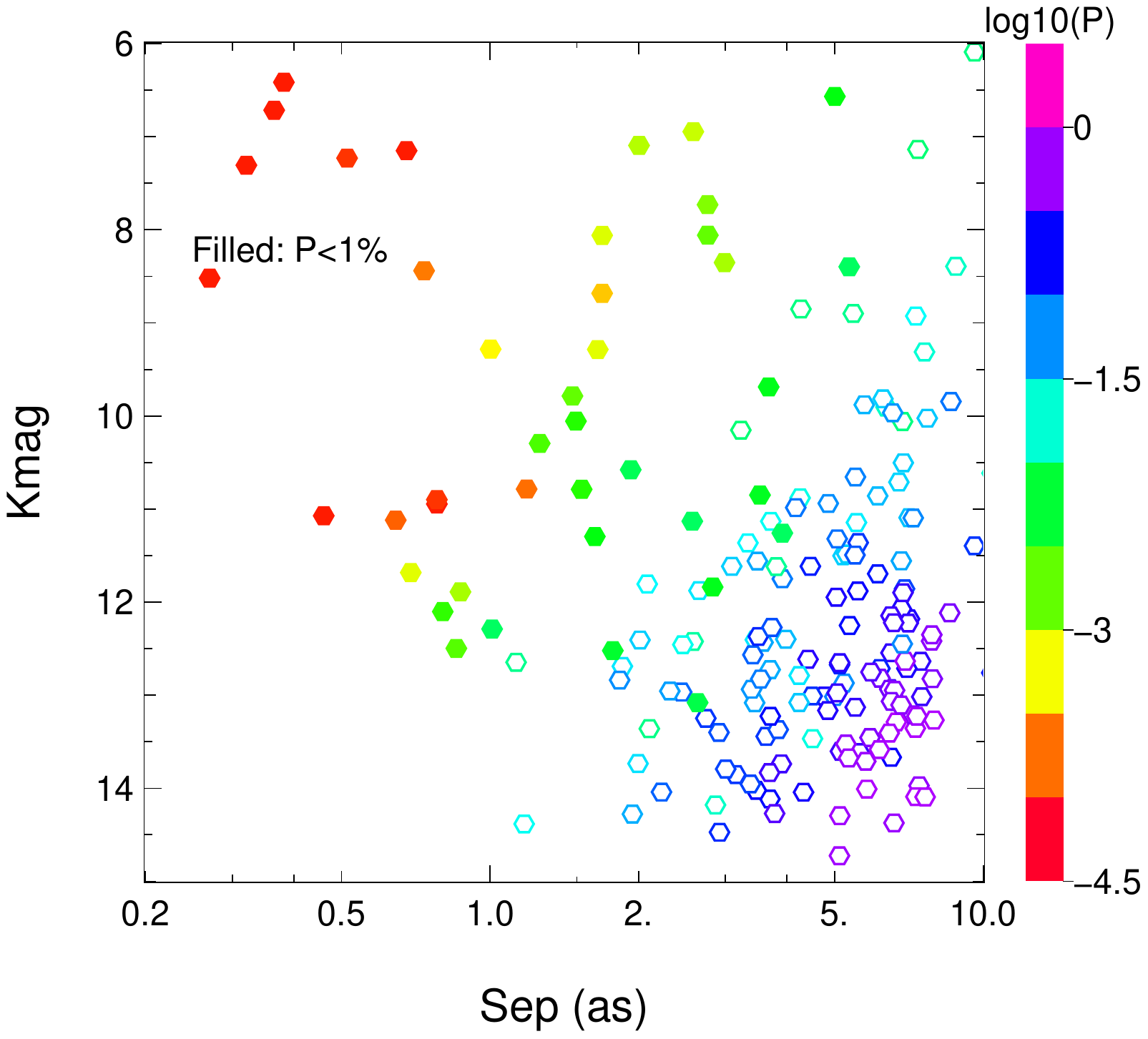}
  \caption{Probability $P_\mathrm{spur}$ for the companions detected in the NACO FOV to result from chance alignment. The probability  $P_\mathrm{spur}$ is color-coded  in the \Ks-band magnitude vs.\ angular separation plane. Filled symbols indicate $P_\mathrm{spur} < 0.01$ while open symbols are used for $P_\mathrm{spur} \geq 0.01$.
}
  \label{fig: Pspur}
\end{figure}

\begin{figure*}
  \centering
  \includegraphics[width=\textwidth]{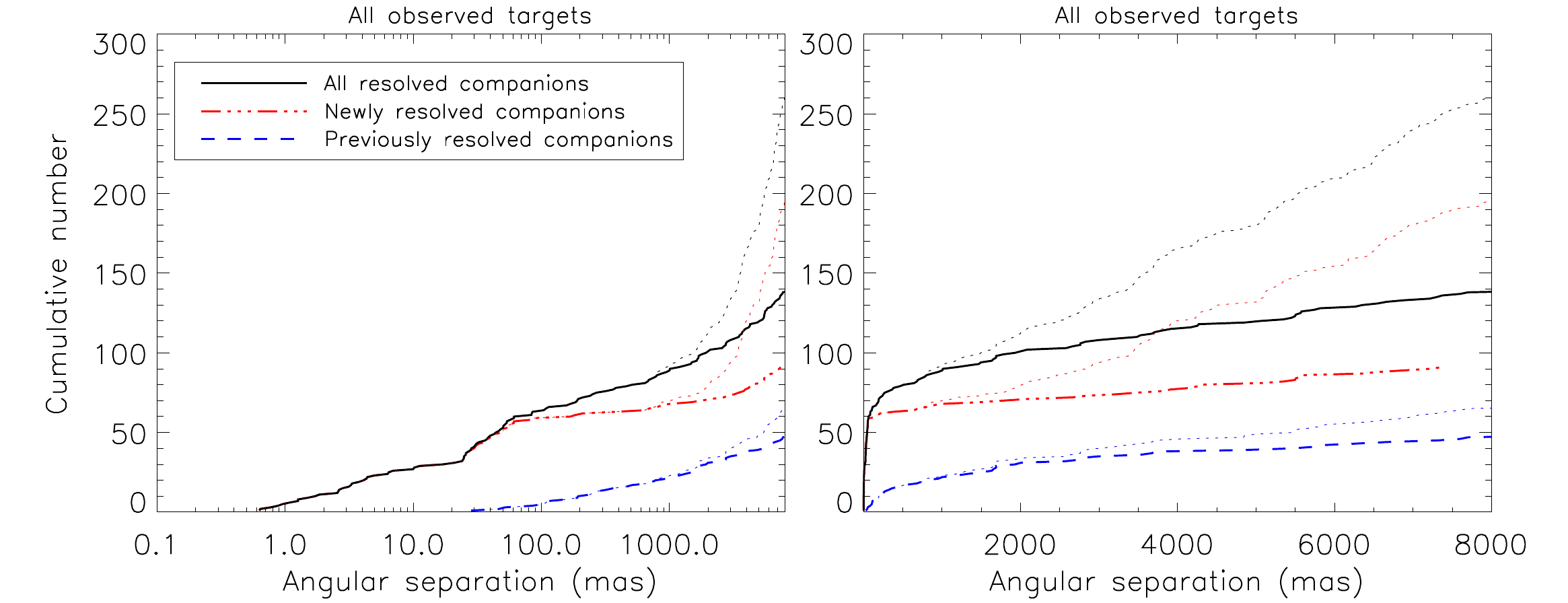}
  \caption{ Cumulative number distributions of the companion separations in logarithmic ({\it left panel}) and linear ({\it right panel}) scales. Dash-dotted, dashed and solid curves indicate the considered samples: companions known before \smash, new companions detected in the course of \smash\ and the combination of both, respectively. Thick lines restrict companions to $\Delta \mathrm{mag} <  5$ whereas thin dotted lines separting at $\rho \gtrsim 1$\arcsec have no contrast selection. }
  \label{fig: sep_cdf}
\end{figure*}

\begin{figure}
  \centering
  \includegraphics[width=0.49\textwidth]{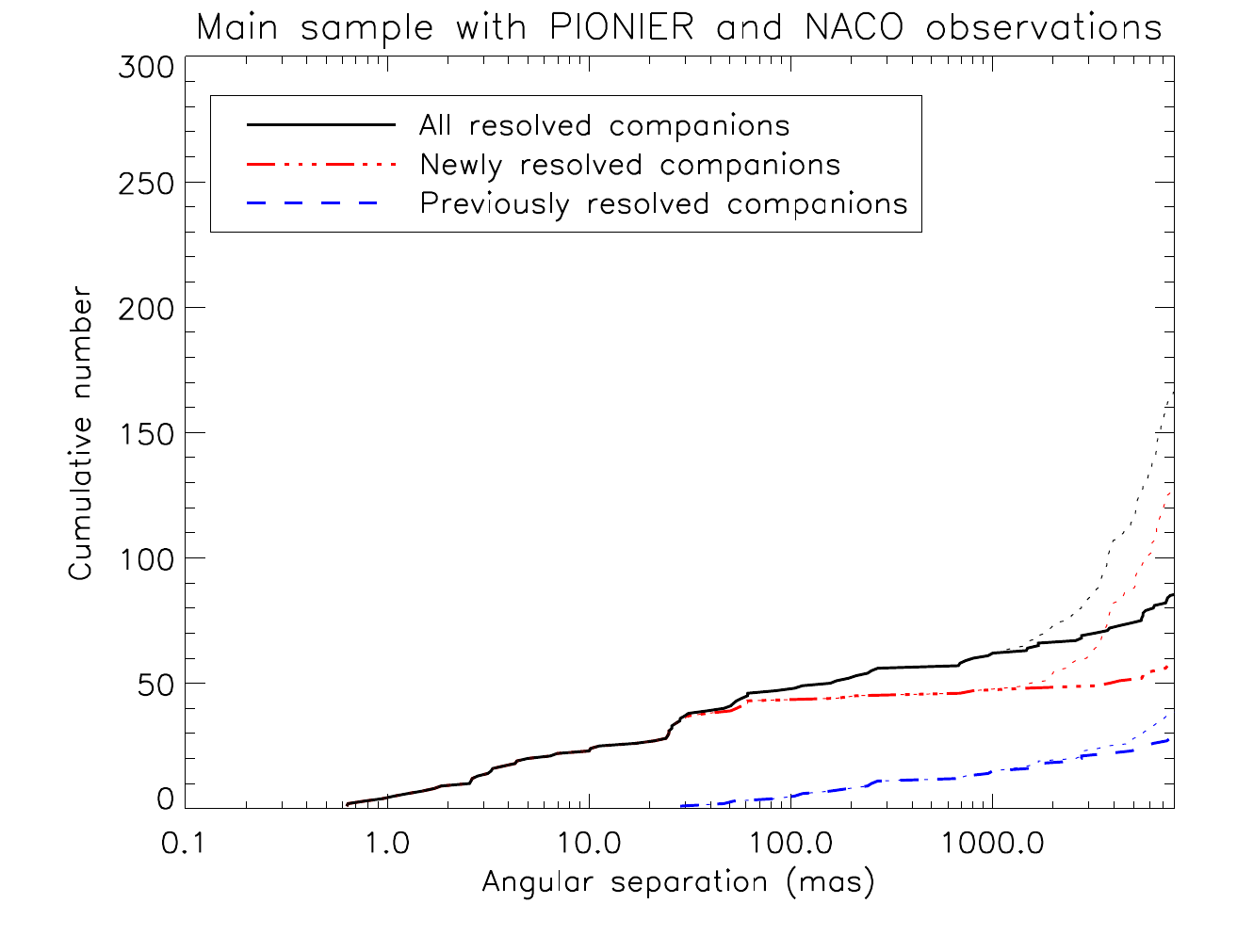}
  \caption{Same as Fig.~\ref{fig: sep_cdf} but restricted to the 96 targets from our main sample that have been observed both with PIONIER and NACO.}
  \label{fig: sep_cdf_main}
\end{figure}

\begin{figure}
  \centering
  \includegraphics[width=.47\textwidth]{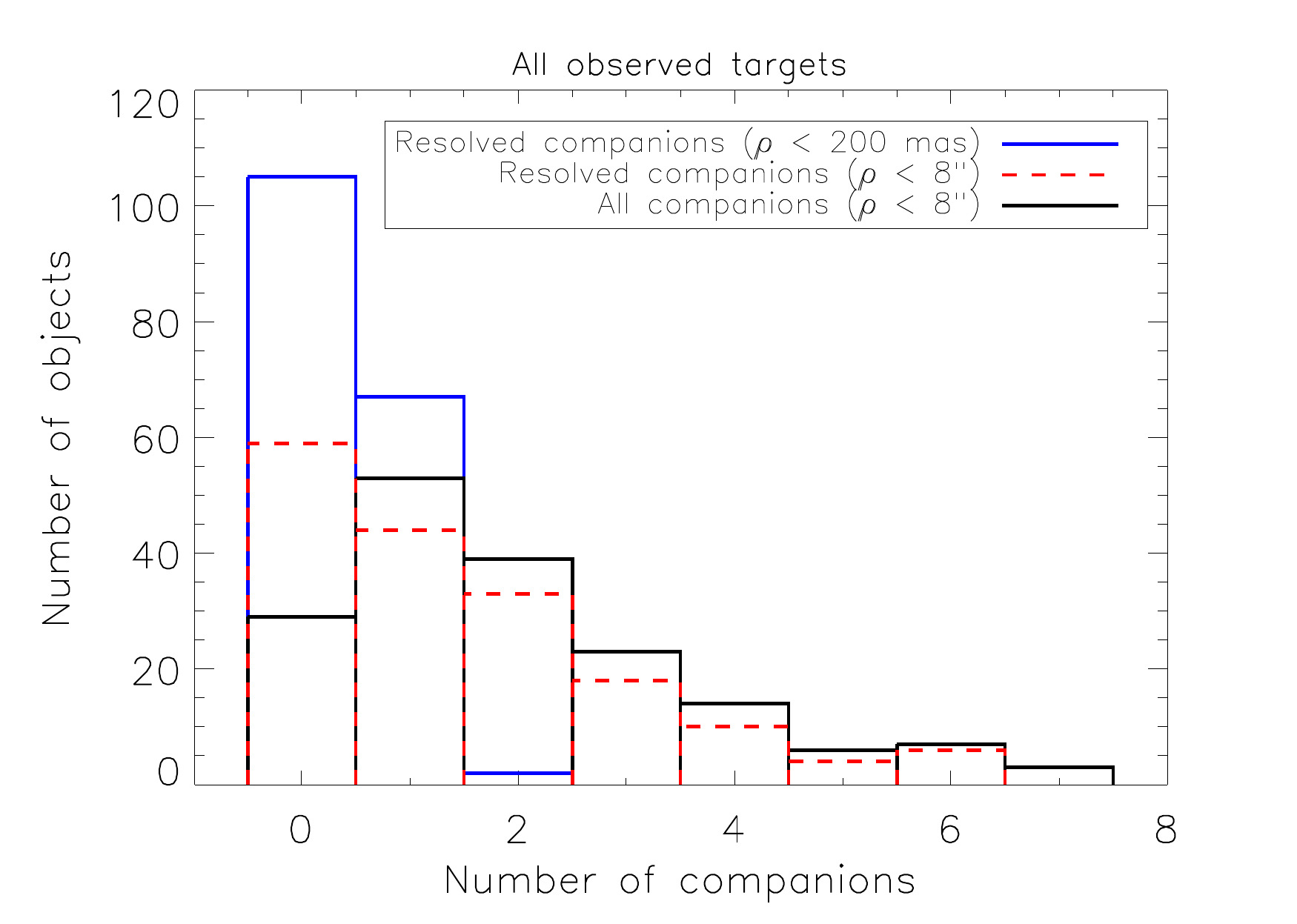}
  \caption{Histogram of the number of companions per target for the full separation range and for separations below 200~mas. All companions account for resolved companions within 8\arcsec\ as well as for known spectroscopic and eclipsing companions.}
  \label{fig: Ncomp}
\end{figure}

\begin{figure}
  \centering
  \includegraphics[width=.47\textwidth]{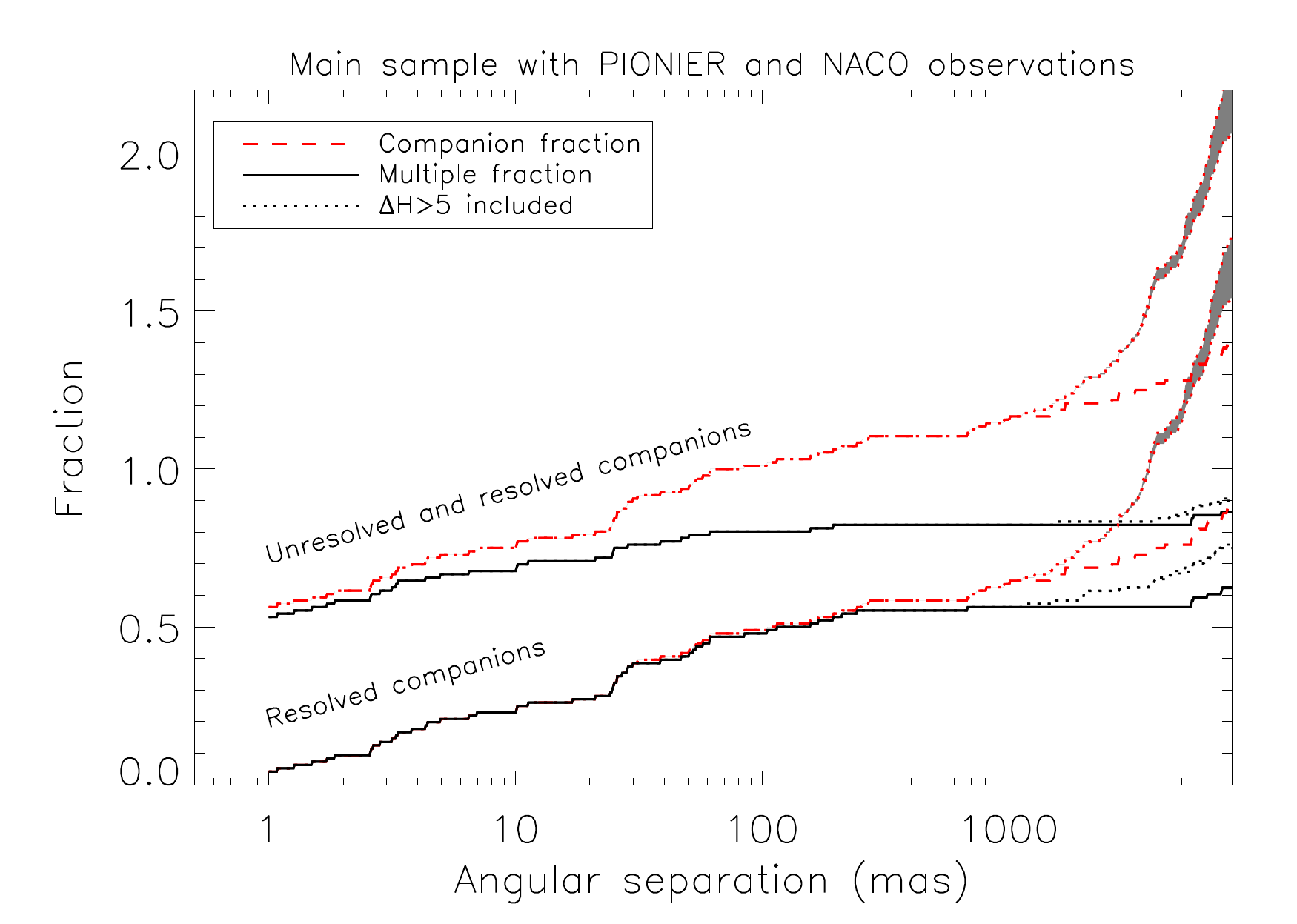}
  \caption{Cumulative fraction of multiple systems (solid line) and average fraction of companions per star (dashed-dotted lines) for increasing angular separations. The upper curves account for the spectroscopic and eclipsing companions whereas the bottom ones do not. Shaded gray areas indicate the statistical contribution of spurious detections due to chance alignment.}
  \label{fig: sep_cdf2}
\end{figure}

\begin{figure}
  \centering
  \includegraphics[width=.47\textwidth]{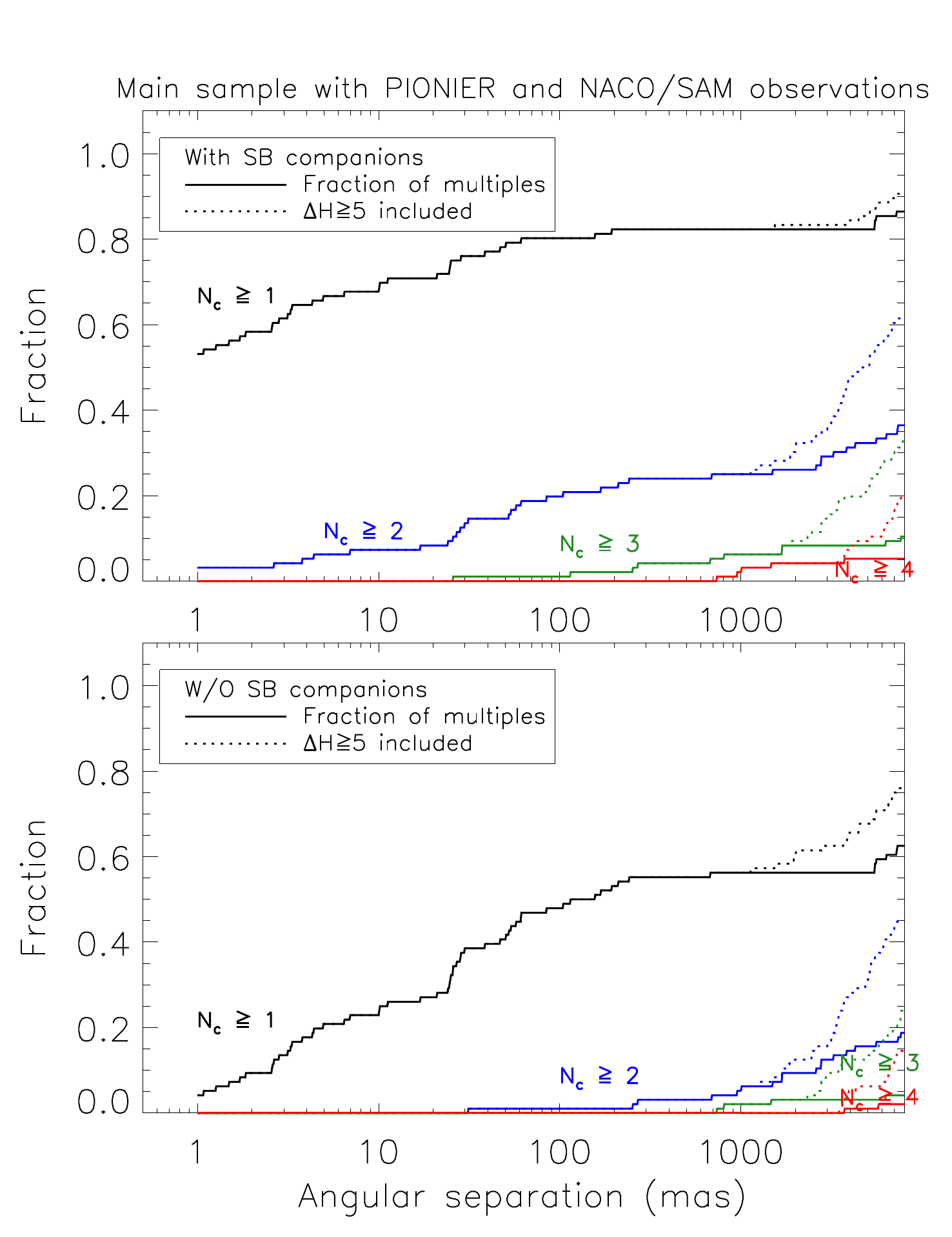}
  \caption{Cumulative fraction of multiple systems for a minimum number of companions of 1, 2, 3 and 4. The top panel includes the unresolved spectroscopic and eclipsing companions whereas the bottom panel does not.}
  \label{fig: triple}
\end{figure}

\begin{figure}
  \centering
  \includegraphics[width=.47\textwidth]{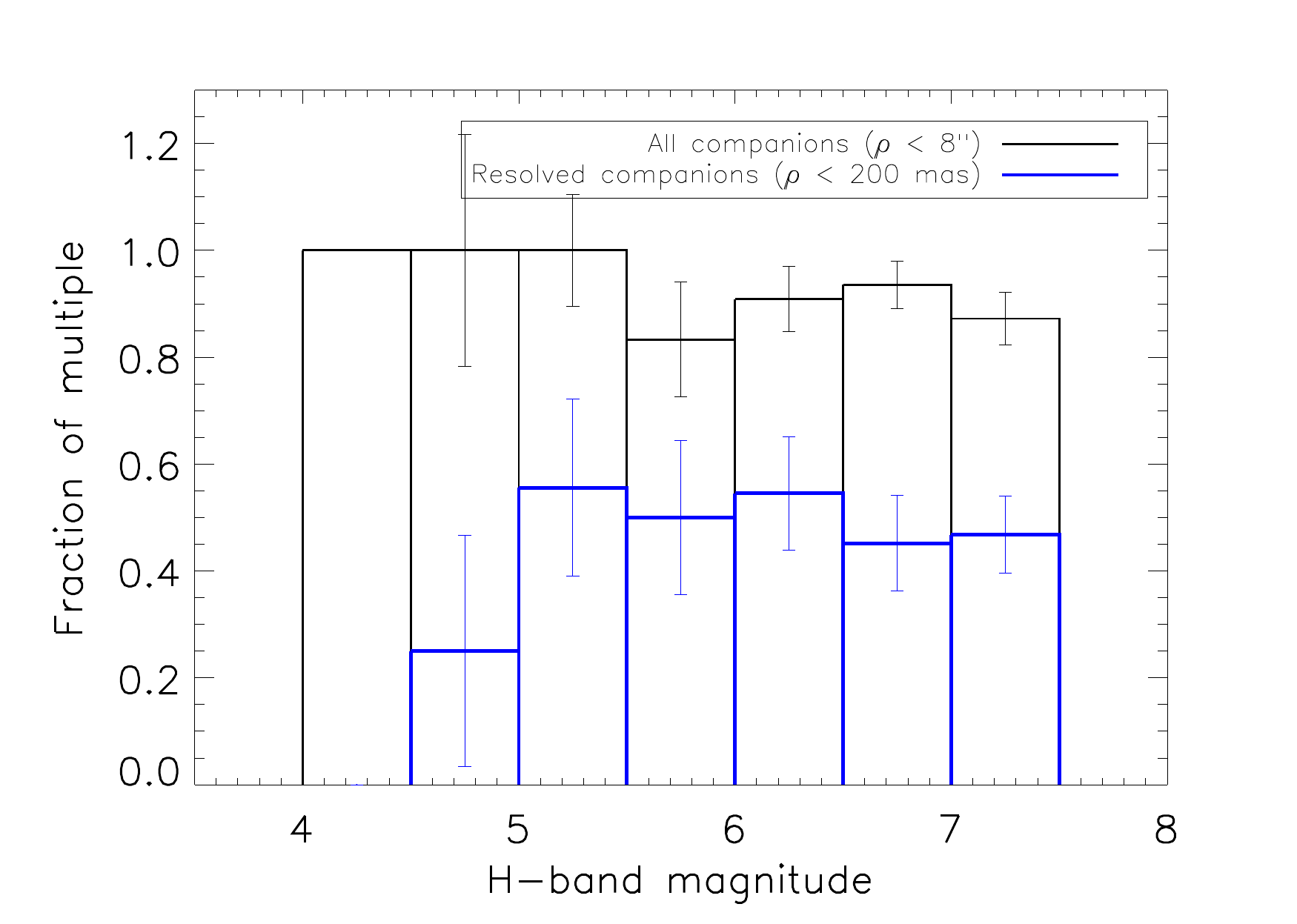}
  \includegraphics[width=.47\textwidth]{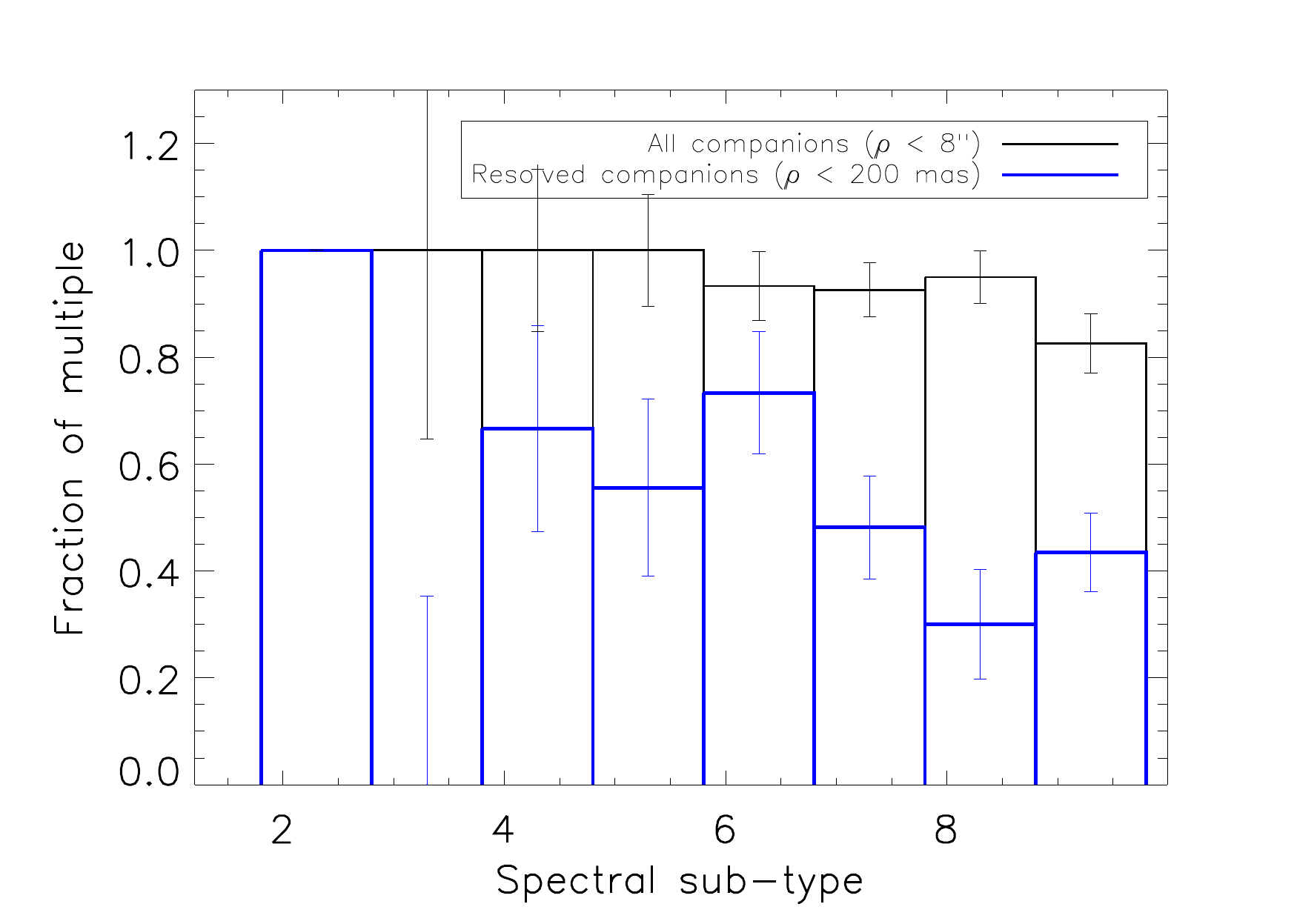}
  \includegraphics[width=.47\textwidth]{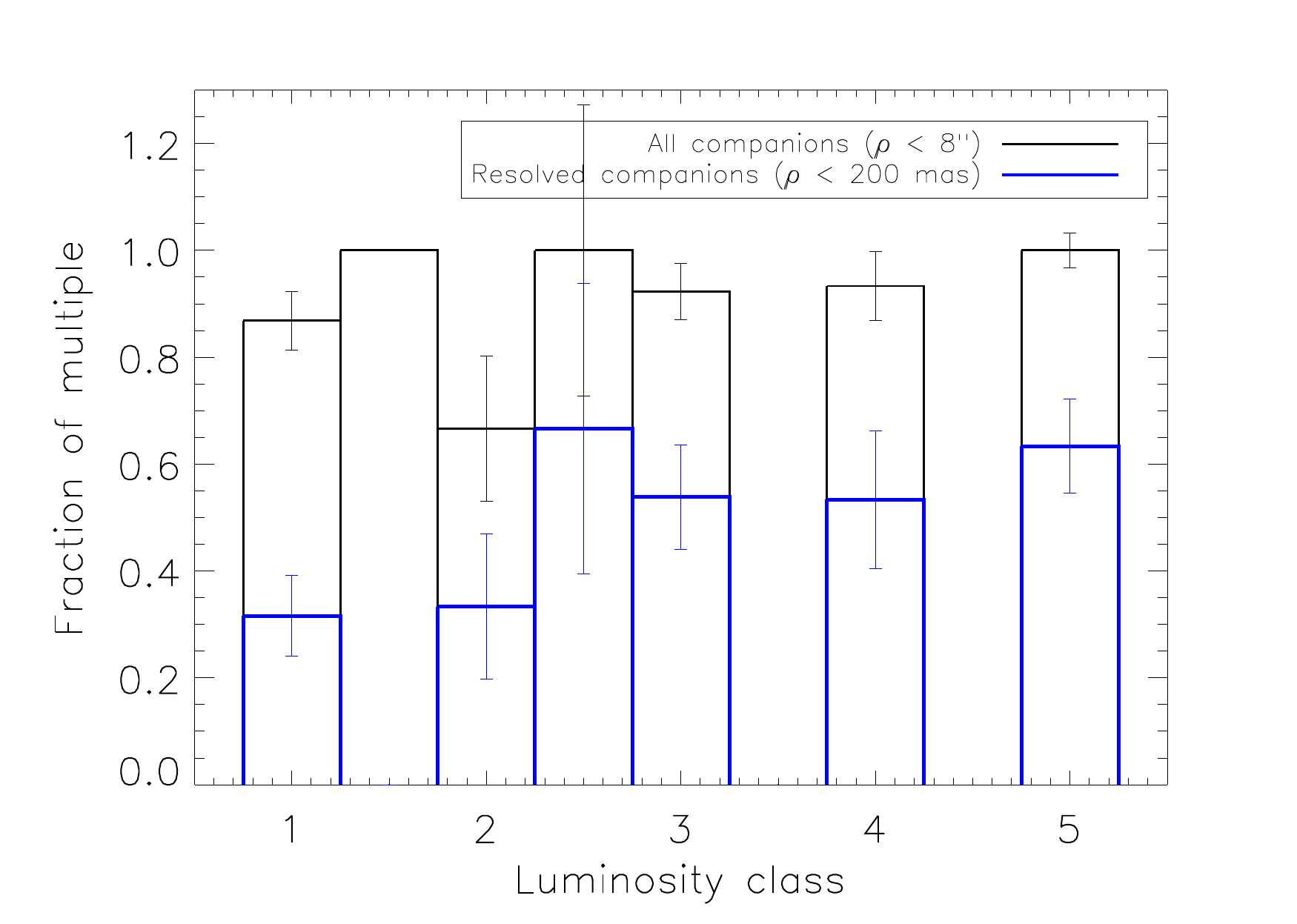}
  \caption{Fraction of multiple systems as a function of their $H$-band magnitude ({\it top panel}), of their spectral sub-type ({\it middle panel}) and of their luminosity class ({\it bottom panel}). }
  \label{fig: spt_kmag}
\end{figure}

\begin{figure}
  \centering
  \includegraphics[width=.47\textwidth]{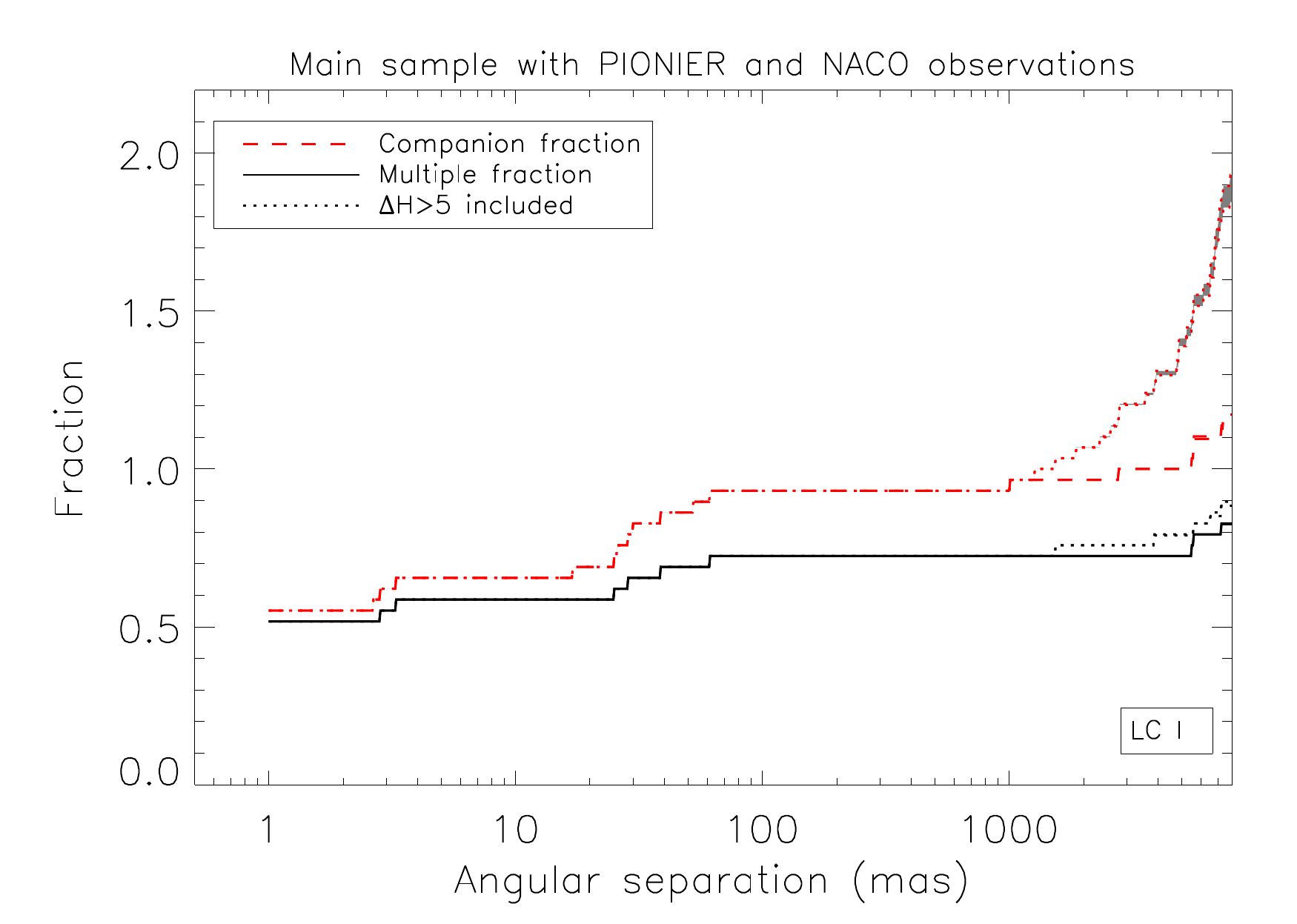}
  \includegraphics[width=.47\textwidth]{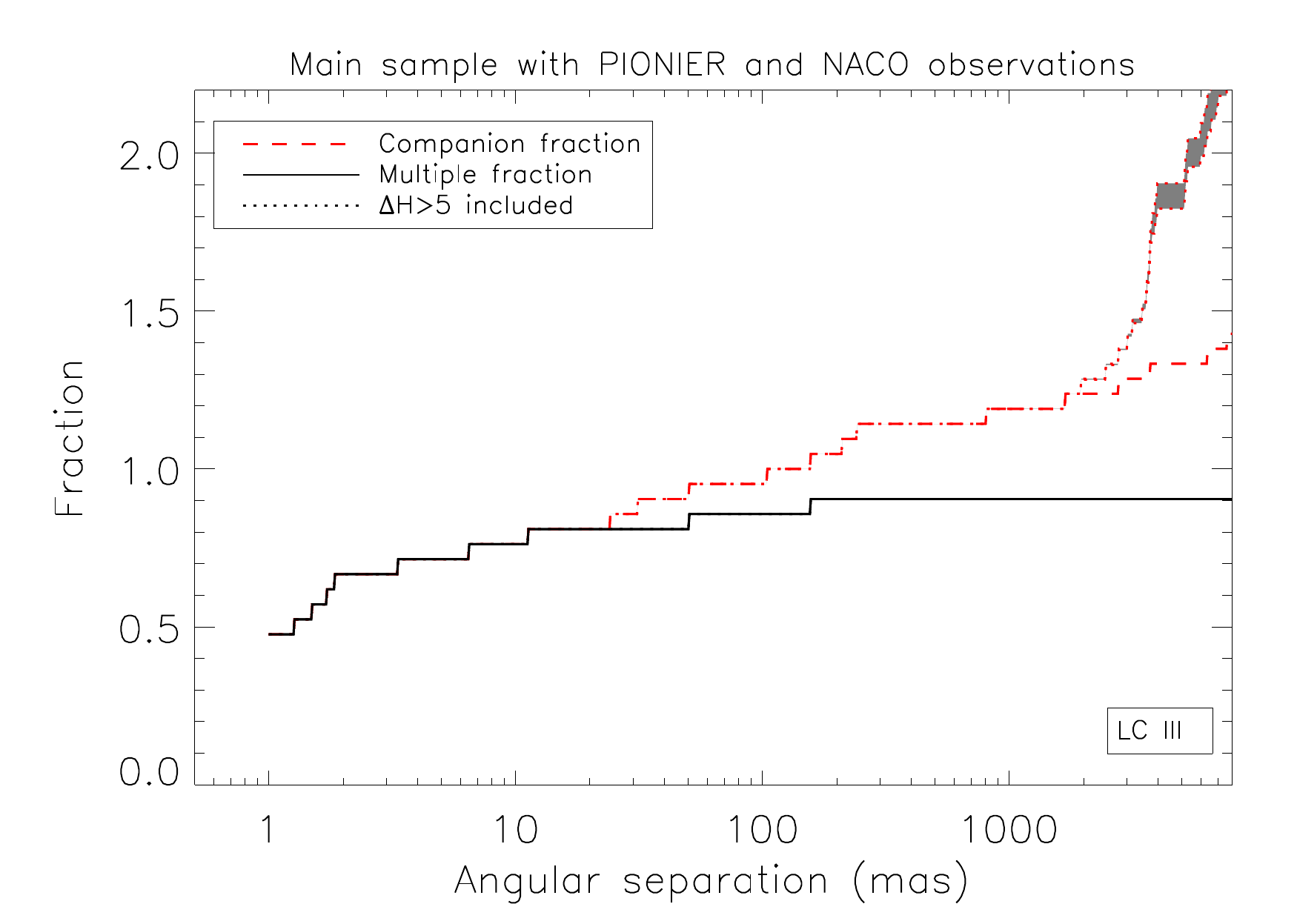}
  \includegraphics[width=.47\textwidth]{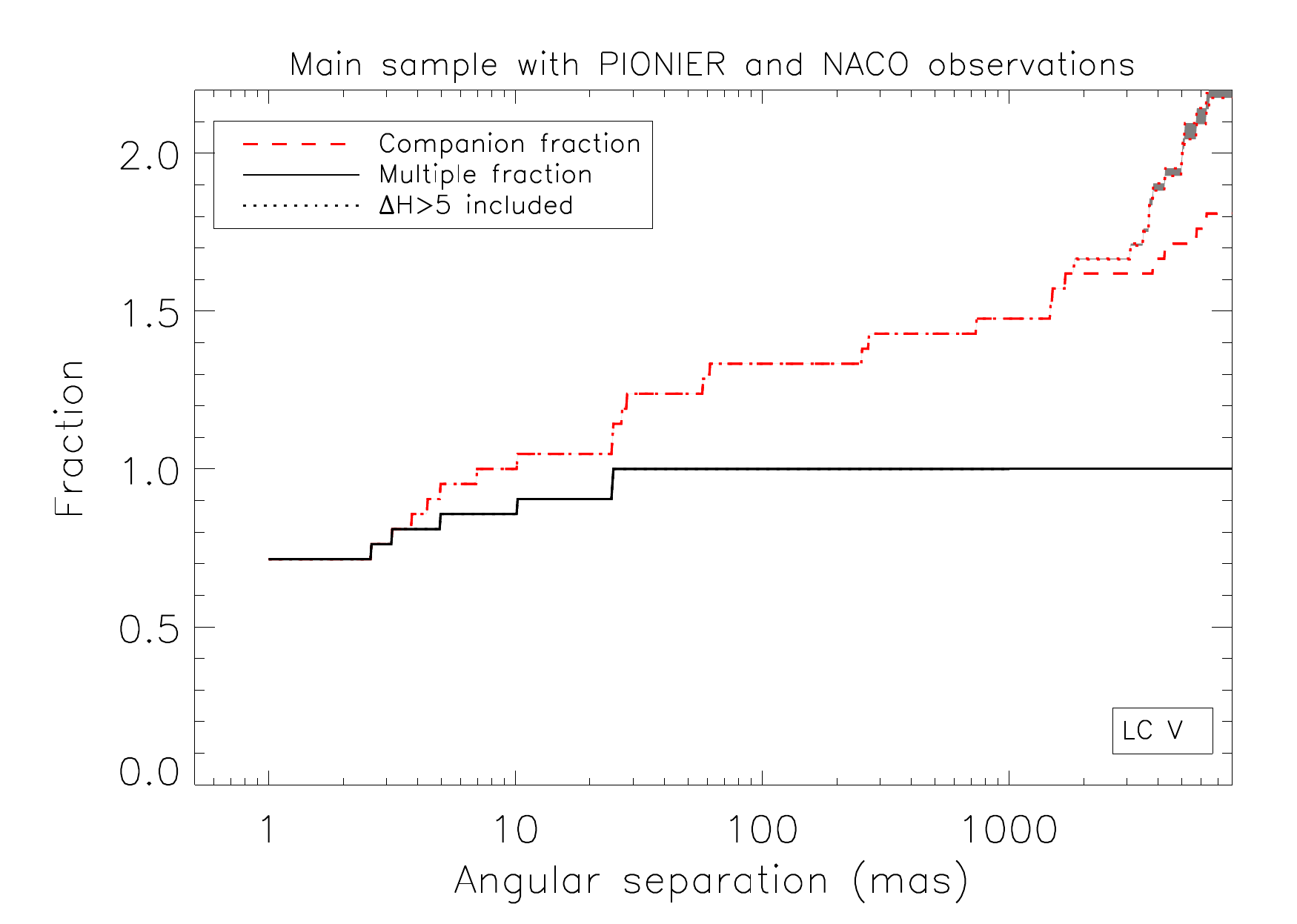}
  \caption{Break down of Fig.~\ref{fig: sep_cdf2}  for  luminosity classes I, III and V. The curves include the unresolved E/SB companions.}
  \label{fig: sep_cdf_lc}
\end{figure}

To check the possible impact of such an observational bias on our results, we perform the following Monte Carlo experiment. We randomly assign to the supergiants in our sample a population of companions with properties drawn from the dwarf sample (LC~V) and we record the impact on the $f_\mathrm{m}^\mathrm{1-200mas}$ and $f_\mathrm{c}^\mathrm{1-200mas}$ fractions accounting for our average detection limits (Fig.~\ref{fig: rho_f}). We obtain that the multiplicity fraction $f_\mathrm{m}^\mathrm{1-200mas}$  will, on average, drop from 0.76 to 0.52 and that the fraction of companions $f_\mathrm{c}^\mathrm{1-200mas}$  will, on average, drop from 0.76 to 0.54. This is in good agreement with the trends observed in Table~\ref{tab: fraction} and may account for an increase of an additional 5\%\ of the overall multiplicity fraction of the entire sample (i.e., from $f_\mathrm{m}^\mathrm{RES}=0.91$ to 0.96 after such bias correction). The observed differences in the multiplicity properties of LC V and I stars are thus fully compatible with the expected increased contrast between supergiants and their nearby companions. This implies that there may not be any significant difference in the multiplicity properties of different luminosity classes for separations $\rho \gtrsim 1$~mas.

\section{Discussion} \label{sect: discuss}

\subsection{Constraints on formation}\label{sect: form}

A detailed comparison with massive star formation theories will follow in subsequent work, as it relies on the bias corrected data, and estimates of period and mass ratio distributions. Due to distance uncertainties we cannot yet provide estimates of the physical separation for all companions. However based on the maximum distances of objects in the sample (3.5~kpc), the angular separations  of 1, 200, and 8000~mas correspond to maximum projected distances of 3.5, 700, and 28\,000~AU. Our results thus indicate that respectively 49, 82 and 91\%\ of our sample have at least one companion at physical distance less than 3.5, 700 and 28\,000~AU. All the dwarfs in our sample have a companion within 105~AU.  Even without bias correction it is clear from the 100\% companion fraction of the dwarfs that massive stars (almost) universally form in binaries or higher order multiples. Moreover, as we describe below, the abundance of dwarf companions found at $<100$AU is compatible with disk fragmentation as a binary formation channel.

Multiplicity is a natural consequence of the high infall rates that are predicted by the theories of massive star formation \citep{MKT03,BVB04}. High infall rates can lead to massive, gravitationally unstable disks, which in turn fragment to produce one or more bound objects that typically grow to stellar masses \citep{KM06,KKmK07,KMK10}. The separation of these companions should be comparable to disk sizes, which are typically 100's of AU. The high accretion rates, which promote binary formation in disks, are consistent with those observed \citep{KTB12}, and those seen in high resolution radiation hydrodynamic models \citep{KKM12}. Massive turbulent core may also fragment on sub-pc scales early in the collapse phase \cite{MKT03}. This prompt fragmentation might be responsible for the wider binaries. Dynamical interactions are also invoked to explain binaries at a range of separations. For recent, broader reviews of theories of massive star formation, see \cite{ZiY07, TBC14}.

\subsection{Runaway stars}\label{sect: rw}

The GOSC-v2 catalog flags 13 of our targets as runaway stars (12 in the main sample and one in the supplement). The bulk of the runaway sample is formed by supergiants (7) and bright giants (2). Only six runaway objects have both PIONIER and SAM observations. Five are missing SAM observations and two could not be observed with PIONIER. Interestingly, none of the runaway stars have companions resolved by PIONIER or SAM. Only two of them (HD\,156212 and HD\,163758) have faint and rather distant  companions  \citep[$\Delta H >7$, $\rho=1.7-7.4$\arcsec]{TMH10}, all of them with a significant spurious detection probability:  0.28 and 0.69 for the two companions of HD\,156212 and 0.07 for HD\,163758). Correcting for the spurious detection probability, this leaves us with a multiplicity fraction in the range 1-8000~mas of $f_\mathrm{m}^\mathrm{R}=0.16\pm0.08$ only, i.e.\ significantly lower than the fraction of  $0.75\pm 0.04$  for the main sample. While our observations are not fully sampling the separation range, the differences are large enough to conclude that wide multiple systems are likely to be disrupted during the event creating the runaway star.

\subsection{The interferometric gap}\label{sect: gap}
Eighteen long-period spectroscopic binaries have been spatially resolved in the course of \smash\ and are discussed separately in the appendix. This data provides an opportunity to obtain three-dimensional orbits  upon continuation of interferometric monitoring with the VLTI. Importantly, PIONIER  has straightforwardly resolved every single of the  known spectroscopic binaries with orbital period ($P_\mathrm{orb}$) longer than 150\,d. This clearly demonstrates that the gap between the period/separation distributions of spectroscopic binaries and visual/astrometric binaries described in \citet{MGH98} has now been bridged for distances typical of our sample. This opens up the study of multiplicity properties across the continuous range of separations from several stellar radii to thousands of AU, including the shape of the period and mass-ratio distributions within the interferometric gap. The next challenge will be to push the detection limits,  both for spectroscopic and visual pairs, in order to probe the regime of faint, lower mass companions, i.e.\ those with a larger magnitude difference.

\subsection{Non-thermal radio emitters}\label{sect: nrt}

Non-thermal  radio emitters display an excess emission compared to the expected power law ($S_\nu \propto \nu^\alpha$) that describes the tail of their thermal spectral energy distribution in the radio domain.  In practice, O stars are reported as non-thermal radio emitters when their spectral index $\alpha$ is smaller than 0.6, which is the value expected for homogeneous winds with a $\beta$-type velocity stratification \citep[for a review, see][]{DeB07}.

This non-thermal radio emission is believed to be synchrotron radiation generated by relativistic electrons in the presence of a magnetic field. The mechanism requires the presence of strong hydrodynamic shocks, in the vicinity of which electrons are accelerated to relativistic velocities through the Fermi mechanism. Furthermore, the synchrotron emission needs to be produced outside of the radiosphere of the star to be observable, or  otherwise it would be re-absorbed by the wind material. This implies distances of tens to hundreds of stellar radii from the star, depending on the considered radio frequency. 

One straightforward scenario to generate such strong shocks at large distances from the star involves the collision of the stellar winds of two massive stars in a sufficiently wide binary system. Indeed many non-thermal radio emitters are spectroscopic binary systems with periods of tens of days to several years.  However, for several non-thermal radio emitters, spectroscopic monitoring has failed to established their binary nature \citep[e.g.][]{RNF09}, raising questions on the universality of the wind-wind collision scenario \citep{vLRB06}.

 \citet{DeB07} listed 16 O-type non-thermal radio emitters, nine of which have been observed by \smash. Remarkably, all of them have been resolved into pairs of bright stars, with separations between 1.5 and 100~mas and magnitude difference  $\Delta H < 1.5$. In particular, the binary status of HD\,168112 and CPD$-$47\degr2963 ($\equiv$ CD$-$47\degr4551) was previously unknown.

The maximum observed magnitude difference between the components of the resolved non-thermal radio emitters corresponds to a flux ratio of 1:4 at most, indicating that the companions have masses that are similar within a factor of two. This is in agreement with the assumption that two massive stars are needed to produce a strong wind-wind collision. The fact that we resolve all  non-thermal radio emitters in our target lists, including two previously unidentified pairs, is an important piece of evidence in favor of the universality of the colliding wind mechanism to produce observable non-thermal radio emission.

\section{Conclusions} \label{sect: ccl}

 We introduce the \smash\ survey, a long baseline and aperture masking interferometric survey designed to probe the visual multiplicity of southern massive stars down to separations of about 1~mas. 117 O stars were observed with the PIONIER four-beam combiner at the VLTI, and 162 O stars where observed with the {\it Sparse Aperture Masking} (SAM) mode of VLT/NACO.  The sample selection is based on the GOSC-v2 applying both a declination selection ($\delta < 0$\degr) and a NIR magnitude cutoff ($H < 7.5$).  All in all, we resolved 240 companions with separations covering almost four orders of magnitude: from about 1~mas to 8\arcsec. We summarize below our main results:\\

(i) The \smash\ survey has increased the number of resolved companions within 100 mas by a factor 17 (from 4 to 66) and within 8\arcsec\ by a factor 4 (from 64 to 260).\\

(ii) None of the companions detected at angular separations below 1\arcsec\ can be explained by foreground/background targets or by chance alignment in a cluster environment. Such close companions are thus expected to be physically linked to their central object.\\

(iii) For the 96 targets in our main sample that have both been observed with PIONIER and NACO/SAM, i.e.\ that have complete observational coverage of the angular separation range, 53\%\ have at least one resolved companion within 200~mas. This fraction increases to  76\%\ when extending the search radius to 8\arcsec\ and to  91\%\ when including the unresolved spectroscopic and eclipsing companions.\\

(iv) Including both resolved and unresolved spectroscopic or eclipsing companions, all the dwarfs in our sample have a $\Delta H <5$ companion within 30~mas. About one third of them have a third companion within 200~mas and are hierarchical triples. \\

(v) The measured fraction of resolved multiple systems is lower for supergiants than for dwarfs. While detailed considerations of observational biases are needed to reach firm conclusions,  initial computations suggest that the observed trend is fully compatible with the larger contrast expected between supergiants and their companions (as a result of the larger brightness of supergiants) and that there may not be any difference between the intrinsic multiplicity properties of dwarfs and supergiants at $\rho > 1$~\mas.\\

(vi) We resolved 17 known spectroscopic binaries, many of them for the first time. In particular, we resolved every single SB system with a (known) orbital period larger than 150\,d. \\

(vii) None of the 13 stars in our runaway sample have a resolved companion in the 1-200~mas separation range.  Only  one has a possibly physical companion at $\rho=1.7''$. Although we only have complete coverage of the 1-8000~mas range for six systems, the fraction of multiple systems with a resolved companion is significantly lower than that of the rest of the sample. \\

(viii) Nine of the 16 known O-type non-thermal radio emitters were observed by \smash. All of them were resolved into a bright pair with separations in the range of 1 to 100~mas and with a magnitude difference $\Delta H<1.5$ (hence a likely mass-ratio of 1:2 at most). Our results strongly support the colliding wind scenario in wide binary systems as a universal explanation of the origin of the non-thermal radio emission of massive O-type stars.\\

As demonstrated by the observational results of the \smash\ survey, the combination of long baseline interferometry and aperture masking techniques  allow us to close the existing gap between spectroscopic and visual companions (the so-called interferometric gap). We can now explore the full separation range of massive O-type binaries, which will be of great value for many aspects of massive stars and binary physics including absolute mass determination, binary formation and stellar evolution.

\acknowledgments

This work is based on observations collected at the European Southern Observatory under programs IDs 086.D-0641, 088.D-0579, 189.C-0644 and 090.C-0672. PIONIER is funded by the Universit\'e Joseph Fourier (UJF), the Institut de Plan\'etologie et d'Astrophysique de Grenoble (IPAG), the Agence Nationale pour la Recherche (ANR-06-BLAN-0421 and ANR-10-BLAN-0505), and the Institut National des Science de l'Univers (INSU PNPand PNPS). The integrated optics beam combiner results from a collaboration between IPAG and CEA/LETI based on CNRS R\&T funding. Support for KMK was provided by NASA through Hubble Fellowship grant \#HF-51306.01 awarded by the Space Telescope Science Institute, which is operated by the Association of Universities for Research in Astronomy, Inc., for NASA, under contract NAS 5-26555. The authors warmly thank the people involved in the VLTI project as well a J. Ma\'{i}z Apell\'aniz and B. Mason for constructive discussions and suggestions. We are also grateful to the referee for comments that improved the quality of the paper. We made use of the Smithsonian/NASA Astrophysics Data System (ADS), of the Centre de Donn\'ees astronomiques de Strasbourg (CDS) and of the Washington Double Star Catalog (WDS) which maintained at the U.S. Naval Observatory. Part of the calculations and graphics were performed with the freeware Yorick.

{\it Facilities:} \facility{VLTI (PIONIER)}, \facility{VLT: Yepun (NACO/SAM)}

\newpage

\clearpage

\clearpage

\clearpage

\clearpage

\clearpage

\begin{figure}
  \centering
  \includegraphics[width=.47\textwidth]{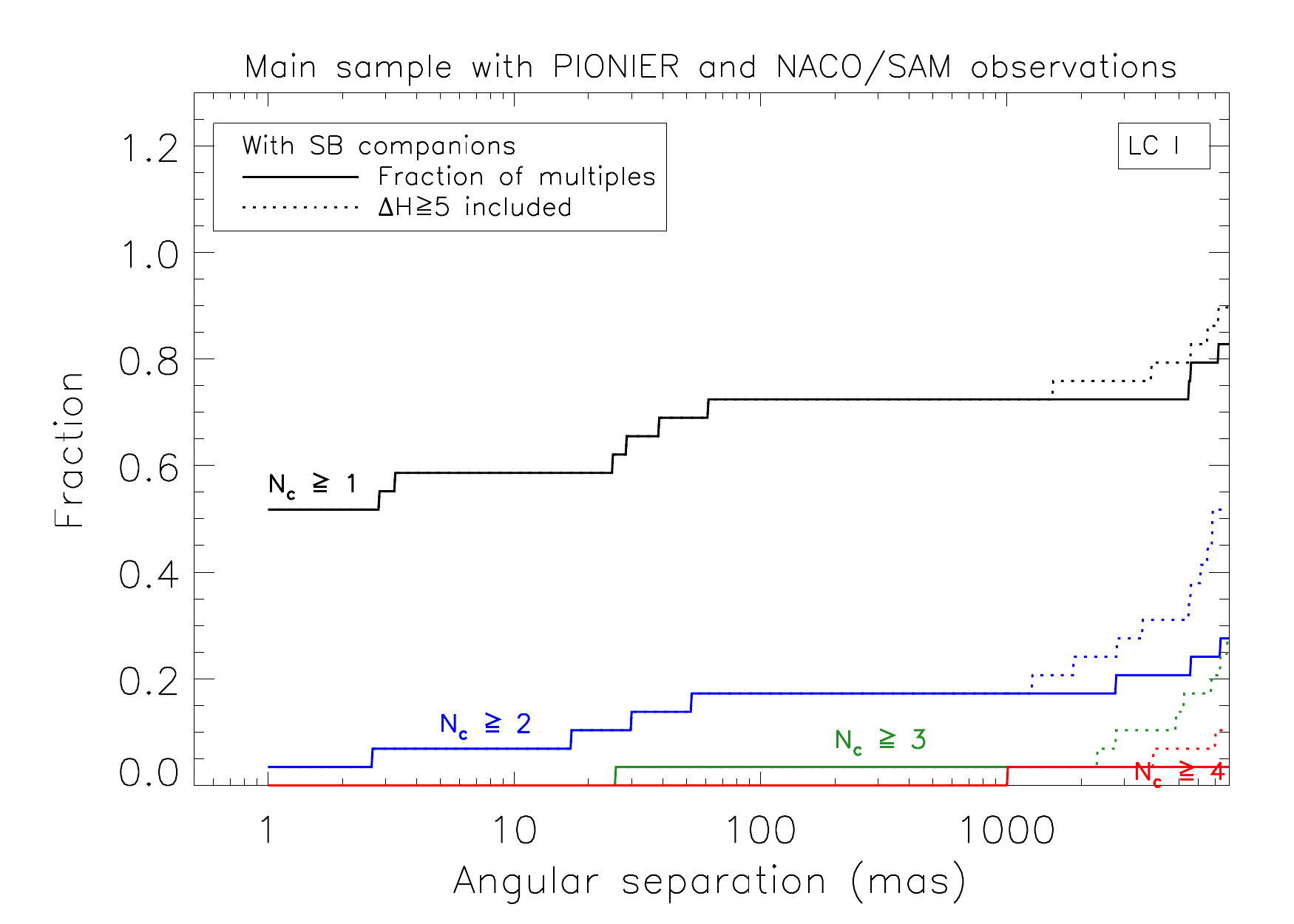}
  \includegraphics[width=.47\textwidth]{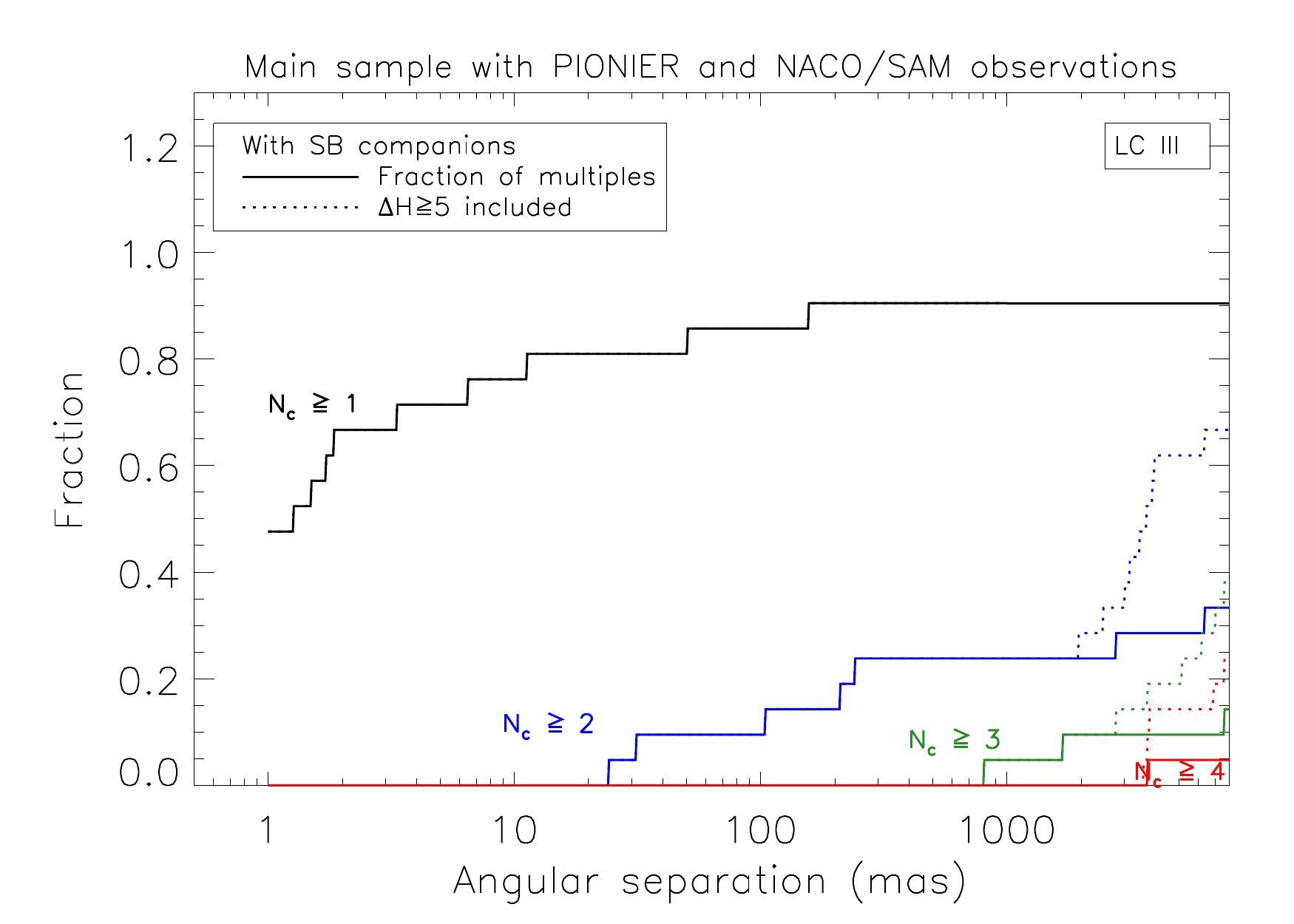}
  \includegraphics[width=.47\textwidth]{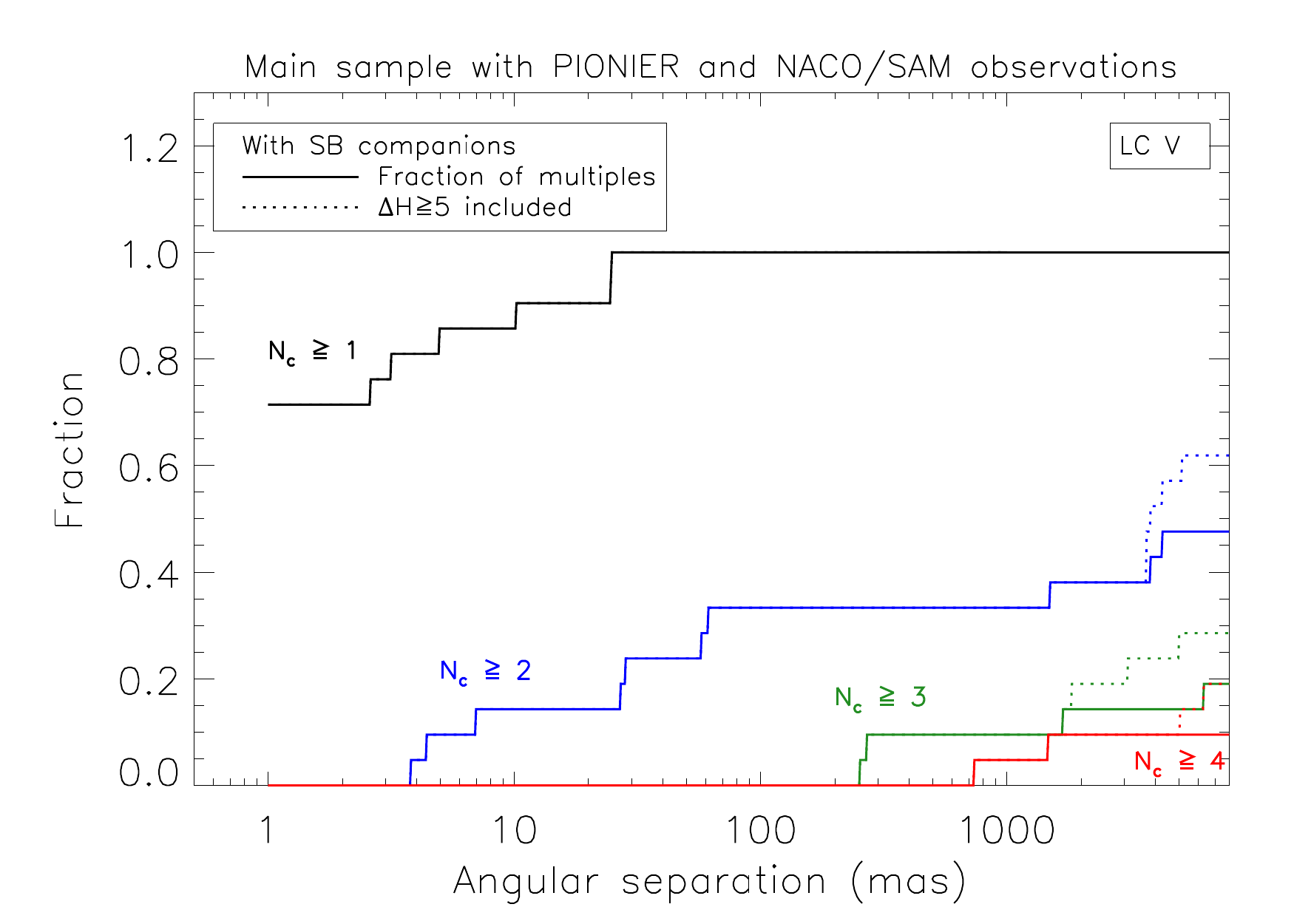}
  \caption{Break down of Fig.~\ref{fig: triple} for luminosity classes I, III and V. The curves include the unresolved E/SB companions.} 
  \label{fig: trip_lc}
\end{figure}

\clearpage

\appendix

\section{Note on individual objects from our main  sample} \label{sect: notes}

This appendix discusses the individual detections for objects in our main target list (Table~\ref{tab: targets}). It  provides background information on each targets, including companion identification, cross-correlation with previous results and adopted naming convention. The nomenclature for multiple systems carries a significant historical weight; in this work, we follow the guidelines outlined in \citet{Mas03}. Figs.~\ref{fig: sam1} to \ref{fig: sam4} provide finding charts for objects with more than three companions detected in the NACO FOV.

\subsection{Newly resolved targets} \label{ssect: new}

\paragraph{HD\,76341} We resolved a $\Delta H=3.7$ companion (A,B) at $\rho=169$~mas with NACO/SAM, in agreement with pre-publication results of Aldoretta et al. mentioned in \citet{SMAM14}. The latter authors noted that the spectrum of HD\,76341 is variable, making it a possible hierarchical triple system.

\paragraph{HD\,76556} PIONIER resolves a new pair (A,B) with $\rho=2.5$~mas and with $\Delta H=3.1$. No companion was mentioned at $\rho>30$~\mas\ by \citet{MHG09} as confirmed by our NACO data. The SB1? status reported by \citet{Crampton:1972} was not confirmed by \citet{Williams:2011}. \citet{CHN12} listed HD\,76556 as SB2 but no period has been published so far. Adopting the same distance as that of HD\,76341 given that both stars are members of the Vel OB1 association yields a projected separation of 2.3~AU. The resolved interferometric companion may be the spectroscopic companion if the spectroscopic period is typically larger than six months. This is a typical example where spatially resolved observations unveil a binary much faster than radial velocity (RV) monitoring.

\paragraph{CPD$-$47\degr2963} We resolve it as a new pair (A,B)) with $\Delta H = 1.4$. The separation is $\rho=1.5$ and $4.1$~\mas\ on our two PIONIER observations separated by 5.6~months, thus indicating a clear orbital motion. No companion at $\rho>30$~\mas\ is reported in \citet{MHG09} but we detect one (A,C) at 5.2\arcsec\ with a magnitude difference of almost 7 in the $H$-band. CPD$-$47\degr2963 has been reported as RV stable in \citet{Denoyelle:1987} but as SB1 with a OWN pre-publication period of 59\,d \citep{SMAM14}, likely a different companion than the one detected by PIONIER. Wind-wind collision in  binary system has been proposed to explain the non-thermal X-ray emission \citep{Benaglia:2001}, a scenario that is clearly confirmed here. \citet{Hubrig:2011} claimed detection of a magnetic field. 

\paragraph{HD\,92206\,A and B} Our NACO/SAM observations were centered on HD\,92206\,A. We resolve a new companion at $\rho=33$~mas (Aa,Ab), though with large uncertainty. With $\Delta H \approx 4$, the new companion is unfortunately too faint for PIONIER. It is unclear whether this newly resolved pair corresponds to the OWN pre-publication SB reported in \citet{SMAM14}. If it does, the orbital period is likely of the order of 5 years at least. We further identified a third faint companion ($\Delta H=5.1$) within 1\arcsec\ of HD\,92206\,Aa (Aa,Ac; $\rho= 0.85$\arcsec). HD\,92206\,B, at 5.3\arcsec\ from HD\,92206\,A, is within the FOV of our NACO \Ks\ band observations, but too close to the detector edge to perform a reliable interferometric analysis of its NACO/SAM data.

\paragraph{HD\,93130 $\equiv$ V661 Car}  Observed once with PIONIER and once with NACO/SAM, we resolve it as a new pair (Aa,Ab) with $\rho=19.8$~\mas. The pair is poorly constrained by SAM as the separation is smaller than SAM's IWA. Most probably, the detected pair does not correspond to the eclipsing binary with $P_\mathrm{orb}=23.9$\,d reported by \citet{Ote06} given the $\approx$2.6~kpc distance to the Cr228/Tr16 complex. No companion at $\rho>30$~\mas\ was reported by \citet{MHG09}, as confirmed by our NACO observations.

\paragraph{HD\,93160} At 12.6\arcsec\ from HD\,93161, HD\,93160 is listed as HD\,93161\,C in \citet{MGH98}. Previously considered to show constant RV, HD\,93160 was reported as SB1 by \citet{CHN12}. PIONIER resolves a close companion at $\rho=6.5$~\mas\ (Ca,Cb).  Without knowledge of the spectroscopic period, one cannot decide whether the newly resolved pair corresponds to the spectroscopic companion. No companion was reported at $\rho>30$~\mas\ \citep{MHG09} but SAM detects at putative pair with $\rho = 30 \pm 14$~mas. The very different position angle and magnitude difference of the SAM pair (Ca,Cc) compared to the PIONIER one, plus the fact that SAM is essentially blind to separations $< 24$~mas (Sect.~\ref{sect: data_analysis}) argued in favor of a third companion. We attempted to fit the PIONIER data using a triple model rather than a binary, and using the SAM measurements as a guess-solution for the third body of the system. The fit converged nicely, yielding $\chi^2=0.98$, significantly better than the binary model. While further observations are certainly desirable to confirm the reality of this tight triple system, we adopt the 3-body solution in Table~\ref{tab: pio_detect}. Another two companions  are seen at separations of 0.8 (Ca,Cd) and 3.7\arcsec, the latter possibly corresponding to the 3.3\arcsec\ pair (C,D) reported by \citet{MGH98}. 

\begin{figure*}  \centering
\includegraphics[width=0.38\textwidth]{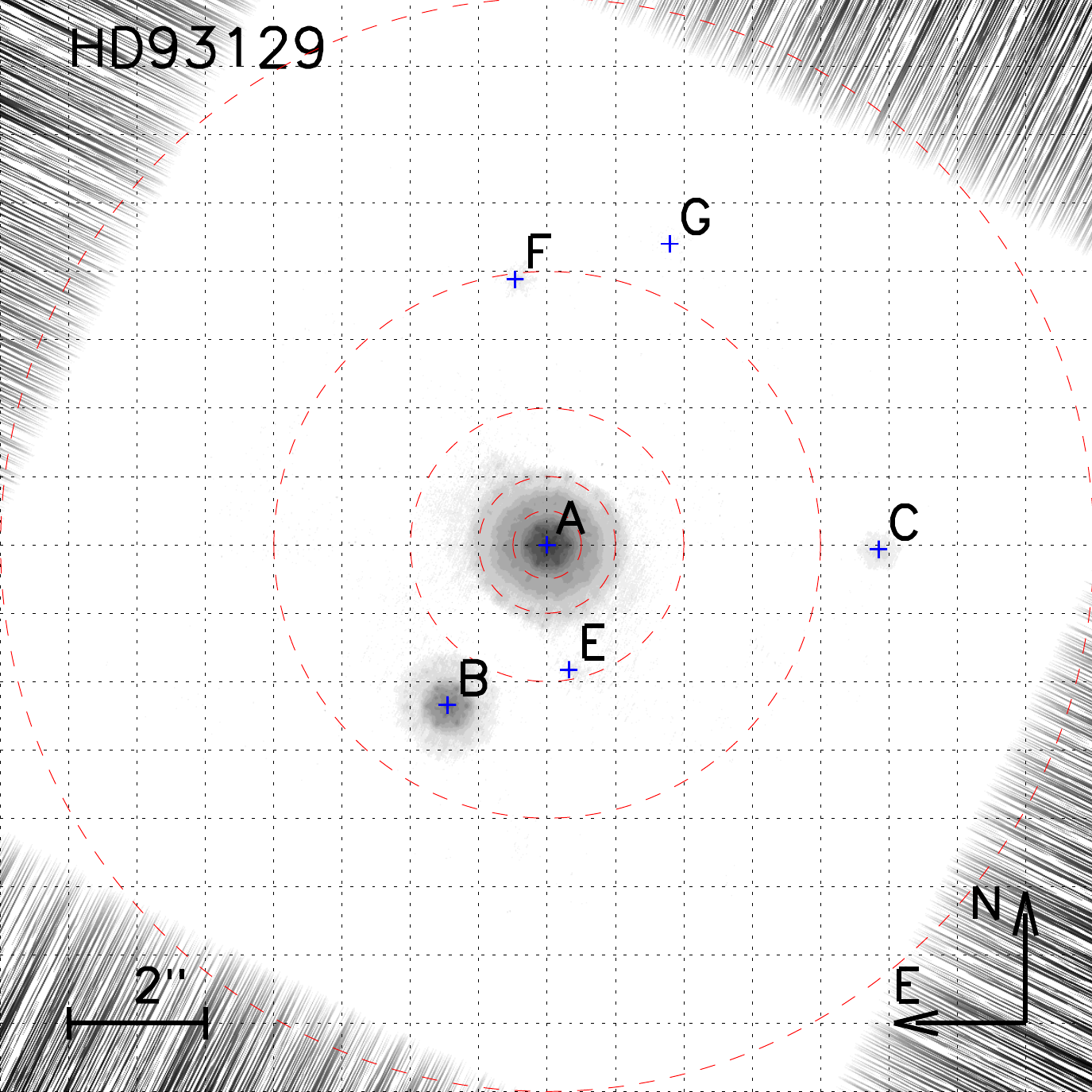} \hspace*{2mm}\vspace*{2mm}
\includegraphics[width=0.38\textwidth]{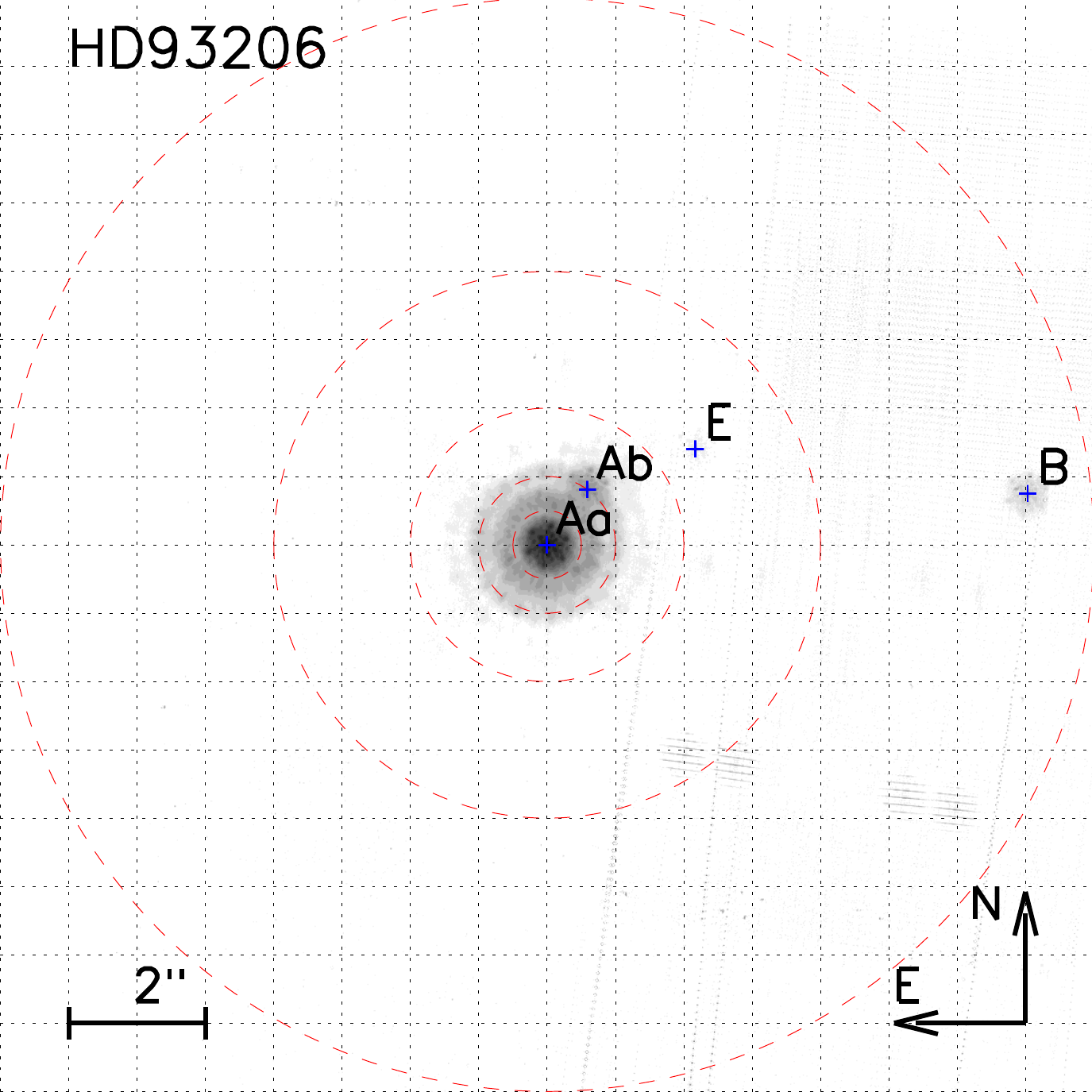} \hspace*{2mm}\vspace*{2mm}
\includegraphics[width=0.38\textwidth]{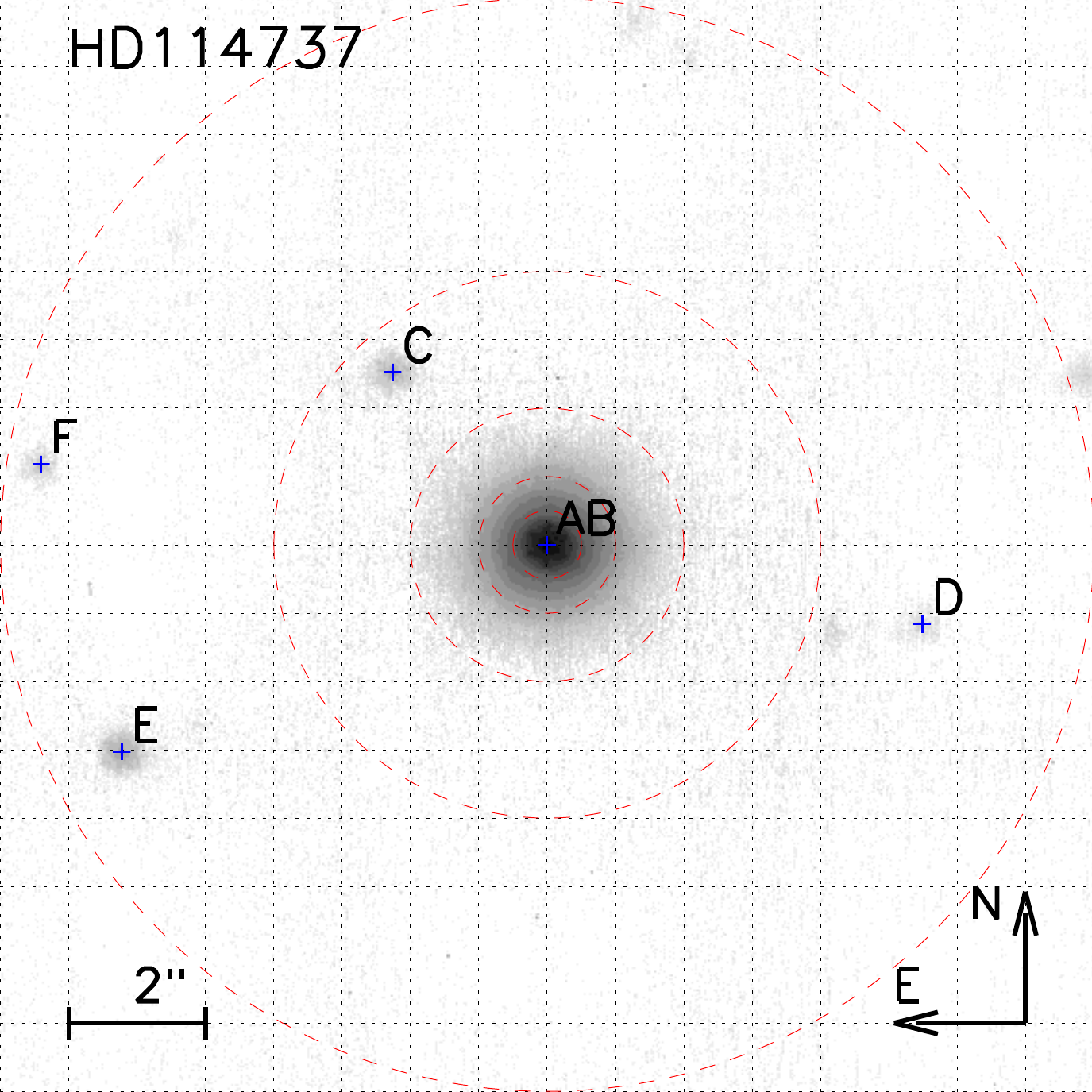} \hspace*{2mm}\vspace*{2mm}
\includegraphics[width=0.38\textwidth]{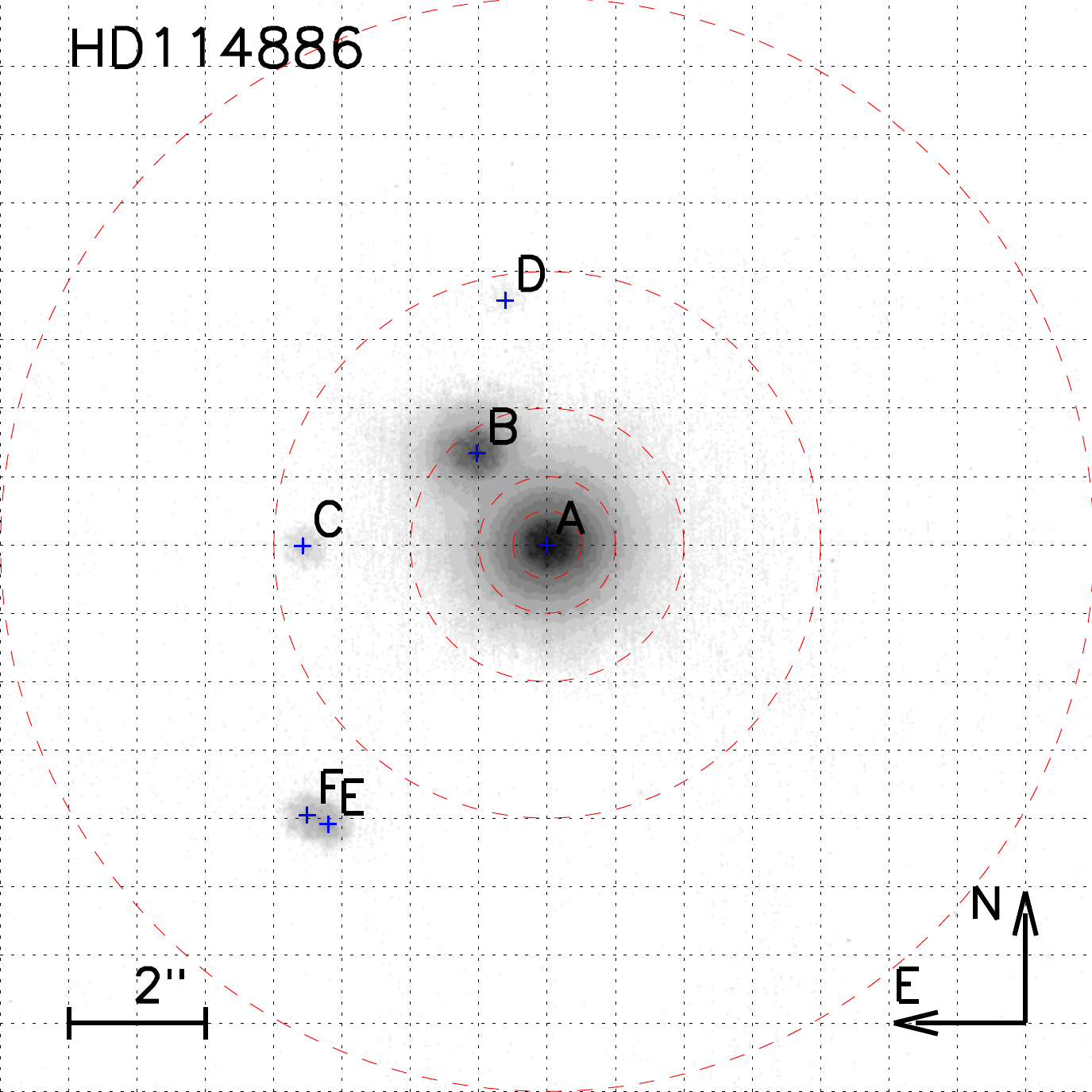} \hspace*{2mm}\vspace*{2mm}
\caption{NACO images of the surroundings of HD\,93129, HD\,93206,  HD\,114737 and HD\,114886. Dotted lines are separated by 1\arcsec. Dashed circles have radii of 0.5\arcsec, 1\arcsec, 2\arcsec, 4\arcsec\ and 8\arcsec. Companions are identified by a letter and their position in the field is marked by a cross-hair ($+$).}
    \label{fig: sam1}
\end{figure*}

\paragraph{HD\,93206 $\equiv$ QZ Car} It is a complex multiple system (Fig.~\ref{fig: sam1}). Two distant visual companions are known at 7.3\arcsec\ (A,B) and 8.8\arcsec\ (A,C) \citep{MGH98} and a 1\arcsec-separation companion (Aa,Ab)  has been detected by Hipparcos and confirmed by \citet{TMH10}. All three visual companions are clearly seen in our NACO image. The central object, HD\,93206 Aa  is itself a quadruple system composed of a pair of spectroscopic binaries: Aa1,Aa2 (O9.7~I + B2~V, $P_\mathrm{orb}=20.7$\,d) and Aa3,Aa4 \citep[O8~III + O9V, $P_\mathrm{orb}=6.0$\,d and eclipsing,][]{MLD01}. We resolve the two binaries Aa12 and Aa34 for the first time with $\rho = 26$~mas and $\Delta H =0.4$.

\paragraph{HD\,93222} Considered as RV stable \citep{LMG90}, we detect two previously unreported companions at separations of 10~mas (A,B) and 3.8\arcsec\ (A,C) with PIONIER and NACO respectively. The inner pair has similar brightness components with $\Delta H=0.28\pm0.25$.

\paragraph{HD\,93250} Our observation represents the third epoch of the long period binary (A,B) discussed in \citet{SLBDB11}. With $\rho=1.5\pm0.1$~mas in our 2013 observations, HD\,93250  is one of the tightest resolved binaries in our sample and a non-thermal radio emitter. No companion were detected  at $\rho>30$~\mas\ by \citet{MHG09} as confirmed by our NACO observations.

\paragraph{HD\,93403} This system is a SB2 binary (A,B) with a 15.1\,d period \citep{RSG00, RVS02}. NACO/SAM resolves a new quite faint companion at 211~mas ($\Delta H=4.2$). PIONIER could not resolved the inner SB2 binary nor any other tight companions.

\paragraph{HD\,93632} Reported as RV stable \citep{LMG90}, we detect a new companion at $\rho=25$~\mas\ and with $\Delta H=2.69$. No companion at $\rho>30$~\mas\ was found by \citet{MHG09} nor by our NACO-FOV data.

\paragraph{HDE\,303492} This object is the O8.5~Iaf spectroscopic standard. No close companion was detected by either PIONIER or SAM. We however report a new faint companion at 6.5\arcsec\ (A,B).The SB2 status reported by \citet{CHN12} may rather trace intrinsic variability due to the strong winds of this Iaf supergiant (similar the case of $\zeta$~Pup).

\paragraph{HD\,96670} A new companion at $\rho=30$~\mas\ is detected both by PIONIER and NACO/SAM. It can hardly be the SB1 companion ($P_\mathrm{orb}=5.5$\,d, $a_1\sin i=6.2$~\Rsun, $a_\mathrm{app}<0.1$~\mas) from \citet{SL01}. We labeled the new pair A,B. No other companion at $\rho>30$~\mas\ was seen by \citet{MHG09}, as confirmed by our NACO image.

\paragraph{HD\,97253} PIONIER reveals a new companion at $\rho=11$~\mas\ and with $\Delta H=2$ (A,B). No companion was detected by SAM in the 30-200~mas range but an additional faint and distant companion is seen in the NACO FOV at 3.4\arcsec\ (A,B). The central object was reported as a possible SB1 by \citet{Thackeray:1973}, and again as SB1 by \citet{CHN12}. Without more information on the spectroscopic period, one cannot decide whether the spectroscopic and the PIONIER companions are identical. Given the separation, and the magnitude difference, it is however a plausible option.

\paragraph{HD\,101131} This system is a known O+O SB2 binary \citep[$P_\mathrm{orb} \approx 9.7$\,d]{GPM02}. NACO/SAM  data reveal an additional component at 61~mas with a $\Delta H$ of about 1~mag (A,B). Given the distance to the IC~2944 cluster \citep{SJG11}, it is not the spectroscopic companion, making HD\,101131 a hierarchical triple system.

\paragraph{HD\,101545\,A and B} HD\,101545\,A, B is a 2.6\arcsec\ pair. Both components are RV stable \citep{SJG11}. Only  component A is an O star while  component B is classified as B0.2 \citep{SMAM14}. We resolve HD\,101545\,A as a close pair (Aa,Ab) with $\rho=2.6$~mas and $\Delta H=0.2$. Given the brightness difference and the fact that the combined spectrum is an O9.2~II star, both Aa and Ab are likely late O stars.

\paragraph{HD\,114737\,A and B} (Fig.~\ref{fig: sam1})  We resolve this previously reported 190~mas pair (A,B) with NACO/SAM. \citet{SMAM14} report pre-publication OWN results indicating a 12.4\,d SB1 system. We further detected an additional four companions in the NACO FOV, with separations of 3.4\arcsec, 5.6\arcsec, 6.9\arcsec\ and 7.5\arcsec. We labeled the new pairs, ordered by increasing separations, A,C to A,F.

\paragraph{HD\,115455} NACO/SAM resolved it into two similar brightness components separated by 48~mas (A,B). This new pair cannot be the 15.1\,d binary reported by \citet[OWN,][]{SMAM14}. HD\,114455 is therefore at least a hierarchical triple system. Unfortunately, the target was not observed with PIONIER.

\paragraph{HD\,117856}  This object is a 27.6\,d spectroscopic binary \citep[OWN,][]{SMAM14}. We did not observe it with PIONIER but we detect two additional companions with NACO/SAM, at separations of 1.6\arcsec\ (A,B) and 7.5\arcsec\ (A,C). The first one was already reported by \citet{MGH98}. 

\paragraph{HD\,120678} We detect no companion in the 30-200~mas range with SAM  but we clearly detect three faint companions at 0.8\arcsec, 4.5\arcsec\ and 6.5\arcsec\ in the NACO FOV, that we labeled B,  C  and D respectively. This object was not observed with PIONIER.

\paragraph{HD\,124314\,A and B} Both components are O stars, separated by about 2.7\arcsec. Only the A component is brighter than our magnitude cutoff for PIONIER. It is marginally resolved, with a best fit formally for a $\rho=1.5$~\mas\ pair (Aa,Ab). This detection possibly corresponds to the newly reported SB2 system \citep[OWN,][]{SMAM14}. We resolve the B component itself as a multiple system. The Ba,Bb separation of $\rho = 209\pm1.5$~mas, $\theta=64.5 \pm 2.3$\degr, $\Delta H = 3.40 \pm 0.22$ and $\Delta K\mathrm{s}=2.70\pm0.12$ is in agreement with the findings of \citet{TMH10}.  We further detect another faint object in the field at 2.46\arcsec\ from component A, which we labeled HD\,124314\,C. 

\paragraph{HD\,125206} We resolve three companions with separations of 40~mas, 1.2\arcsec\ and 6.9\arcsec\ that we labeled B to D, respectively. Given the lack of information, one cannot decide whether the 40~mas companion is also the SB2 system reported from the pre-publication OWN results \citep{SMAM14}.

\paragraph{HD\,148937} This object is one of the few prototypical Galactic member of the Of?p class (together with $\theta^1$~Ori~C, a long period binary seen almost pole-on too). This magnetic O star is resolved by PIONIER as an equal brightness pair (Aa,Ab) with $\rho=20.3$~\mas. Adopting a parallax $\pi=2.35\pm0.79$~\mas\ \citep{van-Leeuwen:2007}, it would correspond to a projected physical separation of 40\,AU. However, the relative error on the Hipparcos parallax is large and the distance would need further confirmation.  While the object is flagged as SB in Simbad, a spectroscopic study by \citet{Naze:2008,Naze:2010} reported no evidence for binarity. The magnetic field and spectral variability with a 7.03\,d period \citep{Wade:2012} constrain the rotation inclination to be $i<30$\degr. The author does not discuss binarity. We further detected a faint ($\Delta H=5.4$) companion at 3.3\arcsec\ that may correspond to the 2.9\arcsec\ B companion reported by \citet{MGH98} if the latter is a high proper motion (possibly foreground) object.

\paragraph{$\mu$~Nor $\equiv$ HD\,149038} We detect two faint companions in the field of view with separations of 1.5\arcsec\ (A,B) and 6.2\arcsec\ (A,C). PIONIER observations are inclusive as already discussed in the main text. 

\paragraph{HD\,149404} This object is a 9.81\,d SB2 system \citep{RNC01}. We detect a previously unreported additional, distant and faint companion (A,B) in the NACO FOV with a separation of 6.8\arcsec\ ($\Delta K\mathrm{s}=7.2$).

\paragraph{HD\,149452} We report the detection of a 2.7\arcsec\ $\Delta K\mathrm{s}=4.4$ companion to this otherwise isolated O star. The companion is undetected in the $H$-band image indicating a strongly reddened, possibly background, object.

\paragraph{HD\,150958\,A and B} Our NACO/SAM data confirmed the previously resolved 0.3\arcsec\ A,B pair \citep{MGH98,MHG09}. We further detect a much fainter ($\Delta H=6.8$) companion at 6.6\arcsec. We named the latter pair A,E owing to the fact that companion C and D are already attributed to stars outside our FOV. 

\paragraph{HD\,151018} We detect two rather faint ($\Delta K\mathrm{s}=4.6$ and 6.2) visual companions with separations of 2.1\arcsec\  (A,B) and 7.3\arcsec\  (A,C) in the NACO FOV. 

\paragraph{HD\,152003} We detect a rather faint ($\Delta K\mathrm{s}=4.8$) companion  (A,B) at about 40~mas with NACO/SAM, although the measurements lack accuracy. The SAM companion is too faint and is not detected in our PIONIER observations. 

\paragraph{HD\,152147} This object is just marginally resolved by PIONIER using well calibrated data. Our best fit is formally obtained for $\Delta H=2.8$ and $\rho=0.77$~mas but with large uncertainties. The object is reported as SB1 by \citet{Williams:2013} with $P_\mathrm{orb}=13.8$\,d and $a_1 \sin i=3.6$~\Rsun\ but \citet{SMAM14} mentioned that OWN obtained a different orbital period. We thus have to wait for the spectroscopic orbit to be clarified before one can decide whether we PIONIER resolved the SB companion or whether HD\,152247 is a triple system. Either way, we label the resolved pair A,B.

\begin{figure*}  \centering
\includegraphics[width=0.38\textwidth]{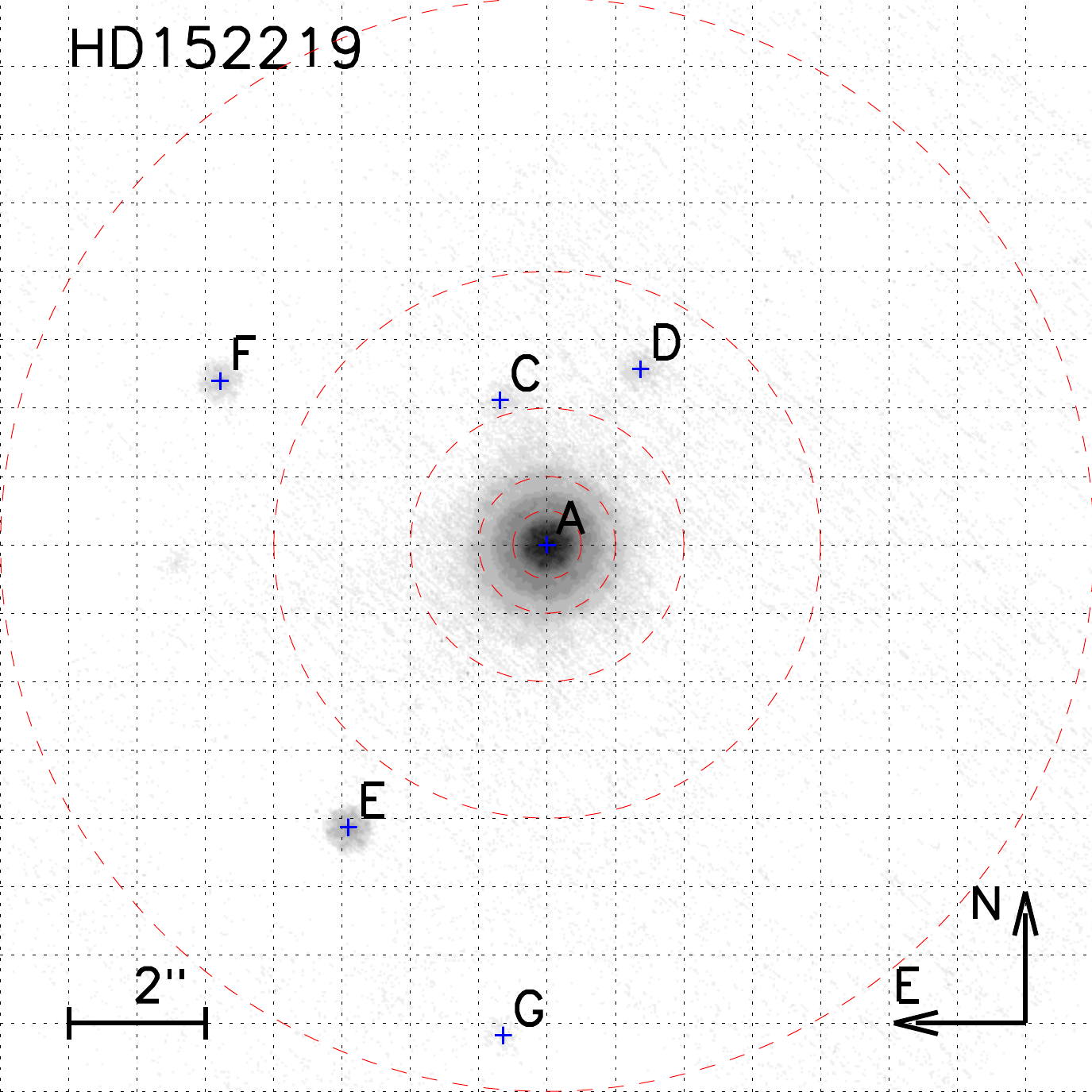} \hspace*{2mm}\vspace*{2mm}
\includegraphics[width=0.38\textwidth]{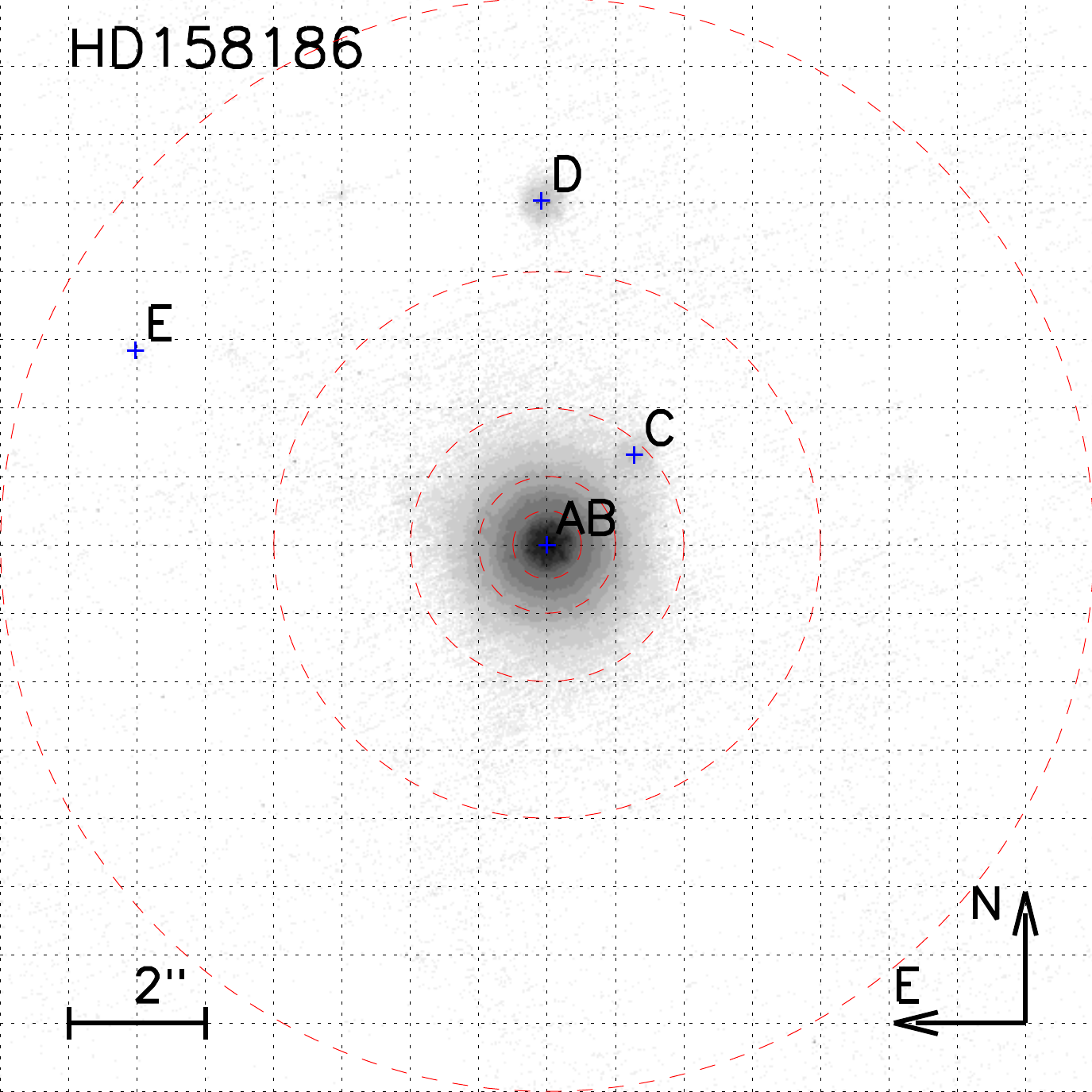} \hspace*{2mm}\vspace*{2mm}
\includegraphics[width=0.38\textwidth]{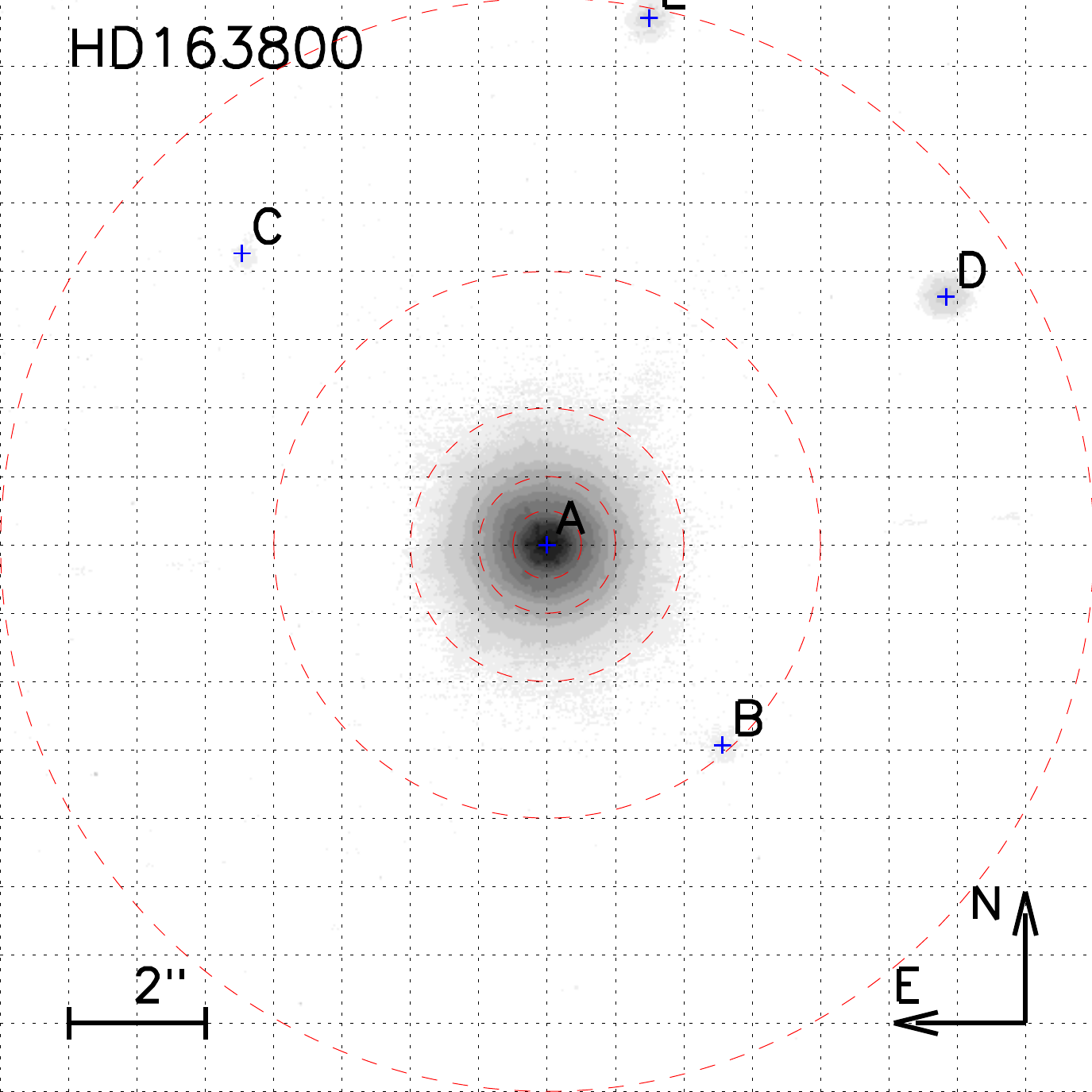} \hspace*{2mm}\vspace*{2mm}
\includegraphics[width=0.38\textwidth]{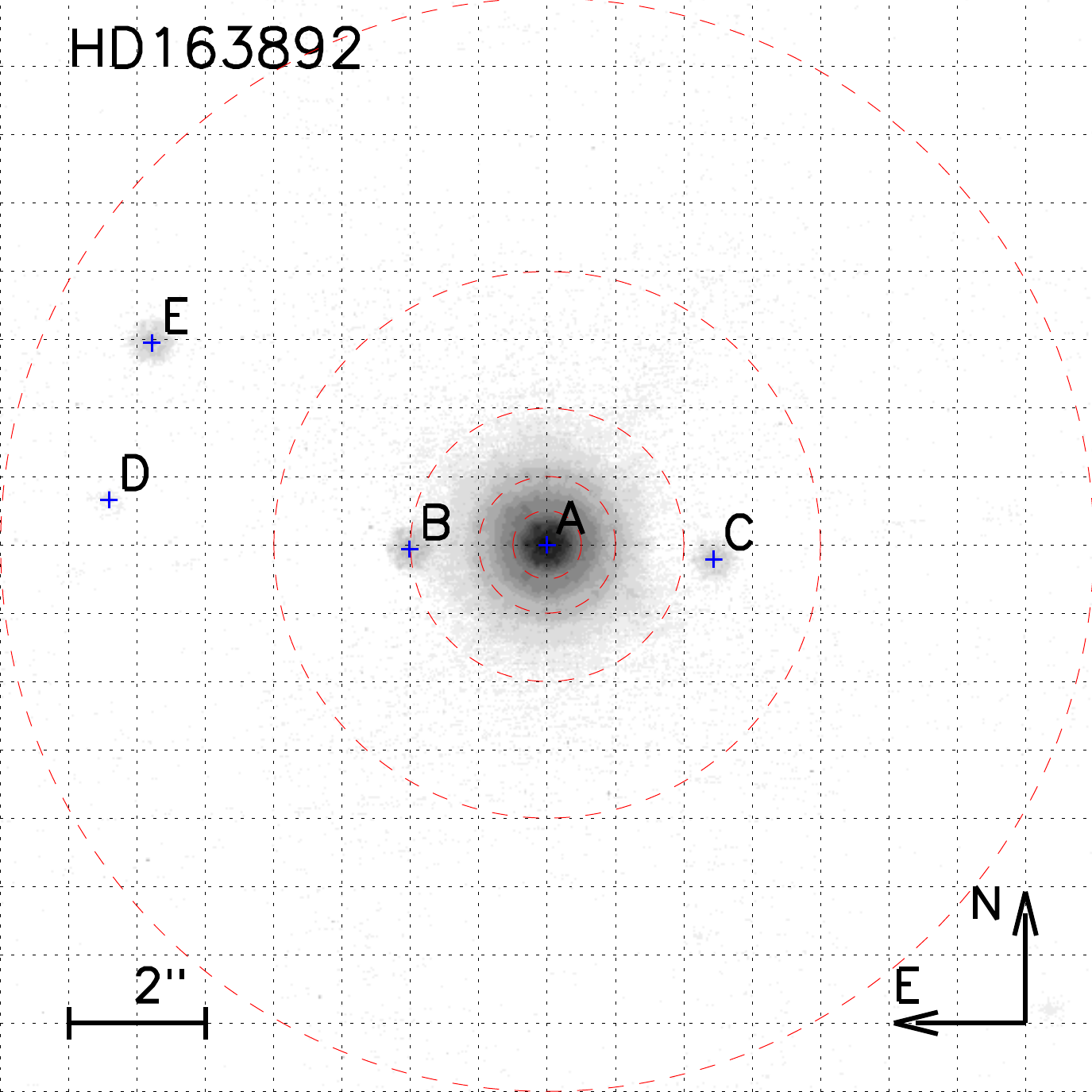} \hspace*{2mm}\vspace*{2mm}
\includegraphics[width=0.38\textwidth]{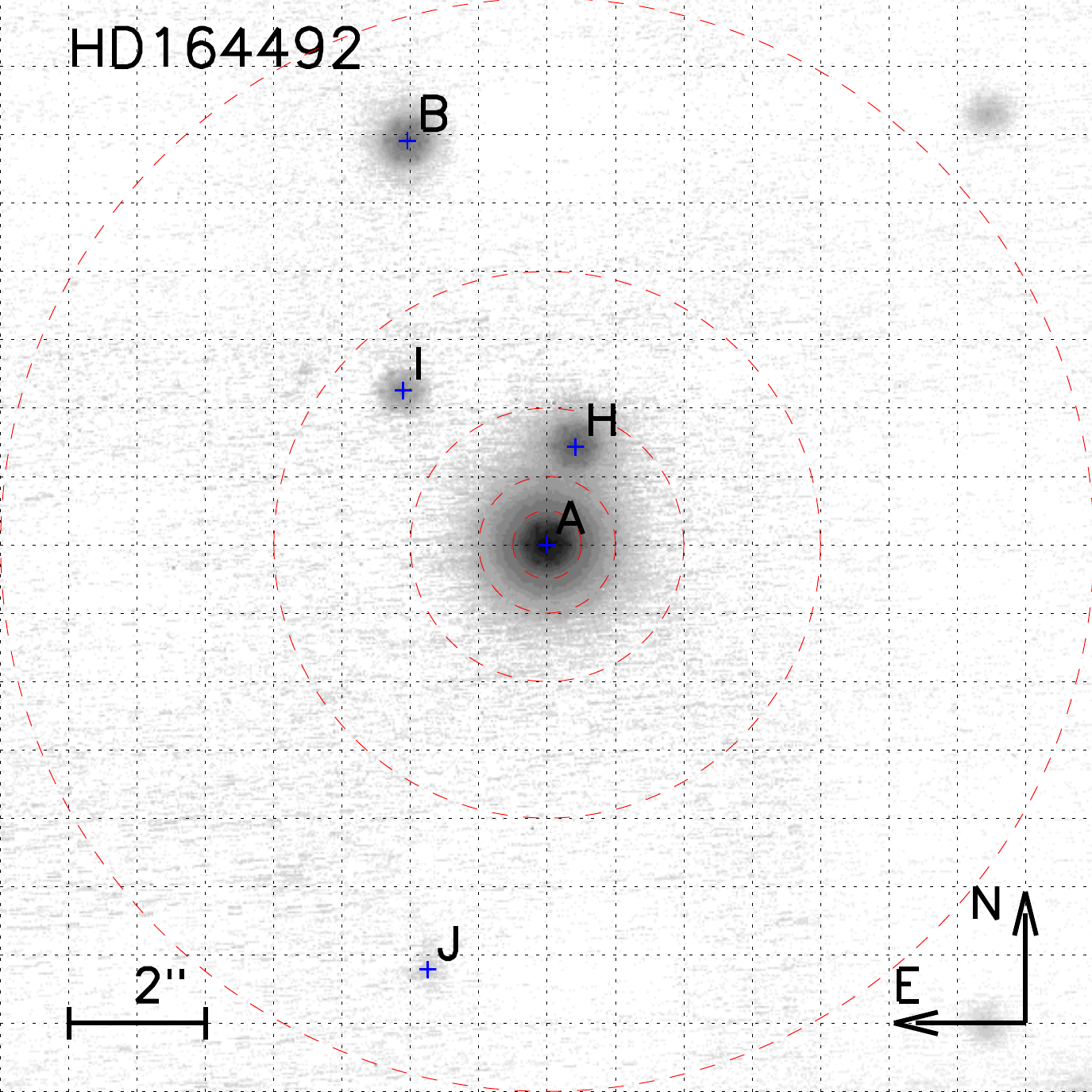} \hspace*{2mm}\vspace*{2mm}
\includegraphics[width=0.38\textwidth]{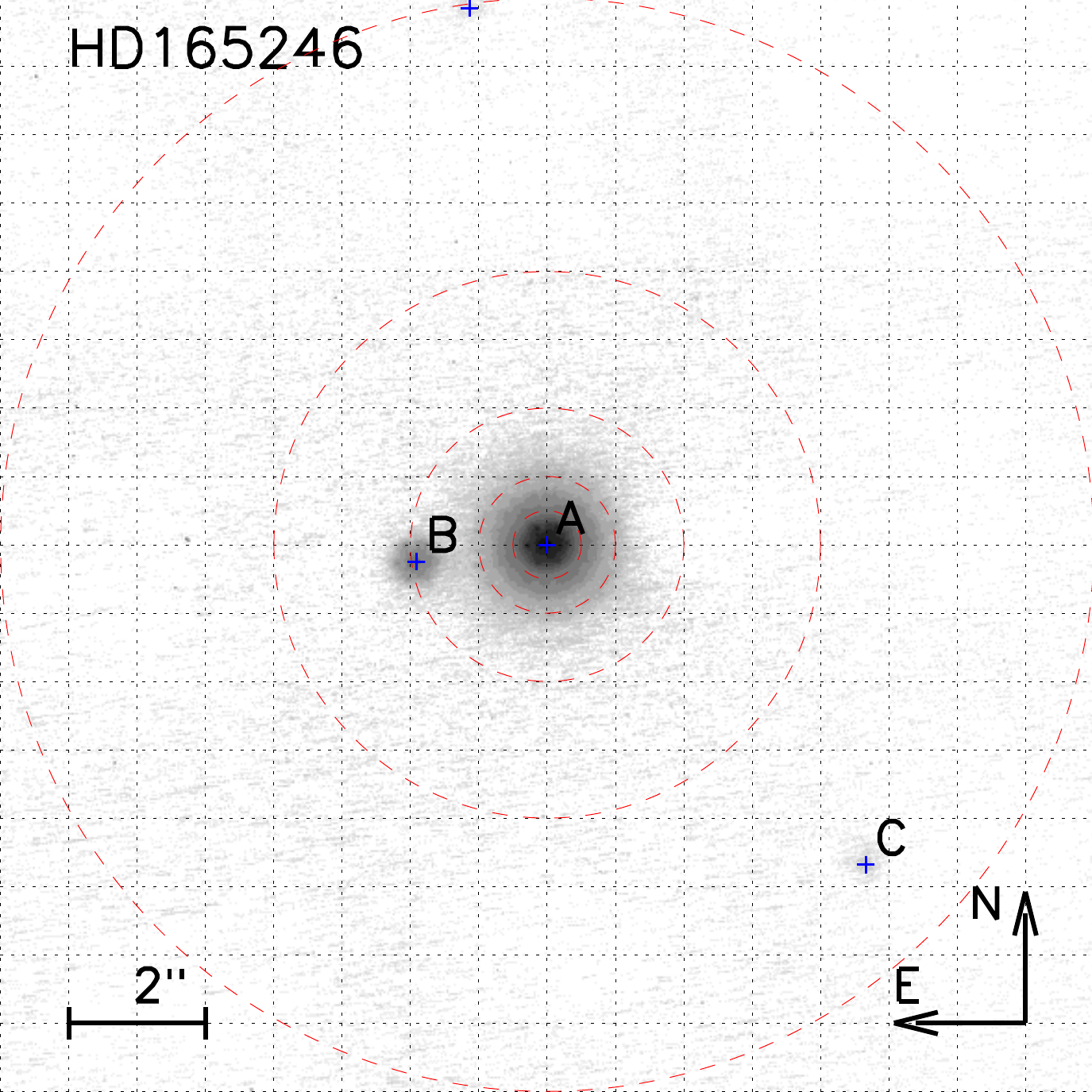} \hspace*{2mm}\vspace*{2mm}
  \caption{Same as Fig.~\ref{fig: sam1} for 
HD\,152219,
HD\,158186,
HD\,163800,
HD\,163892,
HD\,164492 and
HD\,165246.}
  \label{fig: sam2} 
\end{figure*}

\paragraph{HD\,152219} This object is and eclipsing SB2 system with a period of 4.2 d \citep{SGR06, San09}. We resolve a 83~mas companion with NACO/SAM and five other faint companions in the NACO FOV (Fig.~\ref{fig: sam2}), all of them too wide to be the spectroscopic companion. We label the new pairs A,B to A,G  by increasing separation.

\paragraph{HD\,152218} It is a known SB2 system with a period of 5.8\,d \citep{SNoD08}. Our NACO-FOV data further reveal an additional $\Delta K\mathrm{s}=3.8$ companion at 4.3\arcsec\ (A,B).

\paragraph{CPD$-$41\degr7733} It is a known SB2 system with a period of 5.7\,d \citep{SRG07}. We are lacking PIONIER data for this system, but we resolved a third component (A,B) with NACO/SAM at 43~mas, although the weather conditions limited the accuracy of the measurements. A fourth companion (A,C) is detected in the NACO FOV at a 1\arcsec\ separation.

\paragraph{HDE\,326331} Reported as a broad-line fast rotator with line profile variability by \citet{SGN08} and as SB2 in OWN, we detected two visual companions at 1.1\arcsec\ (A,C) and 3.4\arcsec\ (A,D) but we cannot confirm the 7.3\arcsec\ companion (A,B)  reported by \citet{MGH98}. It may lay outside our FOV.

\paragraph{HD\,152405} It is a SB system with an orbital period of 25.5\,d \citep[OWN,][]{SMAM14}. While we are lacking PIONIER measurements, NACO/SAM resolved a third companion at 54~mas (A,B).

\paragraph{HD\,152408} We detect two  companions in the NACO FOV, with separations of 3.8\arcsec\ (A,C) and 5.5\arcsec\ (A,B), the latter one being already reported by \citet{MGH98}.

\paragraph{HD\,152386} We detect a new companion (A,B) with $\rho=56$~\mas\ and $\Delta H=3.3$ using both PIONIER and NACO/SAM. One faint (A,C) and one bright (A,D) companion are further detected in the NACO FOV at separations of 3.5\arcsec\ and 7.4\arcsec, respectively.  \citet{MGH98} reported a 0.55\arcsec\ companion, but the detection remained unconfirmed in \citet{MHG09}. None of our detected companions seems to match the 1998 tentative  detection.

\paragraph{HD\,152623} \citet{MGH98} and \citet{MHG09} both reported a companion (A,B) at $\rho=238$~\mas\ but their listed position angle differs by 180\degr. Our NACO/SAM data confirm that the correct $\theta$ value is 307\degr. A closer, previously unresolved companion (Aa,Ab) is found in the PIONIER data. The best fit model of the PIONIER data is a binary with $\rho=28.24$~\mas\ and $\theta=-75$\degr, plus a background contribution of $\mathrm{bck}=0.15$. This background can be due to the 240~mas companion, especially when accounting for the coupling losses due to its separation. A third companion (A,C) in seen in the VLTI/IRIS FOV at 1.5\arcsec. It is also detected in the NACO FOV with $\Delta H \approx 3.5$. HD\,152623 also reported as a 3.9\,d SB1 system by \citet{Ful90}.

\paragraph{HD\,153426} It is a 22.4\,d period SB1 system according to OWN. The SB pair, probably too tight, is not resolved by PIONIER. Faint companions at 2.0\arcsec\ (A,B) and 3.4\arcsec\ (A,C) are detected in the NACO FOV.

\paragraph{HD\,154368} It is a 16.1\,d period EB system \citep{MGH98}. It has a $\Delta I=6.3$ companion at 2.8\arcsec\ \citep{MGH98, TBR08} but we do not detect it. We however report a $\Delta K\mathrm{s}=5.9$ companion at 6.7\arcsec\ and we label the new pair A,C.

\paragraph{HD\,154643} This object is a 28.6\,d period SB1 system according to OWN. The SB pair is not resolved  by PIONIER, probably because it is too tight. A faint companion at 1.9\arcsec\ (A,B) is detected in the NACO FOV.

\paragraph{V1075 SCO $\equiv$ HD\,155806}  Resolved by PIONIER with $\rho=24.8$~\mas\ (A,B), the star was reported to be single in a RV study by \citet{GCM80}. The star is however reported as SB2 by \citet{CHN12} but, given no period has been published so far, one cannot decide whether the  interferometric companion is the spectroscopic one. A faint companion with a separation of 5.1\arcsec\ (A,C) is further detected in the NACO FOV.

\paragraph{HD\,156738} It is a tight pair (A,B) with $a \rho=45$~\mas\ companion resolved both by PIONIER and NACO/SAM.  RV variability of 7~\kms\ is reported by \citet{Crampton:1972} but not confirmed by \citet{CHN12}.

\paragraph{V1081 SCO $\equiv$ HD\,158186} (Fig.~\ref{fig: sam2}) It is resolved as a close pair (A,B) with $\Delta H=2.1$ and $\rho=26.8$~\mas\ with PIONIER. The pair is also resolved by NACO/SAM although the measurements lack accuracy given the separation is below SAM's \IWA.  It is an Hipparcos eclipsing binary showing apsidal motion \citep{Ote05}, most probably because of the third component that we discovered. The object is reported as SB3 in OWN and we postulate that our detection corresponds to the third component. Three additional faint companions are detected in the NACO FOV at separations of 1.8\arcsec, 5.0\arcsec\ and 6.7\arcsec. We labeled the four discovered companions B to E by increasing separation.

\paragraph{V1036 SCO $\equiv$ HD\,159176} This object is marginally resolved on two PIONIER observations with minimum separation $\rho>0.9$~\mas\ on the first epoch and  with $\rho=10$~mas about one month later. Both detections have several possible minima in the $\chi^2$ map and more observations are needed to better characterize the system. The known SB2 has an equal mass ratio, $P_\mathrm{orb}=3.36$\,d and $a\sin i=14$~\Rsun\ \citep{Stickland:1993, LRS07}. Given a probable distance of around 1.5~kpc \citep{LRS07}, the expected separation of the spectroscopic pair is smaller than $0.2$~mas, so that we probably detected a third fainter component.  \citet{MGH98} reported three other companions at 0.27\arcsec\ (Aa,Ab), 0.74\arcsec\ (Aa,D; also known as HDS2480Aa,Ac in the WDS) and 5.4\arcsec\ (A,B) and 13.3\arcsec\ (Aa,C; outside our FOV). We clearly detect the pairs Aa,D and A,B but not Aa,Ab. This is similar to an unpublished AstraLux NTT result mentioned by \citet{SMAM14}, suggesting that the Ab companion may be a spurious detection (possibly due to the 10~mas pair, denoted Aa1-Aa2) or that it is too faint for both AstraLux and NACO, thus implying $\Delta \mathrm{mag} > 5$. We further detected a faint companion (A,E) at 3.5\arcsec.

\paragraph{HD\,162978} NACO-FOV data reveal a new faint companion (A,B) to this otherwise isolated O star.

\paragraph{HD\,163800} (Fig.~\ref{fig: sam2}) Reported as SB1 by \citet{CHN12}, we detect four faint companions (labeled B to E by increasing separation) in the NACO FOV. 

\paragraph{HD\,163892} (Fig.~\ref{fig: sam2}) It is a 7.83\,d period SB1 system \citep[OWN,][]{SMAM14}. Four faint companions  (labeled B to E by increasing separation) are detected in the NACO FOV. 

\paragraph{HD\,164492\,A} (Fig.~\ref{fig: sam2})
With seven companions reported in the WDS within 40\arcsec, HD\,164492\,A is at the center of a wide multiple system.  Only components B and H are within our FOV. We detected another two faint companions at 3.1\arcsec\ and 6.5\arcsec\ and we  labeled the new pairs A,I and A,J. We further resolved HD\,164492\,A into a  $\rho=25$~\mas\ pair with a rather faint companion ($\Delta H=3.2$). The newly resolved pair, labeled Aa,Ab, is seen both by PIONIER and NACO/SAM although the  latter measurements have limited accuracy given the separation considered. The object was reported as RV variable by \citet{CLL77}, but this has not been confirmed by \citet{CHN12}.

\paragraph{HD\,164816}
A faint but clear companion  separated by 57~\mas\ (A,B) is detected  both with PIONIER and NACO/SAM. Our detection is not the known SB2 system ($P_\mathrm{orb}=3.8$\,d, $a\sin i = 16$~\Rsun) which is separated by $0.07$~\mas\ assuming a distance to the object of 1\,kpc \citep{Trepl:2012}. Moreover the SB2 has nearly equal masses while the resolved pair has $\Delta H=3.4$, pointing to quite different masses. The object is also detected in X-rays. \citet{Trepl:2012}  identified a soft X-ray excess and a 10~s pulsation of the X-ray source, which they interpret as the signature of a neutron star in the system. Our detection is probably an active later-type object, which may provide an alternative explanation to the X-ray excess.

\paragraph{HDE\,313846} (Fig.~\ref{fig: sam4})
We detect three faint companions at separations of 5.6\arcsec\ ($\theta=21$\degr), 5.6\arcsec\ ($\theta=186$\degr) and 7.8\arcsec\ in the NACO FOV (labeled C to E owing to an already assigned B companion at 35\arcsec).

\paragraph{HD\,165246} (Fig.~\ref{fig: sam2})
We detect a third companion to this  SB2 4.6\,d period eclipsing binary system \citep{Ote07, MHP13} and label the resolved system Aa,Ab. With $\rho=30 \pm 16$~mas, the precision of the NACO/SAM measurements is low as expected for a pair at the IWA limit. We unfortunately lack PIONIER observations that would provide a more accurate determination of the separation. The A,B pair reported by \citet{MGH98} is clearly seen in the NACO FOV, together with two additional faint companions at the edge of our FOV.

\begin{figure*}  \centering
  \includegraphics[width=0.38\textwidth]{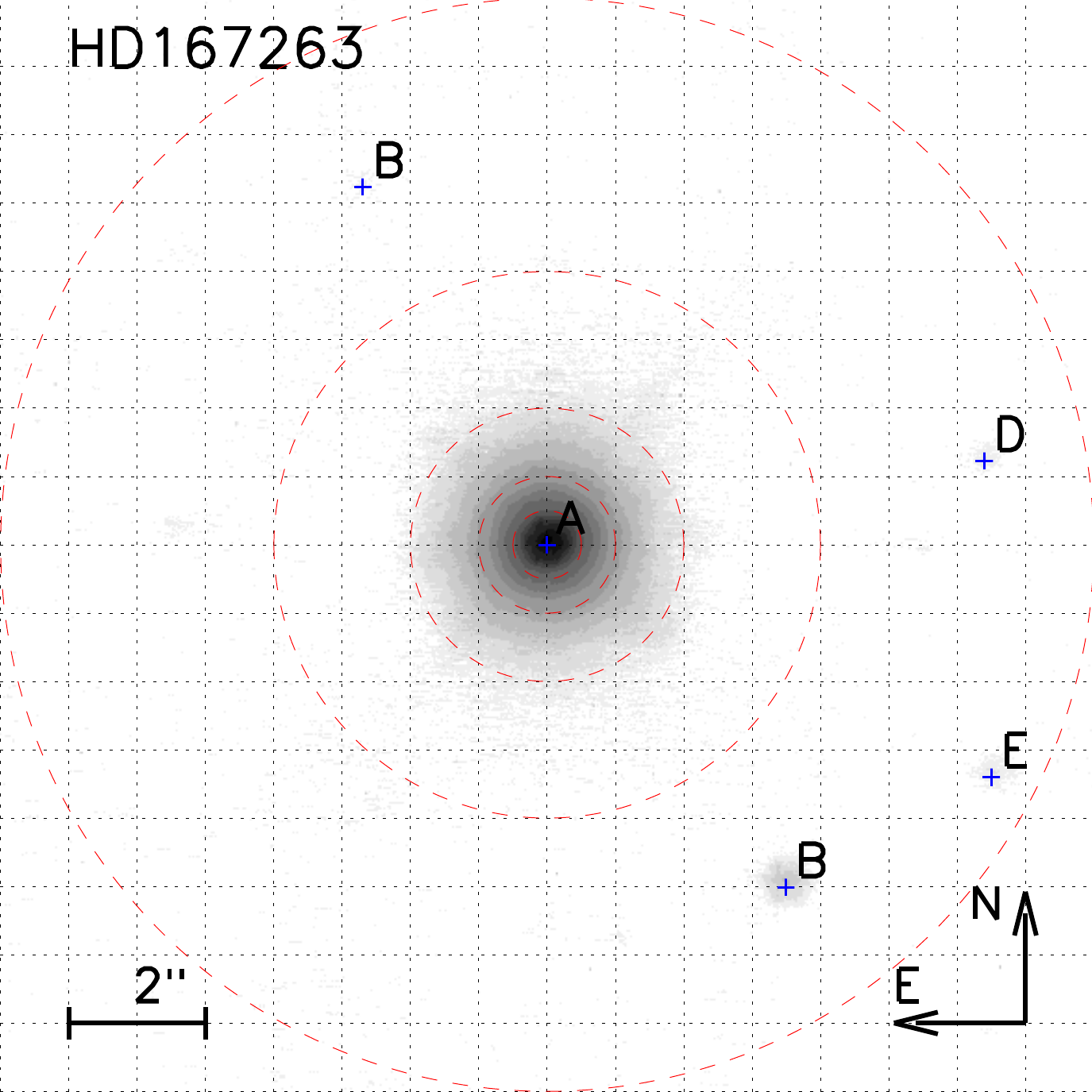} \hspace*{2mm}\vspace*{2mm}
  \includegraphics[width=0.38\textwidth]{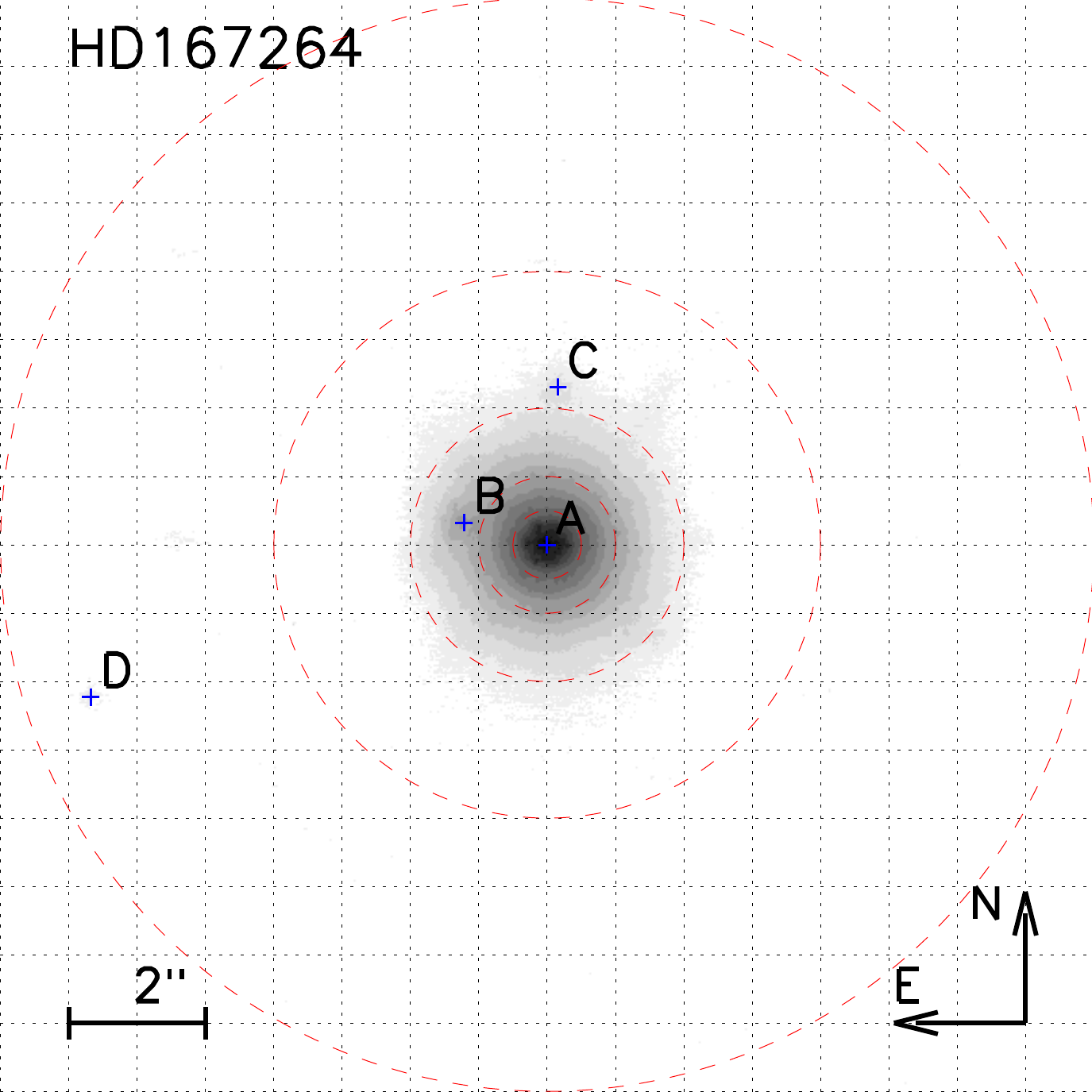} \hspace*{2mm}\vspace*{2mm}
  \includegraphics[width=0.38\textwidth]{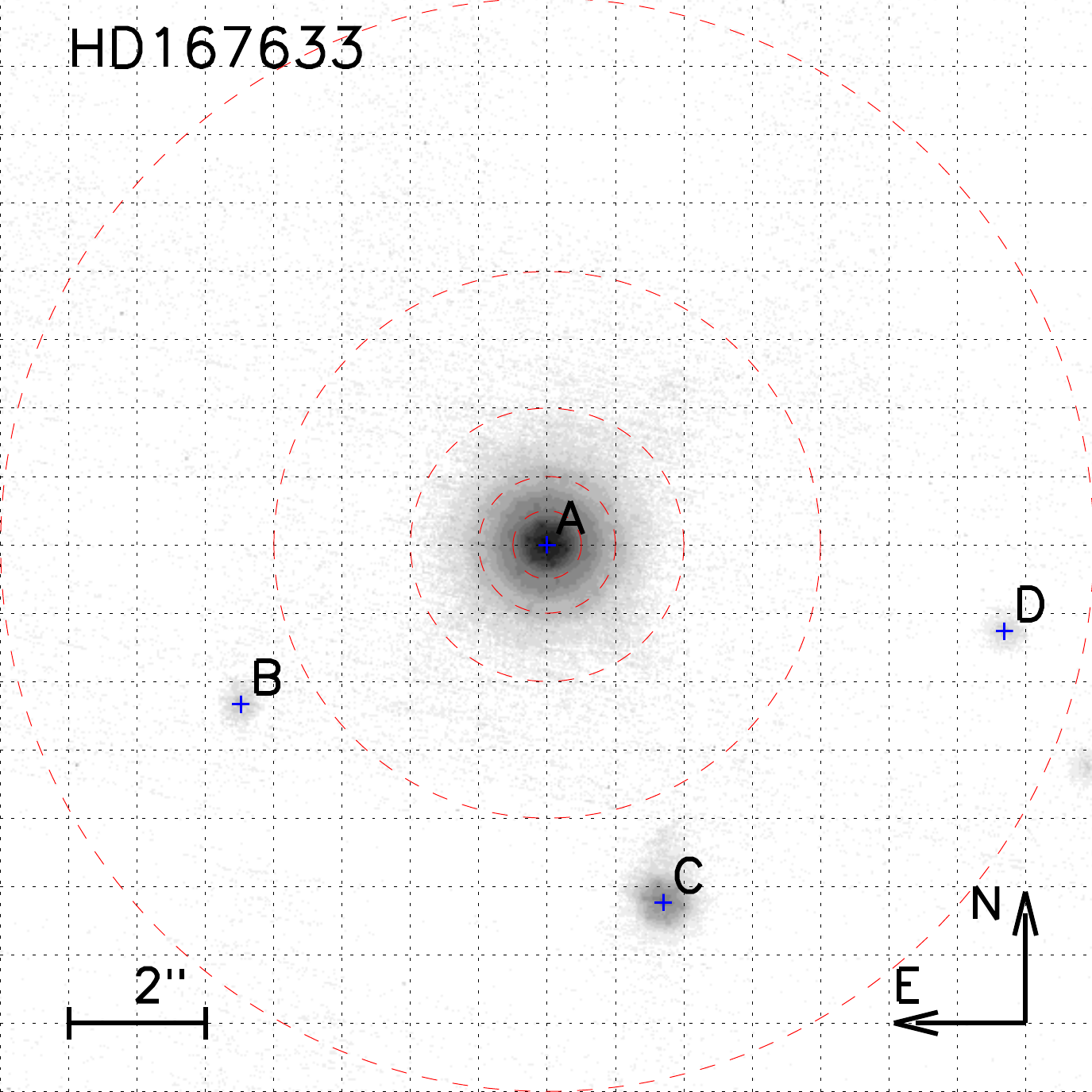} \hspace*{2mm}\vspace*{2mm}
  \includegraphics[width=0.38\textwidth]{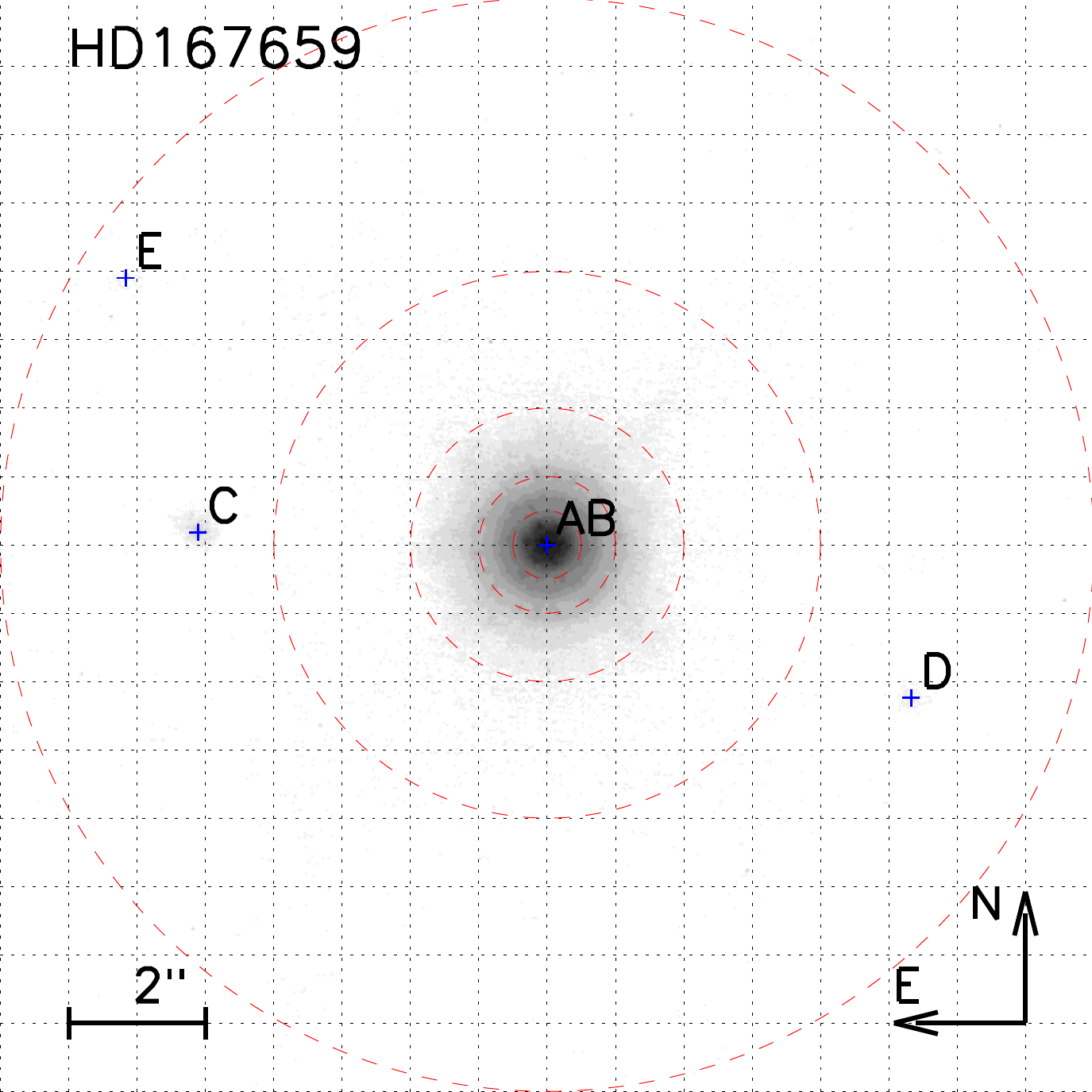} \hspace*{2mm}\vspace*{2mm}
  \includegraphics[width=0.38\textwidth]{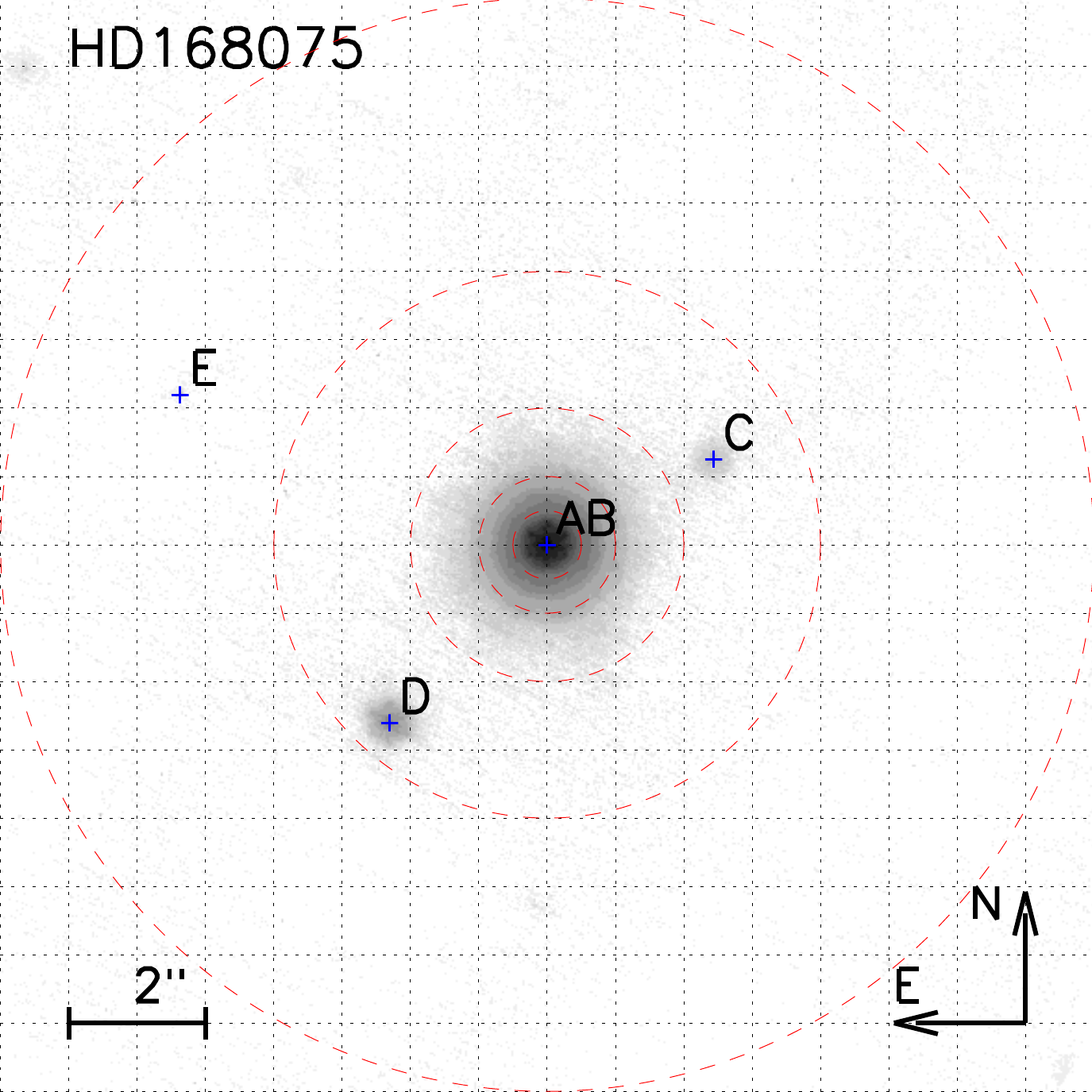} \hspace*{2mm}\vspace*{2mm}
  \includegraphics[width=0.38\textwidth]{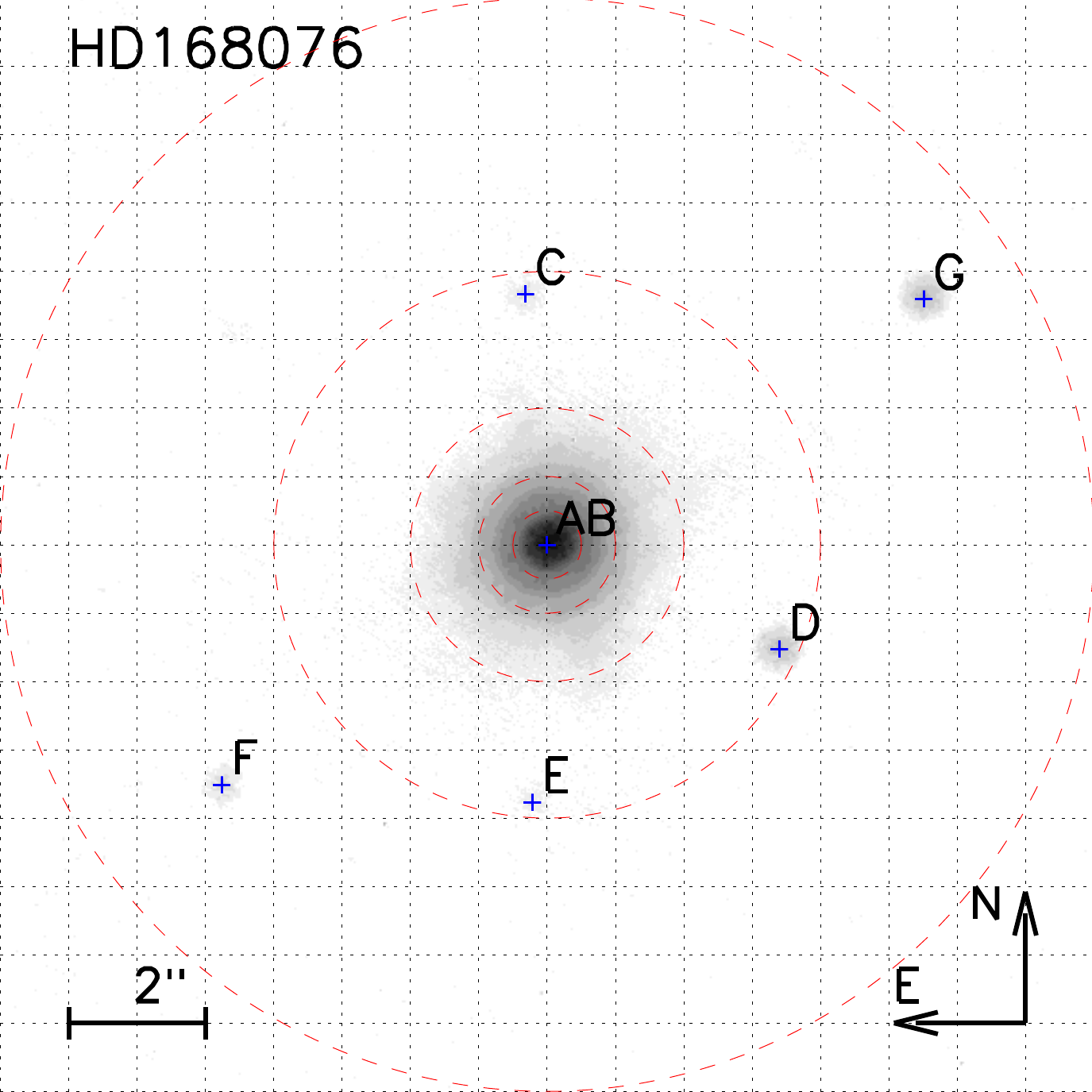} \hspace*{2mm}\vspace*{2mm}
  \caption{Same as Fig.~\ref{fig: sam1} for 
HD\,167263,
HD\,167264,
HD\,167633,
HD\,167659,
HD\,168075 and
HD\,168076.}
    \label{fig: sam3}
\end{figure*}

\begin{figure*}  \centering
\includegraphics[width=0.38\textwidth]{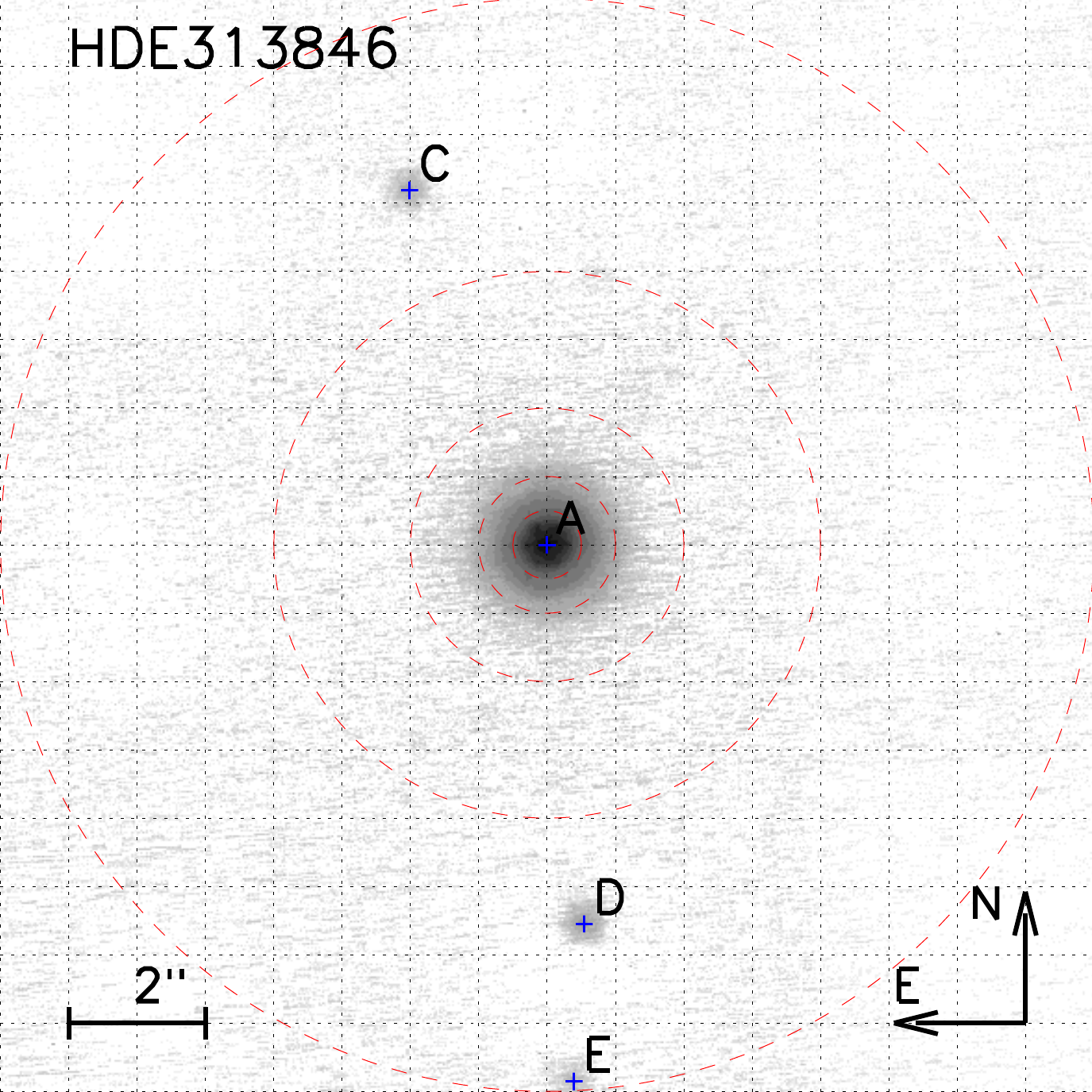} \hspace*{2mm}\vspace*{2mm}
\includegraphics[width=0.38\textwidth]{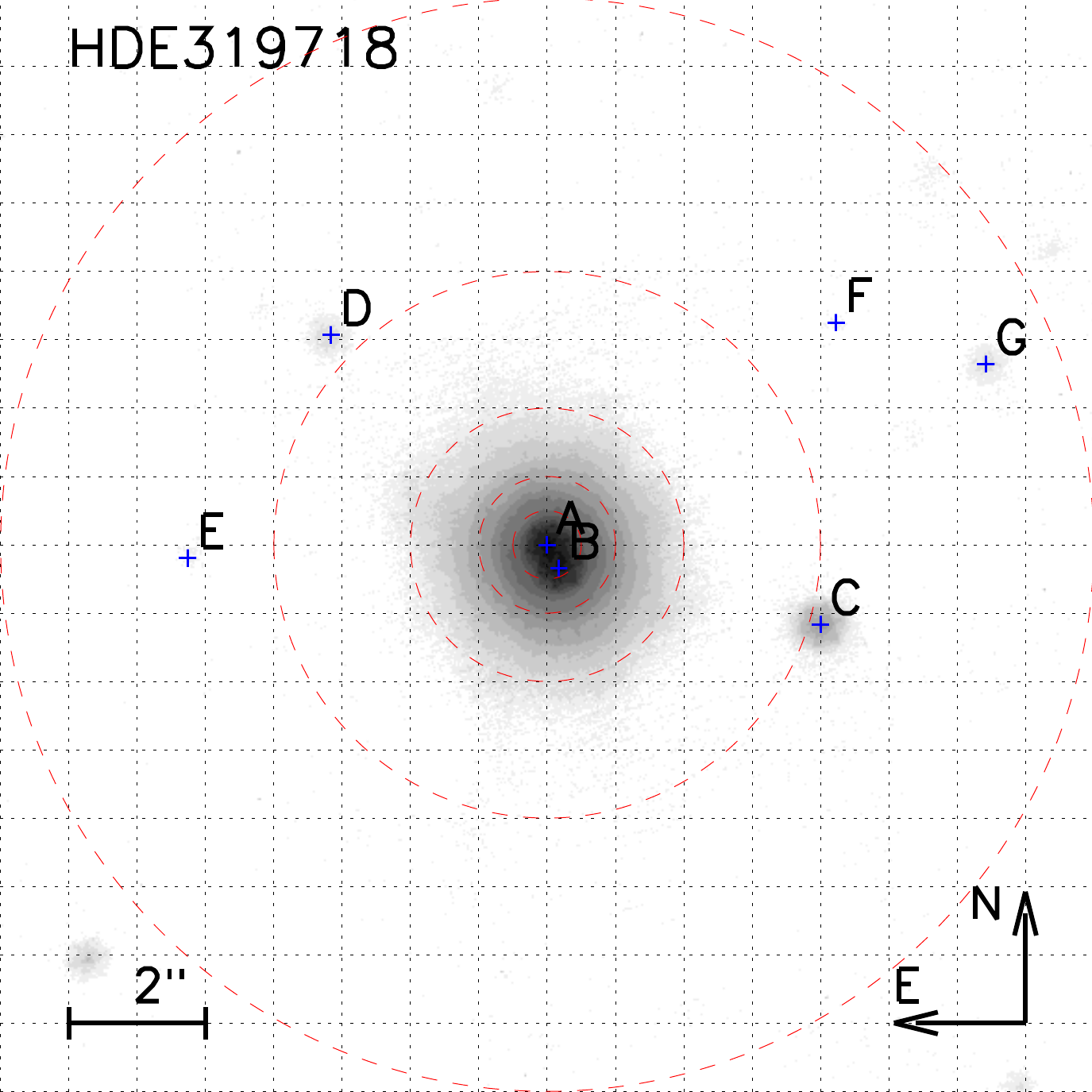} \hspace*{2mm}\vspace*{2mm}
\includegraphics[width=0.38\textwidth]{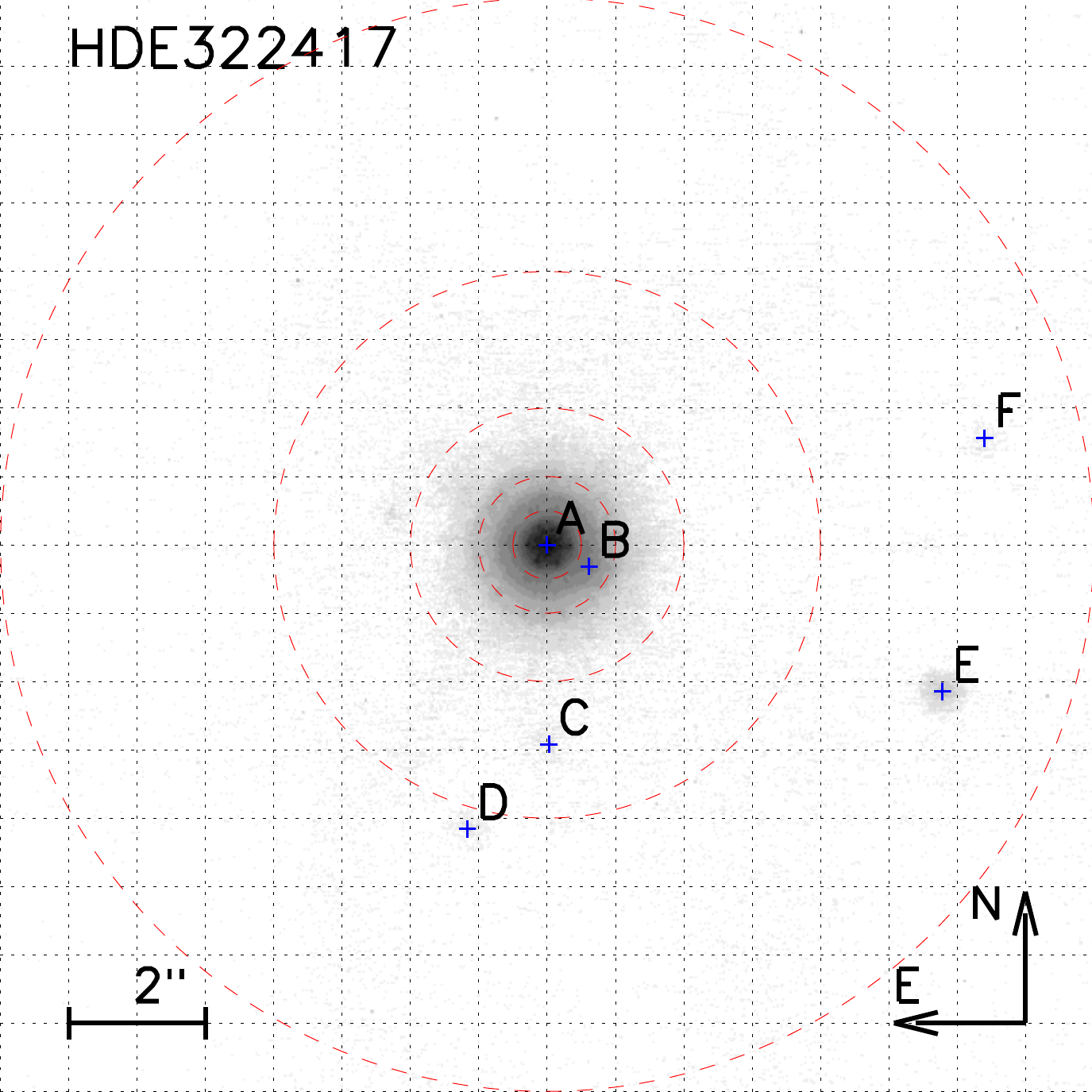} \hspace*{2mm}\vspace*{2mm}
\includegraphics[width=0.38\textwidth]{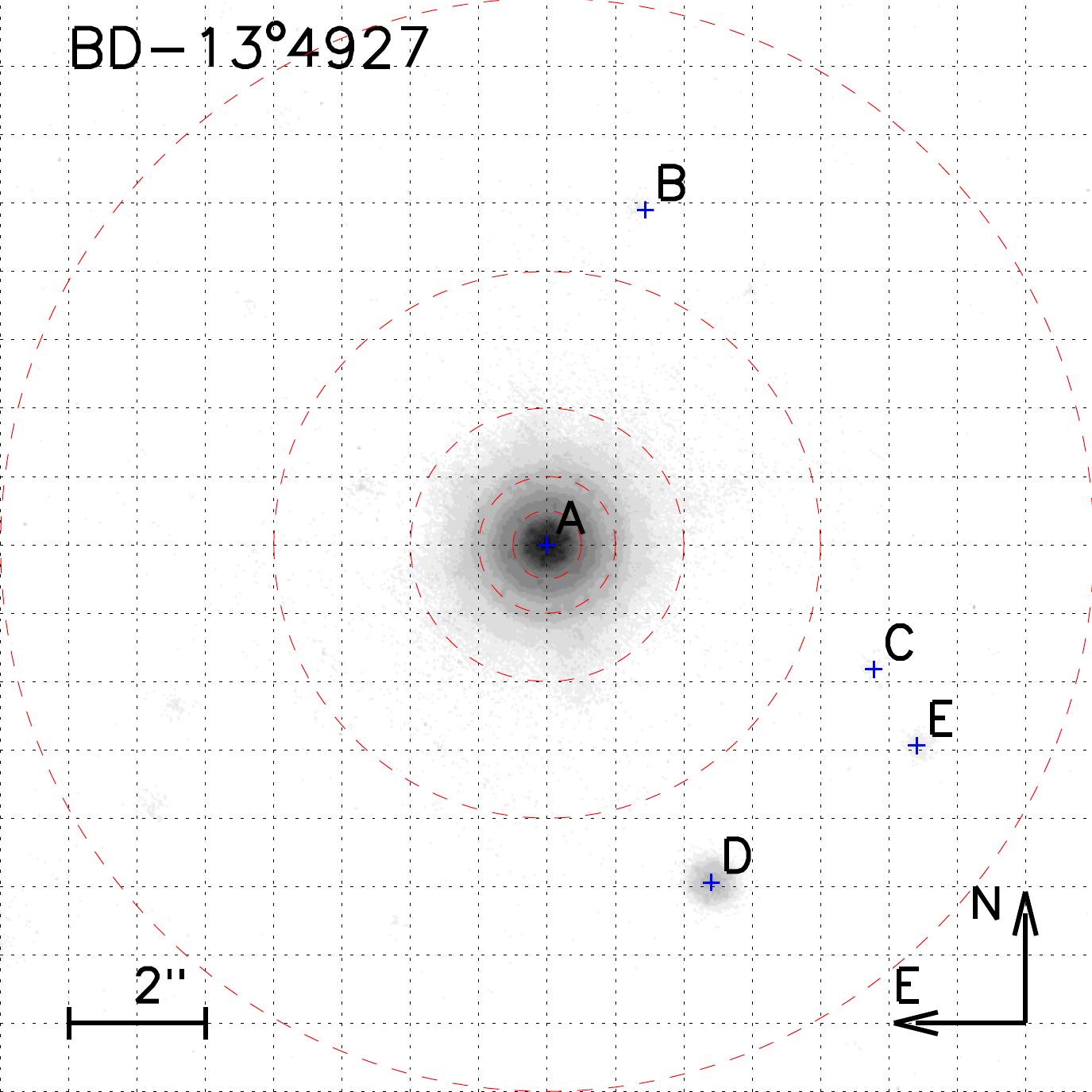} \hspace*{2mm}\vspace*{2mm}
  \caption{Same as Fig.~\ref{fig: sam1} for 
HDE\,313846,
HDE\,319718,
HDE\,322417 and
BD$-$13\degr4927.}
  \label{fig: sam4}
\end{figure*}

\paragraph{HD\,167633} (Fig.~\ref{fig: sam3})
We detect three previously unreported distant companions at 5.1\arcsec\ (A,B), 5.5\arcsec\ (A,C) and 6.8\arcsec\ (A,D) in the NACO FOV but we lack PIONIER observations to investigate the 1 to 30~mas regime. The objects was reported as SB1? by \citet{ALG72}, a fact not confirmed by \citet{CHN12} who prefer a RV stable classification.

\paragraph{HD\,167659} (Fig.~\ref{fig: sam3})
HD\,167659 is a known 17\arcsec\ pair (A,B), with the B companion outside the NACO FOV.  Both  NACO/SAM and PIONIER resolved the A component as a $\rho=50$~\mas\ pair (Aa,Ab) which probably corresponds to the 80~mas pair reported by \citet{MGH98} and detected through occultation. The object may further be an SB1 \citep{Gamen:2008}. Three additional, faint companions (labeled C to E by increasing separation) are seen in our NACO FOV at angular separations from 5.1\arcsec\ to 7.3\arcsec.

\paragraph{BD$-$11\degr4586}
A $\Delta H=4.3$ companion (A,B) is detected at 7.2\arcsec\ in this otherwise isolated O star.

\paragraph{HD\,168075} (Fig.~\ref{fig: sam3}) It is an SB2 binary with a 46\,d period \citep{SGE09, BGA10}. We detect a third companion at 44~mas with NACO/SAM and label the new pair A,B. The measurement however lacks accuracy and is possibly degenerate because  we could only observe the target in a single band. Three other fainter companions are seen in the NACO FOV with separations of 2.7\arcsec\ to 5.8\arcsec\ and are labeled C to E.

\paragraph{BD$-$13\degr4927} (Fig.~\ref{fig: sam4}) We detect four faint companions in the NACO FOV with separations from 5.1\arcsec\ to 6.2\arcsec. They are labeled B to E by increasing separation.

\paragraph{HD\,168112}  This object is a non-thermal radio emitter \citep{DBRB04}. PIONIER clearly resolved the object into a tight pair ($\rho = 3.3$~mas) with almost equal brightness companions ($\Delta H= 0.17\pm 0.19$). Two faint and more distant companions are further detected in the NACO FOV. We labeled the three newly discovered companions B, C and D by increasing separations.

\paragraph{HD\,171589} While we did not acquire NACO/SAM data, we clearly resolved the object with PIONIER as a $\rho\approx1.5$~\mas\ pair (A,B). No companion was reported at $\rho>30$~\mas\ by \citet{MHG09}. The  possible RV variability reported by \citet{CLL77} was not confirmed by \citet{CHN12}.

\subsection{Resolved spectroscopic companions}

\paragraph{HD\,54662} We resolved for the first time the long-period SB2 spectroscopic binary (A,B) unveiled by \citet{BGD07}. \citet{MHG09}  reported no companion at $\rho>30$~\mas\ as confirmed by our SAM measurements. 

\paragraph{HD\,75759} It is marginally resolved with $\rho \approx 0.5$~mas (A,B), although with a  large relative uncertainty given its angular separation is smaller than the PIONIER IWA. Given the distance of 947~pc \citep{SMAM14}, our detection probably corresponds to the known SB2 \citep[$P_\mathrm{orb}=33.1$\,d, $(a_1+a_2)\sin i=0.6\,$AU; ][]{Thackeray:1966}. No outer companion was known at $\rho>30$~\mas\ \citep{MHG09} as confirmed by our NACO data.

\paragraph{HD\,123590} It has a  $\rho=0.7$~\mas\ companion (A,B) marginally resolved  by PIONIER. We may have detected the $P_\mathrm{orb}=60$\,d SB1 reported by \citet{Gamen:2008} which has $a_\mathrm{app}=0.4$~\mas\ assuming $\pi=0.5$~\mas\ and $M=20$~\msun\ \citep{Hohle:2010}.  No outer companion at $\rho>30$\mas\ is detected in our NACO data.

\paragraph{$\delta$ Cir $\equiv$ HD\,135240} \citet{Penny:2001} performed a tomographic decomposition and found $\delta$ Cir to be a triple system, with an eclipsing inner pair (Aa,Ab; $P_\mathrm{orb}=3.9$\,d, $a\sin i=11.44$~\Rsun, $a_\mathrm{app}<0.1$~\mas) and a RV-stable third component (Ac). \citet{MHS14} established the hierarchical nature of the system, obtaining  a 1644\,d period for the outer system. PIONIER clearly resolves the outer system as a $\rho=3.78$~\mas\ pair.  We did not detect the $\Delta V=7.8$ B companion of \citet{MGH98} in the NACO FOV, but it may be just below our detection limit.

\paragraph{HD\,150135} PIONIER marginally resolves the Aa,Ab pair $\rho=0.95$~\mas. It probably corresponds to  the $P_\mathrm{orb}=183$\,d SB2 reported by \citet{Gamen:2008}, assuming $\pi=0.5$~\mas\ and $M=20$~\msun\ \citep{Hohle:2010}. We also report on the detection of a fainter companion at 4.3\arcsec\ (A,B).

\paragraph{HD\,150136} This is a hierarchical triple system known from spectroscopy \citep{MGS12}. The outer pair ((Aa+Ab)+Ac) was resolved for the first time in the course of this survey. The two first observations have been discussed by \citet{SLBM13}. We report here the observation of a third epoch at $\rho=6.9$~\mas. Other distant companions with separations from 1.6\arcsec\ to 20\arcsec\ were further reported in \citet{MGH98}. We clearly detect the 1.6\arcsec\ pair (A,B) in the NACO FOV but the other companions (C to F) are outside our field of investigation.

\paragraph{HD\,151003} This SB2 system (A,B) is resolved by PIONIER with $\rho=1.85$~\mas\ and $\Delta H=1.1$. The object was reported as RV variable by \citet{CLL77} and pre-publication OWN results indicate a 199\,d orbital period, which probably matches the resolved pair. A 4\arcsec\ faint companion (A,C) is further detected in the NACO FOV.

\paragraph{HD\,152233} Reported as HD\,152234 F in \citet{MGH98}, we resolved for the first time this long period SB2 binary discussed in \citet{SGN08,SdMdK12}. We label the resolved pair Fa,Fb.

\paragraph{HD\,152246} It is a long period 470-d hierarchical triple system \citep{CHN12} that PIONIER resolves with a separation of 3~mas (Aa,Ab). A combined spectroscopic and interferometric solution is presented in \citet{NCH14}. We further detected a faint 3.7\arcsec\ companion (A,B) in the NACO FOV.

\paragraph{HD\,152247} We resolve for the first time the long period SB2 binary (Aa,Ab) discussed in \citet{SdMdK12}. Faint ($\Delta K\mathrm{s} > 6$) companions at 3.1\arcsec\ (A,B) and 5.2\arcsec\ (A,C) are also detected in the NACO FOV.

\paragraph{HD\,152314} We resolve for the first time the 3700\,d period SB2 binary discussed in \citet{SGN08, SdMdK12} and label it Aa,Ab in this work. Two additional companions are detected in the NACO FOV with separations of 3.2\arcsec\  (A,B) and 3.5\arcsec\ (A,B) respectively

\paragraph{HDE\,322417} (Fig.~\ref{fig: sam4}) It is a 223\,d, long period SB1 system unveiled by OWN. PIONIER observations reveal a marginal detection ($2.5\sigma$) whose best fit corresponds to $\Delta H=4.3\pm1.8$  and $\rho\approx1.5$~mas (Aa,Ab). Both the large flux difference and the tight separation are compatible with the properties of the spectroscopic companion. Five faint companions, with separations from 0.7\arcsec\ to 6.6\arcsec, are further detected in the NACO FOV. We labeled them B to F by increasing separation.

\paragraph{HD\,164794 $\equiv$ 9~Sgr} PIONIER clearly resolves this long-period SB2 binary  (A,B) discussed by \citet{RSS12} at a separation of about 5~mas.

\paragraph{15 Sgr $\equiv$ HD\,167264} (Fig.~\ref{fig: sam3}) PIONIER resolves a closed pair (labeled Aa,Ab) at $\rho \approx 3$~\mas\ at three epochs, revealing evidence for the orbital motion. The newly resolved pair likely corresponds to the pre-publication 668\,d SB1 system obtained by OWN \citep{SMAM14}. The A,B pair at 1.27\arcsec\ with $\Delta y = 5.2$ \citep{TMH10} is also detected in the NACO FOV, together with two additional companions at 2.3\arcsec\ (A,C) and 7.0\arcsec\ (A,D).

\paragraph{HD\,167971} It is a known hierarchical triple system \citep{DBSA12}. Our observation represents the fifth epoch of the outer pair (Aa,Ab). The 4.7\arcsec\ companion  (A,B) reported by \citet{TBR08} is also seen in the NACO FOV.

\subsection{Previously resolved companions with $\rho < 200$~mas}

\paragraph{HD\,57061 $\equiv \tau$~CMa} The central object Aa is both an eclipsing binary \citep[$P_\mathrm{orb}\sim1.28$\,d]{vLvG97} and a longer period SB1 system \citep[$P_\mathrm{orb}\sim154.9$\,d]{StL98}. The latter authors suggested the eclipsing binary system to correspond to the unseen companion of the SB1 long-period binary, resulting in an hierarchical O9II+(B0.5V+B0.5V) triple system for the Aa component. $\tau$~CMa has two additional known components at $\rho \approx 0.12$\arcsec\ and 0.95\arcsec\  \citep[pairs Aa,Ab and Aa,E respectively;][]{MGH98, MHG09, TMH10}.   We observed the system twice with PIONIER and once with NACO/SAM, measuring  $\Delta H \approx 0.9$, $\rho \approx120$~\mas\ and $\theta\approx308$\degr. The PIONIER value for the position angle is not reliable and subject to a $\pm180$\degr\ uncertainty since the phase closure is not well fitted but the orientation can be constrained thanks to the NACO/SAM value. These measurements most probably correspond to the Aa,Ab pair reported by \citet{MHG09}: $\theta=125.2$\degr\, $\rho=128$~\mas, $\Delta V=0.4$. The SAM detection has a position angle that differs from the one reported by  \citet{MHG09} by 180\degr. Similarly, the NACO-FOV measurements for the 0.95\arcsec-separation Aa,E pair resulted in $\theta = 266$\degr, i.e.\ affected by  180\degr\ compared to the WDS value reported by \citet{MHG09} and the independent value of \citet{TMH10}.  This is in line with a recent footnote in \citet{SMAM14} reporting an independent observations by Aldoretta et al. (in prep.) and by AstraLux for pair Aa,E confirming the probable 180\degr\ offset in some of the position angle measurements listed in the WDS.

\paragraph{HD\,93129\,A and B} (Fig.~\ref{fig: sam1}) This is the closest O2~I star from Earth and has been observed once with PIONIER and three times with NACO/SAM. The Aa,Ab pair is well constrained  at $\rho \approx 30$~\mas\ and $\Delta H \approx 1.3$. Our detections most probably corresponds to the companion first resolved by the HST fine guider sensor \citep{NWW04}. The separation has decreased from 55~mas in 2004 to 43~mas in 2006 and to 27~mas in 2013, indicating that long baseline interferometry will be needed to pursue the monitoring of this extremely long-period system. The position angle value of $\theta=356$\degr\ obtained by \citep{NWW04} seems incompatible with the anti-clockwise rotation of the companion revealed by subsequent measurements ($\theta$ decreasing from 14 to 6\degr\ from 2006 to 2013). 

Two other companions have been reported, with respective separations of 2.8\arcsec\ (Aa,B) and 5\arcsec\ (Aa,C) \citep{MGH98, SMG10}, which we also resolved. The B companion is bright enough that an interferometric analysis of the SAM data can be performed but no close companions were found within the usual 5-mag contrast limit.  We further resolved three previously unreported companions at separations of 1.8\arcsec, 3.9\arcsec\ and 4.8\arcsec\ and  with $\Delta H$ of 6.8, 6.0 and 7.0, resulting in a total of six companions within 5\arcsec\ from the central star. We label the new components E to G by increasing separation. \citet{MGH98} mentioned a (B,D) pair with a separation of 3\arcsec. The location of the D component in the NACO FOV is unclear. The only detection close to B is component E, but (B,E) has a separation of about 2\arcsec, not 3\arcsec. To avoid confusion between misidentified components, we did not assign the D label to any of our detected stars.

\paragraph{HD\,152248} It is a 6\,d period SB2 colliding-wind system \citep{SRG01, SSG04}. \citet{MGH98, MHG09} reported on a $\Delta V=2$ companion at 50~mas, that could not be confirmed by Ladratta et al. (in prep.). We observed the system both with PIONIER and NACO and neither revealed a companion above our adopted significance threshold. The $P_1$ probability of the single star model is 0.92, hence excellent.  We concluded accordingly that a binary model is not needed to explain the data and reported HD\,152248 to be unresolved. Interestingly, the deepest (non-significant) minimum in the $\chi^2$-map of the PIONIER binary model is  at 50~mas but we note tens of  similar minima in the binary-model $\chi^2$-map. If this was still to represent a binary companion, it would have a flux of only 1.3\%\ of that of the central star (hence $\Delta H \sim 4.7$), in stark contrast with $\Delta V = 2$ reported by \citet{MHG09}. 

\paragraph{HD\,152723} It is resolved with  $\rho\approx100$~\mas\ and $\theta\approx 310\degr$ (Aa,Ab). The quality of the PIONIER fit is poor because the separation is larger than the \OWA{}. The companion is most probably the one reported by \citet{MHG09}: $\rho=98$~\mas, $\theta=125.6$\degr, $\Delta V =1.7$. The PIONIER and SAM position angle measurements yield  opposite values to that of \citet{MGH98, MHG09} but it is impossible to obtain a decent fit with a position angle compatible with the value Mason et al.\ value. As for HD\,57061, this suggests a probable $\pm$180\degr\ degeneracy in some of the WDS position angle values. The B, C and D components reported by \citet{MGH98} are outside our FOV. Finally, the object  is also reported as an 18.9\,d period SB1 system is early OWN results.

\paragraph{HD\,155889} It is a  $\rho\approx 190$~\mas\ pair (A,B). The quality of the PIONIER fit is poor because the separation is larger than the \OWA{}  but the SAM data are very good and confirm that our detection corresponds to the companion already reported by \citet{MHG09}. A faint companion at 7\arcsec\ (A,C) is further seen in the NACO FOV. The object is reported as SB2 (possibly SB3) by OWN.

\paragraph{HDE\,319703\,A} OWN indicates a 16.4\,d SB system. SAM resolved it as a 185~mas pair (A,B). A distant faint companion (A,C)  is also seen in the NACO FOV, indicating a total number of three companions for HDE\,319703. 

\paragraph{16 Sgr $\equiv$ HD\,167263\,A and B} (Fig.~\ref{fig: sam3}) The A component is a known pair (Aa,Ab) that we could resolve both with  PIONIER and NACO/SAM. The measured separation of $\rho\approx80$~\mas\ is slightly larger than the 2006 measurements of \citet{MHG09} ($\rho=70$~mas).  The PIONIER position angle is not well constrained (and affected by a $\pm180\degr$ degeneracy) since the phase closure is poorly fitted. However, NACO/SAM allows us to fix $\theta = 333$\degr, i.e.\ with a 180\degr\ offset compared to the  \citet{MHG09} observations ($\theta\approx 150\degr$). The much smaller magnitude difference between the Aa,Ab components in the H-band than in the $V$ and \Ks-band suggests that Ab is a rather red object. The known B component at 5.9\arcsec\ as well as three additional companions (C, D and E) with separations between 5.8\arcsec\ and 7.3\arcsec\ are all well detected in our NACO FOV.

\paragraph{HD\,168076\,A and B} (Fig.~\ref{fig: sam3}) We resolve the A,B pair at $\rho\approx 150$~\mas\ reported by \citet{MHG09} twice with PIONIER and once with NACO/SAM. The NACO/SAM measurements are the most reliable. The pair is visible as SB2 in spectroscopy, although no orbital motion was detected \citep{SGE09} in agreement with the large separation, hence very long period. Five other faint companions are detected in the NACO FOV with separations between 3.7\arcsec\ and 6.6\arcsec, which we labeled C to G by increasing magnitudes.

\subsection{Previously known companions with $\rho>200$~mas}

\paragraph{NX Vel = HD\,73882} It is a quadruple system formed by a known 0.65\arcsec\ pair (A,B) that we resolved in the NACO-FOV data and that contains an eclipsing system \citep[$P_\mathrm{orb}\sim2.9$]{Ote03} and a SB one \citep[$P_\mathrm{orb}\sim20.6$]{SMAM14}. The agreement with previous  measurements is excellent.

\paragraph{LM Vel $\equiv$ HD\,74194} This O8.5~Ib-II(f) star presents RV variations \citep{BGM06} and is a SB candidate. It has been suggested as the possible counterpart of the fast X-ray transient IGR~J08408-4503 \citep{MBB06}. We detect a $\Delta H$ and \Ks\ $=6$~mag companion at 4.5\arcsec\ (A,B) in the NACO FOV.

\paragraph{HD\,93161\,A and B} Both the A and B components are O-type stars, separated by 2\arcsec\ and are clearly resolved in the NACO- FOV. HD\,93161\,A is a SB2 system with a period of 8.6\,d \citep{NAS05}. HD\,93161\,B is noted as RV variable by the same authors and as SB1 by \citet{CHN12}. Both the A and B components are bright enough for an interferometric analysis of the SAM data, but no companion was found. The object was not observed with PIONIER.

\paragraph{HD\,93205\,A} Components A and B of HD\,93205 form a pair separated by almost 20\arcsec, the B companion being outside our FOV. HD\,93205 A itself is an eclipsing SB2 binary. \citet{SMAM14} reported on the previously unpublished companion (noted C in the present work) at 3.7\arcsec\ with $\Delta V=9.3$ and a $\theta = 272$\degr. We also detect it in the NACO-FOV data, and obtain $\rho=3.70\pm0.05$\arcsec, $\Delta H=5.8\pm0.1$, $\Delta K\mathrm{s}=5.3 \pm 0.2$ and $\theta = 270.4\pm1.3$\degr, in perfect agreement with the position reported by \citet{SMAM14}. This demonstrates the accuracy of the astrometry obtained for companions detected in the NACO FOV despite the unusual shape of the PSF. It also demonstrates the fact that it is easier to detect companion  in the NIR given the more favorable flux contrast resulting either from the color of the central O-type object or from the more limited extinction affecting background objects. 

\paragraph{HD\,101205}
This is another complicated multiple system with three visual pairs previously resolved at separations of 0.36\arcsec\ (A,B), 1.7\arcsec\ (AB,C) and 9.6\arcsec\ (AB,D). The outer pair falls outside our FOV, but we detected the B and C components in our NACO images. The A,B pair further contains an eclipsing binary \citep[$P_\mathrm{orb}=2.08$\,d]{Ote07} and a  spectroscopic binary with a period of 2.8\,d \citep{SJG11}, bringing the total number of stars within 10\arcsec\ to six. It is currently not possible to decide which component of the A,B pair is the eclipsing one and which is the spectroscopic one.

\paragraph{HD\,113904\,B $\equiv$ $\theta$~Mus B}
The $\theta$~Mus A,B pair is separated by $\approx$ 5.5\arcsec. The A component is a WR+O binary (WR48) and the B component is an O9~III star. While both components fall within the NACO FOV, HD\,113904 B was our prime target given the \smash\ focus on O stars.  $\theta$~Mus B is reported as SB by both \citet{CHN12} and \citet{SMAM14}. Unfortunately, the star was not observed with PIONIER. We however reported a new pair (B,C) with a separation of 3.45\arcsec\ and a magnitude difference of 5.5 in the $H$-band.   

\paragraph{HD\,114886\,A} (Fig.~\ref{fig: sam1})
is a high-order multiple system. The central pair, Aa,Ab, is separated by 0.24\arcsec\ \citep{MHG09, TMH10}, one of the components being a 13.6\,d period SB1 \citep[OWN,][]{SMAM14}. The B component, separated by 1.7\arcsec\ was already reported by \citep{MGH98}. We detected four other components in the NACO FOV -- labeled C to F in order of increasing separation --, yielding  total  of seven companions. 

\paragraph{HD\,135591} The 5.5\arcsec\ known A,B pair is clearly resolved in our NACO-FOV data.

\paragraph{HDE\,319718\,A and B $\equiv$ Pismis 24-1\,AB} (Fig.~\ref{fig: sam4}) This object was previously resolved by \citet{MaWM07} with a separation of 0.36\arcsec. \citet{BDCPN13} recently reported a 2.36\,d photometric period indicating that one of the components of the pair is an eclipsing binary system. The A and B components are clearly resolved by NACO/SAM. We further detected five additional faint companions (labeled C to G) in the NACO FOV. 

\section{Notes on supplementary targets}\label{sect: notes_supp}
\subsection{Newly resolved targets}

\paragraph{HD\,46150} Located in NGC~2244, HD\,46150 is a probable long period binary \citep{MNR09}. The WDS lists 15 companions (B to P), 11 of them (B to L) within 75\arcsec\ \citep{MGH98, MAp10}. Only companions B and C are within our field of investigation. We clearly detected the B component but not the C one. 
With $\Delta V \approx 5$ and $\Delta z \approx 6$, the C component magnitude may fall below our detection threshold in $H$ and \Ks. We however detected a new faint companion (A,Q) at 2.1\arcsec\ with $\Delta H=7.2$.

\paragraph{HD\,46202} This object is identified as HD\,46180 D in the WDS. A 3.7\arcsec\ companion to HD\,46202 (D,E) was identified by \citet{MAp10}, which we confirm. The companion is about 1~mag brighter in the NIR than in the $z$-band. We detect an additional star at 86~mas from  HD\,46202 (Da,Db) and with 1.9~\Ks-mag difference.

\paragraph{HD\,46966} This object is a 3.2\arcsec\ pair (A,B) with an extremely faint companion $\Delta I=10$ \citep{TBR08}, i.e.\ well below the detection limit of our NACO observations. We however resolve a relatively bright nearby companion. The new pair (Aa,Ab) has a separation of 50~mas and a magnitude difference of $\Delta H=1.1$.

\paragraph{V640~Mon $\equiv$ HD\,47129} 
Plaskett's star is a known SB2 system with $P_\mathrm{orb}\sim14.4$\,d \citep{LRM08} and the only O-type binary known with a magnetic star. In addition to the two known visual companions at 0.78\arcsec\ and 1.12\arcsec\ \citep{TBR08}, NACO/SAM resolved a new faint companion at 36~mas with $\Delta H \approx 4.0$, i.e.\ too faint to be confirmed by PIONIER. Uncertainties on the separation are large, calling for new measurements.

\paragraph{HD\,51533} It has five identified companions (B to F) in the WDS. With a separation of 2.6\arcsec, only the A,B pair falls within the NACO FOV. Beside companion B, we detected two new pairs: Aa,Ab with a separation of 0.6\arcsec\ and Aa,G at 2.9\arcsec.

\paragraph{HD\,76535} We detect a previously unreported $\Delta H=4.4$ companion at 2.8\arcsec.

\paragraph{HD\,93128} We detect two  companions: A,B with  $\Delta H=2.1$ and $\rho=6.6$\arcsec\ and A,C with $\Delta H=5.4$ and $\rho=3.7$\arcsec. A,C was previously unreported.

\paragraph{HD\,93190} We detect a previously unreported pair of companions at 4.2\arcsec, separated by a fraction of an arcsec. We labeled them Ba and Bb according to their brightness. Hence Bb ($\Delta H_\mathrm{A,Bb}=5.45$) is a couple of mas closer to A than Ba ($\Delta H_\mathrm{A,Ba}=5.31$).

\paragraph{HDE\,306097} We detect a previously unreported bright companion at 38~mas with $\Delta H=1.0$ (A,B).

\paragraph{HD\,100099} It is an 21.6\,d period SB2 system \citep{SJG11}. We detect an additional companion at 0.9\arcsec\ with $\Delta H= 4.2$. We label the new visual pair A,B.

\paragraph{HD\,100444} We detect a previously unreported companion at 3.9\arcsec\ with $\Delta H=3.55$(A,B).

\paragraph{HD\,101413} It is a 3 to 6~month-period SB2 system \citep{SJG11}, with the spectroscopic companion being a likely mid-B star. Our NACO/SAM data reveal a rather nearby companion (A,B) at 54~mas. This companion is too far away and too bright ($\Delta H =2.6$) to be associated with the spectroscopic companion, so that HD\,101413 is a likely hierarchical triple system. A third component (A,C) is detected in the NACO FOV at 1.8\arcsec.

\subsection{Resolved spectroscopic companions}

\paragraph{HD\,47839 $\equiv$ 15 Mon} With a period close to 25~yr \citep{GMB97}, HD\,47839 Aa,Ab is the prototypical O-type SB system that has been resolved by high resolution imaging techniques \citep{GMH93}. Given the long timescales involved, the exact orbit is still debated \citep{CvV10, TMH10, MAp10} with each new measurement adding its contribution to estimate the orbital motion of the companion. Our 2011.2 measurement indicates $\rho=108.5\pm3.5$~mas and $\theta=258\pm3$\degr. Our measured position is more in agreement with the  2008.8 and 2009.2 measurements of \citet{TMH10} than with the contemporaneous 2008.0 measurement of \citet{MAp10}. The 3.0\arcsec\ A,B pair reported by \citet{MGH98} is also detected in the NACO FOV.

\paragraph{HD\,152234} It is a 125\,d period SB2 system \citep{SdMdK12} that we label Aa,Ab. The spectroscopic companion is marginally resolved in our PIONIER observations with $\rho=0.9 \pm 1.9$~mas and an magnitude difference of $\Delta H=1.37$. HD\,152234 has two more distant companions (A,B and A,C) at 0.5\arcsec\ and 5.5\arcsec\ \citep{MGH98}. Unfortunately, we are lacking NACO data for this system so that we cannot confirm their presence.

\paragraph{HD\,168137} It was resolved as a 2\arcsec\ pair (A,B) by Hipparcos (WDS) but we lack NACO observation for this system. HD\,168137A itself is an O7~V + O8~V 912\,d period SB system \citep[Aa,Ab][]{SdMdK12} that we marginally resolve with PIONIER with a 6~mas separation. 

\clearpage

\newpage

\bibliographystyle{apj}
\bibliography{/home/hsana/Dropbox/literature}

\clearpage

\begin{deluxetable}{cclllccllll}
\tabletypesize{\scriptsize}
\tablewidth{0pt}
\tablecolumns{11}
\tablecaption{\label{tab: targets}\smash\ survey main target list}
\tablehead{\multicolumn{2}{c}{Instrum.}& \multicolumn{2}{c}{\hspace*{3mm} Object}              &\colhead{Sp.\ Type} &\colhead{R.A.} &\colhead{DEC.} &\colhead{$H$}&\colhead{\Ks}&\colhead{$V$\tablenotemark{a}}\\
            \colhead{PIO}&\colhead{SAM} &\colhead{HD/BD/CPD}& \colhead{Name}&\colhead{}          &\colhead{hh:mm:ss.sss}&\colhead{dd:am:as.ss}}
\startdata
y      &       y  &           HD\,52266 &               \nodata &   O9.5    III & 07:00:21.077&$-$05:49:35.95&  7.237&  7.265 & 7.213\\
y      &       y  &           HD\,53975 &               HR 2679 &   O7.5      V & 07:06:35.964&$-$12:23:38.23&  6.814&  6.819 & 6.473\\
y      &       y  &           HD\,54662 &               HR 2694 &     O7      V & 07:09:20.249&$-$10:20:47.64&  6.172&  6.206 & 6.212\\
y      &       y  &           HD\,55879 &               HR 2739 &   O9.7    III & 07:14:28.253&$-$10:18:58.50&  6.452&  6.502 & 6.018\\
y      &       y  &           HD\,57060 &                29 CMa &     O7     Ia & 07:18:40.378&$-$24:33:31.32&  5.190&  5.146 & 4.966\\
y      &       y  &           HD\,57061 &            $\tau$ CMa &     O9     II & 07:18:42.487&$-$24:57:15.78&  4.769&  4.786 & 4.390\\
y      &       y  &           HD\,68450 &               \nodata &   O9.7     II & 08:11:01.683&$-$37:17:32.55&  6.444&  6.474 & 6.442\\
y      &       y  &           HD\,71304 &               \nodata &     O9     II & 08:24:55.790&$-$44:18:03.01&  7.021&  6.927 & 8.187\\
\enddata
\tablenotetext{a}{Values in italic are taken from SIMBAD (\anchor{http://simbad.u-strasbg.fr}{http://simbad.u-strasbg.fr}).}
\end{deluxetable}

\newpage

\begin{deluxetable}{cclllccllll}
\tabletypesize{\scriptsize}
\tablewidth{0pt}
\tablecolumns{11}
\tablecaption{\label{tab: rw}\smash\ survey runaway target list}
\tablehead{\multicolumn{2}{c}{Instrum.}& \multicolumn{2}{c}{\hspace*{3mm} Object}              &\colhead{Sp.\ Type} &\colhead{R.A.} &\colhead{DEC.} &\colhead{$H$}&\colhead{\Ks}&\colhead{$V$}\\
            \colhead{PIO}&\colhead{SAM} &\colhead{HD/BD/CPD}& \colhead{Name}&\colhead{}          &\colhead{hh:mm:ss.sss}&\colhead{dd:am:as.ss}}
\startdata 
y      & y        &            HD\,57682 &               \nodata &   O9.5     IV & 07:22:02.053&$-$08:58:45.77&  6.966&  6.939 & 6.417\\
--     & y        &HD\,60848\tablenotemark{a} &           BN Gem &    O8:     V: & 07:37:05.731&$+$16:54:15.29&  7.071&  6.965 & 6.850\\
y      & y        &            HD\,66811 &           $\zeta$ Pup &     O4      I & 08:03:35.047&$-$40:00:11.33&  2.955&  2.968 & 2.249\\
y      & y        &            HD\,75222 &               \nodata &   O9.7    Iab & 08:47:25.137&$-$36:45:02.68&  6.493&  6.403 & 7.415\\
--     & y        &           HD\,105056 &                GS Mus &  ON9.7     Ia & 12:05:49.879&$-$69:34:23.00&  7.136&  7.051 & 7.437\\
y      & --       &           HD\,148546 &               \nodata &     O9    Iab & 16:30:23.312&$-$37:58:21.15&  6.901&  6.811 & 7.711\\
y      & y        &           HD\,149757 &           $\zeta$ Oph &   O9.5     IV & 16:37:09.530&$-$10:34:01.75&  2.667&  2.684 & 2.565\\
y      & y        &           HD\,153919 &              V884 Sco &     O6     Ia & 17:03:56.773&$-$37:50:38.91&  5.639&  5.496 & 6.546\\
\enddata
\tablenotetext{a}{Not listed as part of the main GOSC catalogue but known runaway within our magnitude limits.}
\end{deluxetable}

\newpage

\begin{deluxetable}{cclllccllll}
\tabletypesize{\scriptsize}
\tablewidth{0pt}
\tablecolumns{11}
\tablecaption{\label{tab: sup}\smash\ survey supplementary target list}
\tablehead{\multicolumn{2}{c}{Instrum.}& \multicolumn{2}{c}{\hspace*{3mm} Object}              &\colhead{Sp.\ Type} &\colhead{R.A.} &\colhead{DEC.} &\colhead{$H$}&\colhead{\Ks}&\colhead{$V$\tablenotemark{a}}\\
            \colhead{PIO}&\colhead{SAM} &\colhead{HD/BD/CPD}& \colhead{Name}&\colhead{}          &\colhead{hh:mm:ss.sss}&\colhead{dd:am:as.ss}}
\startdata 
--     &       y  &        HD\,46056\,A &               \nodata &     O8      V & 06:31:20.862&$+$04:50:03.85&  7.835&  7.820 & 8.245 \\
--     &       y  &           HD\,46149 &               \nodata &   O8.5      V & 06:31:52.533&$+$05:01:59.19&  7.251&  7.251 & 7.601  \\
--     &       y  &           HD\,46150 &               \nodata &     O5      V & 06:31:55.519&$+$04:56:34.27&  6.470&  6.436 & 6.739  \\
--     &       y  &           HD\,46223 &               \nodata &     O4      V & 06:32:09.306&$+$04:49:24.73&  6.703&  6.676 & 7.262  \\
--     &       y  &           HD\,46202 &               \nodata &   O9.5      V & 06:32:10.471&$+$04:57:59.79&  7.779&  7.720 & 8.182 \\
--     &       y  &           HD\,46485 &               \nodata &     O7      V & 06:33:50.957&$+$04:31:31.61&  7.511&  7.446 & 8.243  \\
--     &       y  &           HD\,46573 &               \nodata &     O7      V & 06:34:23.568&$+$02:32:02.94&  7.167&  7.128 & 7.933  \\
--     &       y  &           HD\,46966 &               \nodata &   O8.5     IV & 06:36:25.887&$+$06:04:59.47&  6.970&  7.018 & 6.876  \\
\enddata
\tablenotetext{a}{Values in italic are taken from SIMBAD (\anchor{http://simbad.u-strasbg.fr}{http://simbad.u-strasbg.fr}).}
\end{deluxetable}

\newpage

\begin{deluxetable}{ccc}
\tablewidth{0pt}
\tablecolumns{3}
\tablecaption{\label{tab: campaign}\small{The \smash\ observational campaign}}
\tablehead{\colhead{Instrument}& \colhead{Epoch} & \colhead{Nbr. of nights}}
\startdata
NACO/SAM & Mar 2011 & 3  \\ 
NACO/SAM & Feb 2012 & 3  \\ 
NACO/SAM & Jun 2012 & 3  \\ 
NACO/SAM & Jan 2013 & 2  \\ 
NACO/SAM & Jul 2013 & 0.8  \\ 
VLTI/PIONIER & Jun 2012 & 5  \\
VLTI/PIONIER & Aug 2012 & 2.5   \\
VLTI/PIONIER & Sep 2012 & 2.5  \\
VLTI/PIONIER & Nov 2012 & 2.5  \\
VLTI/PIONIER & Jan 2013 & 6  \\
VLTI/PIONIER & Mar 2013 & 2  \\
\enddata
\end{deluxetable}

\newpage

\begin{deluxetable}{lcrrrrrrr}
\tabletypesize{\scriptsize}
\tablewidth{0pt}
\tablecolumns{9}
\tablecaption{\label{tab: pio_detect} \smash\ companion detections with PIONIER and NACO/SAM}
\tablehead{\colhead{Target} &\colhead{Pair} & \colhead{Instr.} & \colhead{Obser. epoch} & \colhead{$\theta$} & \colhead{$\rho$} &\colhead{$\Delta H$}&\colhead{$\Delta K\mathrm{s}$}& \colhead{$\Delta L'$} \\
                \colhead{}  & \colhead{}    & \colhead{}       & \colhead{(b.y.)}       & \colhead{ (\degr)} & \colhead{ (mas)} &\colhead{ }& \colhead{} }
\startdata
           HD\,54662 &       A-B  &   PIO &            2012.9073 &  127.09 $\pm$   7.23 &     2.60 $\pm$     0.24 &  0.23 $\pm$  0.04 &           \nodata &           \nodata \\
\vspace*{-1mm}\\
           HD\,57061 &     Aa-Ab  &   PIO &            2013.0575 &  142.68 $\pm$   0.35 &   120.62 $\pm$     0.44 &  0.69 $\pm$  0.33 &           \nodata &           \nodata \\
             \nodata &     Aa-Ab  &   PIO &            2013.0603 &  313.08 $\pm$   0.42 &   115.17 $\pm$     0.53 &  0.55 $\pm$  0.19 &           \nodata &           \nodata \\
             \nodata &     Aa-Ab  &   SAM &            2013.0848 &  307.80 $\pm$   1.76 &   114.02 $\pm$     2.02 &  1.07 $\pm$  0.11 &  0.89 $\pm$  0.10 &           \nodata \\
\vspace*{-1mm}\\
           HD\,75759 &       A-B  &   PIO &            2012.4437 &   66.31 $\pm$  70.30 &     0.63 $\pm$     1.17 &  1.25 $\pm$  1.75 &           \nodata &           \nodata \\
             \nodata &       A-B  &   PIO &            2013.0604 &   35.52 $\pm$  62.53 &     0.39 $\pm$     0.58 &  0.20 $\pm$  1.75 &           \nodata &           \nodata \\
\enddata
\end{deluxetable}

\begin{deluxetable}{lcrrrrrr}
\tabletypesize{\scriptsize}
\tablewidth{0pt}
\tablecolumns{8}
\tablecaption{\label{tab: naco_detect} \smash\ companion detections in the NACO FOV}
\tablehead{\colhead{Target} &\colhead{Pair} & \colhead{Obser. epoch} & \colhead{$\theta$} & \colhead{$\rho$} &\colhead{$\Delta H$}&\colhead{$\Delta K\mathrm{s}$}& \colhead{$P_\mathrm{spur}$}\\
                \colhead{}  & \colhead{}    &  \colhead{(b.y.)}     & \colhead{ (\degr)} & \colhead{ (\arcsec)} &\colhead{ }& \colhead{} }
\startdata
           HD\,57061 &      Aa-E  &            2013.0848 &  265.81 $\pm$   2.21 &     0.95 $\pm$     0.03 &  4.47 $\pm$  0.21 &  4.39 $\pm$  0.14 &   0.000  \\
\vspace*{-1mm}\\
           HD\,73882 &       A-B  &            2012.1238 &  254.26 $\pm$   3.02 &     0.68 $\pm$     0.03 &  1.19 $\pm$  0.36 &  1.13 $\pm$  0.35 &   0.000  \\
\vspace*{-1mm}\\
           HD\,74194 &       A-B  &            2013.0822 &  178.62 $\pm$   1.71 &     4.51 $\pm$     0.11 &  6.63 $\pm$  0.11 &  6.58 $\pm$  0.16 &   0.017  \\
\vspace*{-1mm}\\
   CPD$-$47\degr2963 &       A-C  &            2013.0824 &  109.99 $\pm$   1.32 &     5.22 $\pm$     0.08 &  6.95 $\pm$  0.19 &  6.81 $\pm$  0.15 &   0.066  \\
\vspace*{-1mm}\\
\enddata
\end{deluxetable}

\begin{deluxetable}{rlrrrrrr}
\tabletypesize{\scriptsize}
\tablewidth{0pt}
\tablecolumns{7}
\tablecaption{\label{tab: fraction} Multiplicity and companion fractions in our main sample (96 stars)}
\tablehead{
\colhead{Category} &\colhead{Notation}  &\multicolumn{6}{c}{Luminosity classes} \\
\colhead{}         &\colhead{}          &\colhead{V} &  \colhead{IV} &  \colhead{III} &  \colhead{II} &  \colhead{I} &  \colhead{V-I} 
}

\startdata
\\
 \multicolumn{8}{c}{Fraction of multiple systems ($f_\mathrm{m}$)} \\
\\        
    Unresolved E/SB         & $f_\mathrm{m}^\mathrm{ES}$       & $0.57_{-0.10}^{+0.10}$ & $0.45_{-0.18}^{+0.18}$ & $0.48_{-0.10}^{+0.10}$ & $0.33_{-0.11}^{+0.11}$ & $0.48_{-0.10}^{+0.10}$ & $0.49_{-0.05}^{+0.05}$ \\
\vspace*{-2mm}\\
    $1 < \rho < 200$~mas    & $f_\mathrm{m}^\mathrm{1-200mas}$ & $0.76_{-0.10}^{+0.10}$ & $0.55_{-0.18}^{+0.18}$ & $0.52_{-0.10}^{+0.10}$ & $0.33_{-0.11}^{+0.11}$ & $0.41_{-0.10}^{+0.10}$ & $0.53_{-0.05}^{+0.05}$ \\
\vspace*{-2mm}\\
    $200 < \rho < 8000$~mas & $f_\mathrm{m}^\mathrm{0.2-8as}$  & $0.52_{-0.10}^{+0.10}$ & $0.64_{-0.18}^{+0.18}$ & $0.62_{-0.10}^{+0.10}$ & $0.44_{-0.11}^{+0.11}$ & $0.52_{-0.10}^{+0.10}$ & $0.57_{-0.05}^{+0.05}$ \\
\enddata
\tablenotetext{a}{Statistically corrected for spurious detections due to chance alignment. Unresolved E/SB systems and systems with $\rho < 200$~mas are unaffected by this correction.}
\end{deluxetable}
\end{document}